\documentclass[%
 reprint,
superscriptaddress,
nofootinbib,
 amsmath,amssymb,
 aps,prd
floatfix,
]{revtex4-2}

\usepackage[symbol]{footmisc}  % Use symbols instead of numbers/letters for footnotes
\usepackage{hyperref}
\usepackage[T1]{fontenc}
\usepackage{lineno}
\usepackage{xcolor}
\usepackage{color, colortbl}
\usepackage{graphicx}
\usepackage{dcolumn}
\usepackage{bm}
\usepackage{upgreek}
\usepackage{txfonts} % for \muup
\usepackage{textcomp}
\usepackage{adjustbox}
\usepackage{multirow}
\usepackage{makecell}
\catcode`\|=12\relax % added  <<<<<<<<<<<<<<<<<
\usepackage{orcidlink} % added <<<<<<<<<<<<

\usepackage{rotating} % For sidewaystable
\usepackage{booktabs} % For professional table formatting

\usepackage{lipsum}  

\hypersetup{colorlinks=true, citecolor=blue, urlcolor=blue, linkcolor=blue}

\definecolor{Gray}{gray}{0.9}
\definecolor{Red}{rgb}{1,0.9,0.9}

\begin{document}
% \begin{CJK}{UTF8}{mj}

\preprint{APS/123-QED}

\title{Signal processing and spectral modeling for the BeEST experiment}

\newcommand{\llnl}{Lawrence Livermore National Laboratory, 7000 East Ave, Livermore, CA 94550, USA} % [6] 
\newcommand{\triumf}{TRIUMF, 4004 Wesbrook Mall, Vancouver, BC V6T 2A3, Canada} % [3]
\newcommand{\nova}{LIBPhys-UNL, Departamento de F\'{i}sica, Faculdade de Ci\^{e}ncias e Tecnologia, NOVA FCT, Universidade Nova de Lisboa, 2829-516 Caparica, Portugal} %[4]
\newcommand{\umr}{Universit\'e de Strasbourg, CNRS, Institut de Physique et Chimie des Mat\'eriaux de Strasbourg,UMR 7504, F-67000 Strasbourg, France} % 5
\newcommand{\mines}{Department of Physics, Colorado School of Mines, 1500 Illinois St, Golden, 80401, Colorado, USA} % 1
\newcommand{\starc}{STAR Cryoelectonics LLC, Santa Fe, NM 87508, USA} % 7 
\newcommand{\sifmines}{Shared Instrumentation Facility, Colorado School of Mines, Golden, CO 80401, USA.} % 8
\newcommand{\xia}{XIA LLC, Oakland, CA 94601, USA} % 9 
\newcommand{\caen}{LPC Caen, ENSICAEN, Université de Caen, CNRS/IN2P3, Caen, France} % 10
\newcommand{\frib}{Facility for Rare Isotope Beams, Michigan State University, 640 S Shaw Lane, East Lansing, 48824, Michigan, USA} % [2]
\newcommand{\lnhb}{Universit\'e Paris-Saclay, CEA, List, Laboratoire National Henri Becquerel (LNE-LNHB), F-91120, Palaiseau, France}
\newcommand{\pnnl}{Pacific Northwest National Laboratory, Richland, WA 99354, USA}
\newcommand{\mcmaster}{Department of Physics and Astronomy, McMaster University,
Hamilton, Ontario L8S 4M1, Canada}%[11]

\author{Inwook~Kim \orcidlink{0000-0002-8394-6613}}~\email[Contact author: ]{kim124@llnl.gov} \affiliation{\llnl} 
% \author{Inwook~Kim~(김인욱) \orcidlink{0000-0002-8394-6613}}~\email{kim124@llnl.gov} \affiliation{\llnl} 
\author{Connor~Bray \orcidlink{0000-0002-7109-3326}}\affiliation{\mines}\affiliation{\llnl}
\author{Andrew~Marino}\affiliation{\mines}\affiliation{\llnl}
\author{Caitlyn~Stone-Whitehead}\affiliation{\mines}
\author{Amii~Lamm}\affiliation{\mines}
\author{Ryan~Abells}\affiliation{\triumf}
\author{Pedro~Amaro~\orcidlink{0000-0002-5257-6728}}\affiliation{\nova}
\author{Adrien~Andoche~\orcidlink{0000-0002-5910-4205}}\affiliation{\umr}
\author{Robin~Cantor}\affiliation{\starc}
\author{David~Diercks~\orcidlink{0000-0002-5138-0168}}\affiliation{\sifmines}
\author{Spencer~Fretwell~\orcidlink{0000-0002-0640-4853}}\affiliation{\mines}
\author{Abigail~Gillespie}\affiliation{\mines}
\author{Mauro~Guerra~\orcidlink{0000-0001-6286-4048}}\affiliation{\nova}
\author{Ad~Hall}\affiliation{\starc}
\author{Cameron~N.~Harris}\affiliation{\mines}
\author{Jackson~T.~Harris}\affiliation{\xia}
\author{Calvin~Hinkle~\orcidlink{0009-0008-0597-2470}}\affiliation{\mines}
\author{Leendert~M.~Hayen~\orcidlink{0000-0002-9471-0964}}\affiliation{\caen}
\author{Paul-Antoine~Hervieux~\orcidlink{0000-0002-4965-9709}}\affiliation{\umr}
\author{Geon-Bo~Kim~\orcidlink{0000-0002-0523-5496}}\affiliation{\llnl}
% \author{Geon-Bo~Kim~(김건보)}\affiliation{\llnl}
\author{Kyle~G.~Leach~\orcidlink{0000-0002-4751-1698}}\affiliation{\mines}\affiliation{\frib}
\author{Annika~Lennarz~\orcidlink{0000-0002-8138-0126}}\affiliation{\triumf}\affiliation{\mcmaster}
\author{Vincenzo~Lordi~\orcidlink{0000-0003-2415-4656}}\affiliation{\llnl}
\author{Jorge~Machado \orcidlink{0000-0002-0383-4882}}\affiliation{\nova}
\author{David~McKeen}\affiliation{\triumf}
\author{Xavier~Mougeot~\orcidlink{0000-0001-6161-8208}}\affiliation{\lnhb}
\author{Francisco~Ponce~\orcidlink{0000-0002-9655-8994}}\affiliation{\pnnl}
\author{Chris~Ruiz~\orcidlink{0000-0001-6468-6097}}\affiliation{\triumf}
\author{Amit~Samanta}\affiliation{\llnl}
\author{Jos\'e~Paulo~Santos~\orcidlink{0000-0002-5890-0971}}\affiliation{\nova}
\author{Joseph~Smolsky}\affiliation{\mines}
\author{John~Taylor~\orcidlink{0009-0008-8530-0868}}\affiliation{\mines}
\author{Joseph~Templet}\affiliation{\mines}
\author{Sriteja~Upadhyayula~\orcidlink{0000-0001-6918-2488}}\altaffiliation[Present Address: ]{\llnl}\affiliation{\triumf}
\author{Louis~Wagner~\orcidlink{0009-0000-9954-9658}}\altaffiliation[Present Address: ]{\frib}\affiliation{\triumf}
\author{William~K.~Warburton}\affiliation{\xia}
\author{Benjamin Waters}\affiliation{\mines}
\author{Stephan~Friedrich~\orcidlink{0000-0001-8393-3516}}\affiliation{\llnl}

% ----------------------------------------------

\collaboration{BeEST Collaboration}
\date{\today}

\begin{abstract}

The Beryllium Electron capture in Superconducting Tunnel junctions (BeEST) experiment searches for evidence of heavy neutrino mass eigenstates in the nuclear electron capture decay of $^7$Be by precisely measuring the recoil energy of the $^7$Li daughter. In Phase-III, the BeEST experiment has been scaled from a single superconducting tunnel junction (STJ) sensor to a 36-pixel array to increase sensitivity and mitigate gamma-induced backgrounds. Phase-III also uses a new continuous data acquisition system that greatly increases the flexibility for signal processing and data cleaning. We have developed procedures for signal processing and spectral fitting that are sufficiently robust to be automated for large data sets. This article presents the optimized procedures before unblinding the majority of the Phase-III data set to search for physics beyond the standard model.

\end{abstract}

\maketitle

\section{Introduction}\label{sec:intro}
The standard model of particle physics~(SM) stands as one of the most successful theoretical frameworks in modern science, cataloging the fundamental particles and their interactions that constitute the universe. 
Yet, despite its remarkable successes, the SM is widely acknowledged as being incomplete. 
The existence of gravity, dark matter, and the matter-antimatter asymmetry, for instance, require physics beyond the standard model~(BSM) to fully comprehend the universe~\cite{patrignani2016darkmatter}.
Notably, the neutrino sector, arguably one of the least explored areas of the SM, already exhibits signs of new physics, such as the observation of nonzero neutrino mass~\cite{skk1998evidence,sno2001oscillation,bilenky2015neutrino} and the chirality asymmetry of neutrinos~\cite{wu1957parity}.

The simplest extension of the SM requires the inclusion of non-zero neutrino masses through the addition of  right-handed neutrino flavor eigenstates, known as \textit{sterile neutrinos}~\cite{pontecorvo1968neutrino,gariazzo2015light,giunti2019annual,boser2020status,diaz2020where,seo2021review,dasgupta2021sterile}.
Unlike their SM counterparts, sterile neutrinos do not interact via the weak force, and interact only through gravity and mixing.
This elusive nature allows sterile neutrinos to play important roles in explaining the above anomalies while eluding the three-flavor constraints from experiments~\cite{janot2019improved,DELPHI2003photon,L32003single,aleph2006precision,Planck2018results,Fields2019bigbang,Ivanov2019cosmological}. 

The Beryllium Electron capture in Superconducting Tunnel junctions~(BeEST) experiment utilizes low temperature superconducting tunnel junction~(STJ) sensors to search for signatures of heavy neutrino mass eigenstates in the electron capture~(EC) decay of $^7$Be ~\cite{beest2021phase2}. 
High doses of $^7$Be are implanted directly into the high-resolution STJ sensors to precisely measure the nuclear recoil energy of the $^7$Li daughter. 
Neutrinos from the EC decays do not interact with the sensor, but their masses can be directly studied, independent of any BSM physics model, owing to the two-body nature of the decay and energy-momentum conservation.
If heavy sterile neutrino mass eigenstates exist, transition to the heavy states would reduce the recoil energy and generate additional lower-energy peaks as a distinctive signature.

In Phases I and II, the BeEST experiment accumulated a high statistics $^7$Be EC spectrum and examined it with unprecedented precision. 
In 2021, the first search for neutrino transition to a heavy state was performed using data collected for one month with a single STJ pixel~\cite{beest2021phase2}.
The BeEST experiment then transitioned to Phase-III, which involved scaling to a 36-pixel array and improvements in data acquisition hardware, analysis tools, and the underlying theories that determine the electron capture spectra~\cite{guerra2024shaking}.

This article introduces the analysis procedures and the spectral modeling developed for Phase-III of the BeEST experiment. 
First, we introduce the new data acquisition system, which records and stores a continuous stream of data and thus enables detailed pulse shape analysis. We then detail the signal processing steps to extract the nuclear recoil spectra and discuss the data cleaning processes adopted to minimize the systematic uncertainties that may affect the final spectra. Finally, we fit the cleaned spectra to a refined model spectrum, assess the variations between different STJ pixels and compare the extracted values to the known underlying standard model physics.

\section{Experimental Details}\label{sec:daq}

\begin{figure}
    \centering
    \includegraphics[width=\columnwidth]{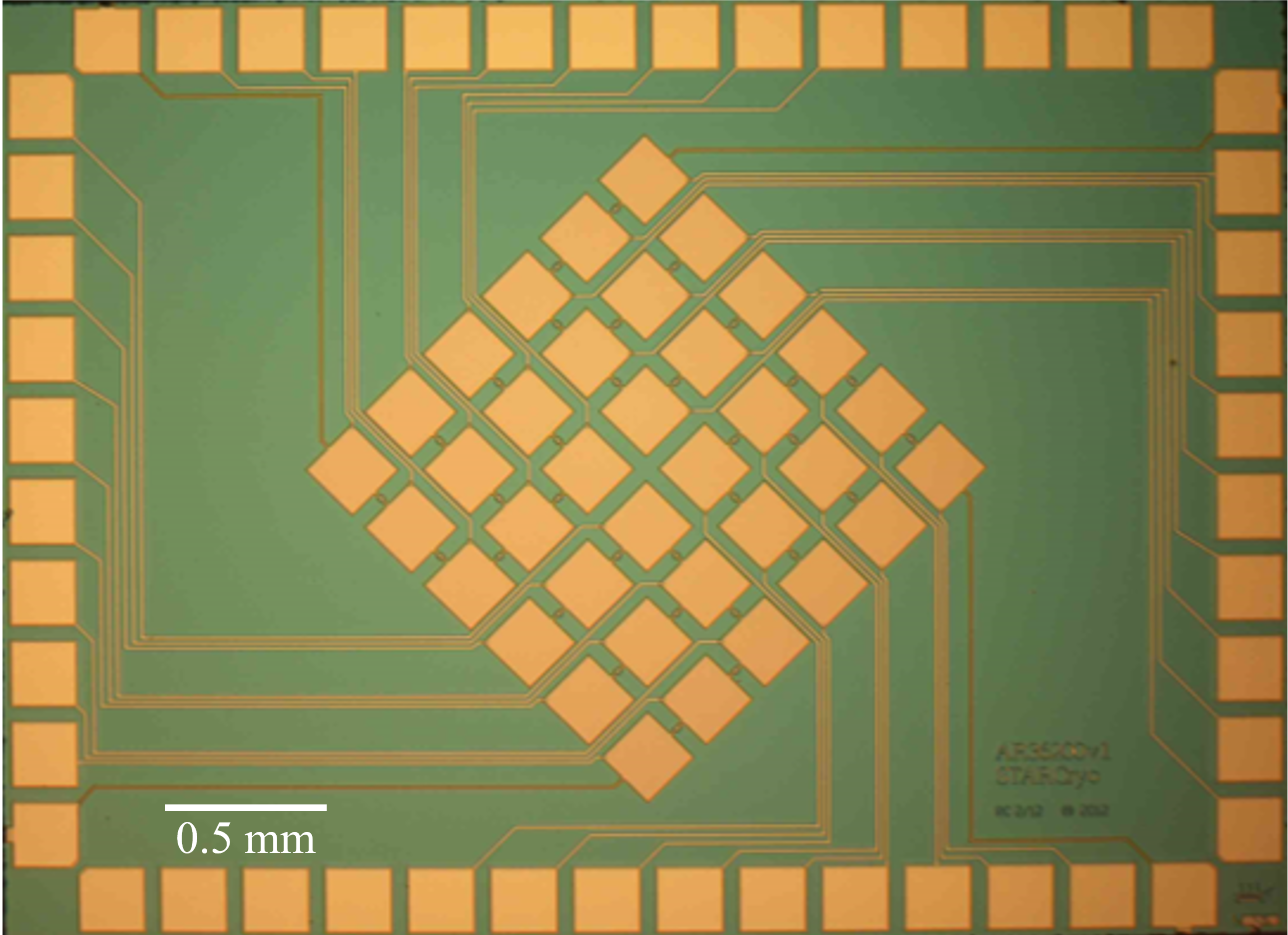}
    \caption{Photo of the 36-pixel STJ array employed in BeEST Phase-III prior to implantation. Groups of 9~pixels in the array shared a common ground wire~(dark traces).}
    \label{fig:STJPhoto}
\end{figure}

Phase-III of the BeEST experiment used a 36-pixel array of Ta~(265~nm)-Al~(50~nm)-AlOx-Al~(50~nm)-Ta~(165~nm) STJs fabricated at STAR Cryoelectronics LLC~(Fig.~\ref{fig:STJPhoto})~\cite{starcryo}. 
The pixel area of (208~$\upmu$m)$^{2}$ was larger than the (130~$\upmu$m)$^{2}$ STJ used in Phase-II~\cite{beest2020lkratio,beest2021phase2}, but both chips were from the same wafer. 
Groups of 9 pixels shared a single ground wire to reduce the number of wires from room temperature to the cryostat cold stage.

For the Phase-III physics run, $^7$Be$^+$ was implanted at an energy of 30~keV into the top Ta absorber film at TRIUMF’s isotope separator and accelerator (ISAC) facility in Vancouver, Canada~\cite{dilling2014isac}. 
The Ion Guide Laser Ion Source (IG-LIS)~\cite{raeder2014iglis,mostamand2020production}  was used to generate a high-purity $^7$Be$^+$ beam by selectively ionizing $^7$Be and suppressing isobaric $^7$Li$^+$. 
The measured Be:Li ratio was 7:1, a factor of >500 improvement over Phase-II when IG-LIS was not available~\cite{lennarz2023implantation}. 
Simulations of the $^7$Be$^+$ distribution using Stopping and Range of Ions in Matter~(SRIM)~\cite{srim2010} show a mean implantation depth of 58~nm for $^7$Be nuclei in the Ta layer and a straggle of 30~nm. 
A photolithographic Si mask was installed $\approx$100~$\upmu$m in front of the STJ to reduce the amount of $^7$Be$^+$ implanted into the substrate between pixels, and the chip was rinsed with ethanol after implantation to remove  $^7$Be on the surface due to beam scattering.
The initial $^7$Be EC decay rate ranged from 10 to 50~Bq, depending on the pixel.

The $^7$Be EC decay spectrum was measured at a temperature of $\approx$0.1~K in an adiabatic demagnetization refrigerator~(ADR) with liquid N$_2$ and He~precooling. 
The $^7$Li~recoil and the subsequent Auger electron were stopped in the Ta~film and their energy was measured with the STJ detector. 
This energy breaks Cooper pairs and excites single charges~(quasiparticles) above the superconducting energy gap $\Delta_\textrm{Ta}\approx0.7$~meV in proportion to the deposited energy. 
The quasiparticles then generate a tunneling current across the insulating barrier that is read out directly with a custom-designed preamplifier at room temperature with a gain of 10$^6$~V/A from XIA LLC~\cite{warburton2015xia}. 
The small energy gap in Ta allows measuring phonon signals from the recoil and enables an energy resolution of $\approx1-2$~eV FWHM in the energy range of interest~(ROI) up to 120~eV~\cite{kurakado1982stj,ponce2018u235,beest2021phase2}.
Throughout the recoil measurements, the sensor array was also exposed to a pulsed frequency-tripled Nd:YVO$_4$ calibration laser at 100~Hz to produce a comb of peaks at integer multiples of the single-photon energy of 3.49865(15) eV~\cite{ponce2018u235,beest2020lkratio,beest2021phase2}. 
The calibration spectra were separated from the recoil spectra by time-tagging the events coinciding with the laser triggers. 
For Phase-III of the BeEST experiment, we employed a new data acquisition~(DAQ) system capable of continuously sampling voltage signals. 
This system records data in an uninterrupted stream, allowing us to perform advanced offline analysis by revisiting the entire data stream to detect subtle signals and apply refined processing techniques~\cite{bray2024daq}.
This data taking was based on two 8-channel NI PXIe-6356 cards that each continuously recorded 16-bit voltage samples from 8 STJ preamplifiers at a rate of 1.25 MSa/s. 
We connected the two 8-channel cards to the 16~STJ pixels with the best signal-to-noise ratios, and wrote the data to disc in 10-minute increments of 12~GB each, which lost $\approx0.5$\% of the data stream and produced $\approx$3~TB of raw data per day. 
Both PXIe-6356 cards received their clock signals from the same chassis clock for accurate time synchronization across all pixels at the level of 0.2~$\upmu$s~\cite{bray2024daq}. 
One of the preamplifier outputs was split and also read out with an Ortec 927 multi-channel analyzer~(MCA) for direct comparison with the Phase-II data. 
Four additional STJ pixels were read out by a 50~MSa/s MPX-32D digitizer from XIA~LLC operated in list mode~\cite{warburton2015xia}. 
16 other pixels could not be read out due to a grounding problem, shorts, or poor performance.

Data were collected over 50~days from 11/03/2022 to 12/28/2022 at Lawrence Livermore National Laboratory~(LLNL), with a break in late November during which the ADR was allowed to warm up. 
The cryostat had a hold time of approximately 22~hours below 150~mK and was cycled once a day. 
At the beginning and the end of each ADR cycle, the I(V) curves for all channels were measured to monitor flux trapping in the STJ pixels, which varied slightly from day to day.
Data from three days were released from the full dataset to develop and test the signal processing and analysis pipeline. Dataset~1 from 11/06/2022 was released at an early stage of the run to test the data quality and develop the signal processing procedures. 
In this dataset, one of the 16~channels exhibited anomalous pulse shapes and was excluded from analysis.
Dataset~2 from 11/28/2022 was released after the measurement break to confirm the consistency of data collection after warming up the ADR.
On 12/12/2022, we conducted a test on varying the calibration laser intensity and rate to examine the impact of laser-induced substrate heating.  The test data was released as Dataset~3.
The released data comprise $\approx6.6$\% of the full dataset, and by developing the analysis pipeline on a small fraction of our data, we limit potential sources of bias in our analysis and maintain the objectivity of the final search results. 
This paper summarizes the signal processing routines we have developed for the continuous DAQ using Dataset~1, and explains data cleaning and spectral modeling in detail.

\section{Signal Processing}\label{sec:processing}

\begin{figure}
    \centering
    \includegraphics[width=\columnwidth]{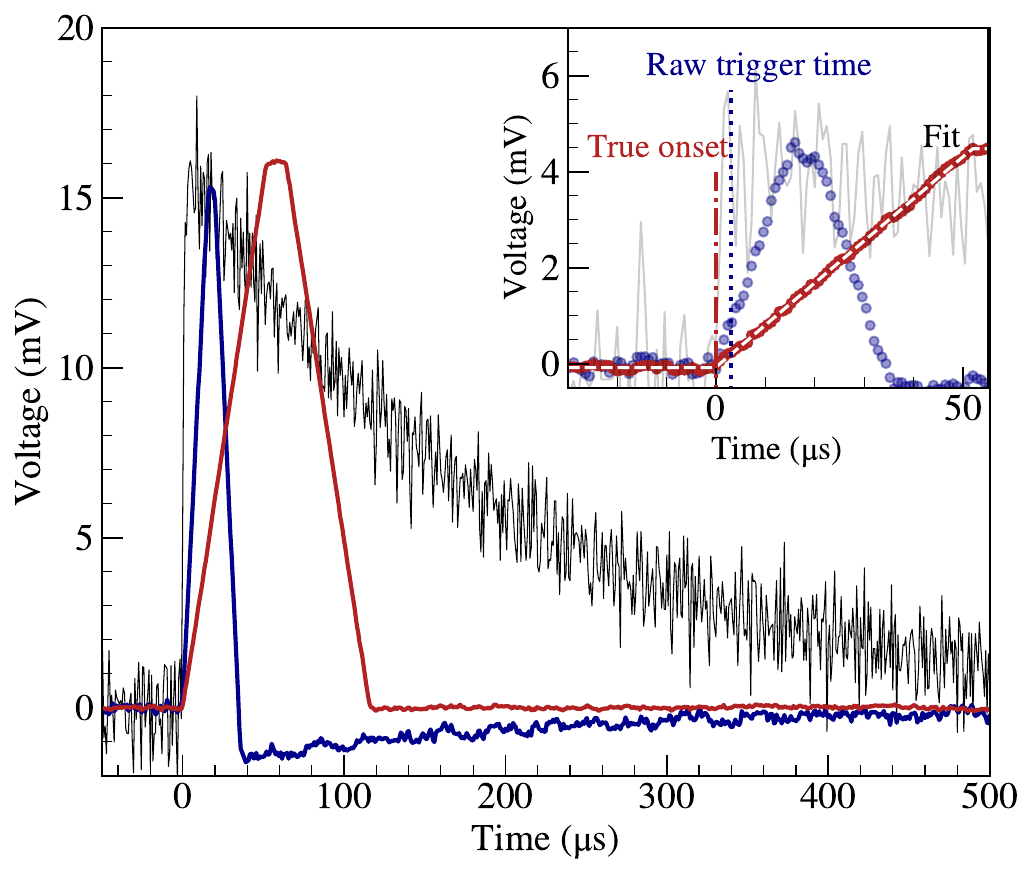}
    \caption{Raw STJ signal (black) and waveforms for long (red) and short (blue) trapezoidal filtering. The signal rise times are on the order of 1~$\upmu$s. (Inset)~The raw trigger time differs from the true pulse onset extracted from a fit to the filtered waveform (white), especially for small signals. Outputs of 1~mV correspond to 1~nA signals.}
    \label{fig:pulse_shape}
\end{figure}

Signals from the Ta-based STJ detectors used in Phase-III had amplitudes of $\approx$150~nA/keV and rise times ($\tau_\textrm{rise}$) of order $\approx1~\upmu$s.
Their decay times ($\tau_\textrm{decay}$) were set by the charge recombination times in the junction electrodes and were on the order of $\approx$100~$\upmu$s for Ta-based STJs (Fig.~\ref{fig:pulse_shape}). Their value depends on the number of magnetic flux vortices trapped in the STJ electrodes during the ADR cycle, which reduce the superconducting gap locally so that quasiparticles can be trapped and recombination is enhanced. The decay times therefore differed slightly for different pixels and different ADR cycles but were constant for each pixel during each cycle~\cite{wigmore2004fluxtrapping}. 

In the first stage of the analysis, two trapezoidal filters with different time constants were applied to the raw data traces. 
The short trapezoidal filter did not use any pole-zero~(PZ) correction and produced an output waveform that was used for triggering and waveform analysis. 
The long trapezoidal filter did use PZ corrections and its output was used for the energy estimate. 
The next sections describe the algorithms used to optimize the filter coefficients and the pulse parameters extracted from the filtered waveforms for further processing.

\subsection{Trapezoidal filter optimization}\label{sec:PZ}

The optimal values of the filter parameters depend on the signal shape and the noise conditions. In STJ detectors, the noise is determined by statistical fluctuations in the number of quasiparticles and electronic noise of the preamplifier.
For constant noise, the optimal shaping time $\tau_\textrm{shape}$ is roughly equal to the decay time of the pulse. 
For STJs, however, there are time-variations in the statistical noise that favor shorter shaping times~\cite{hiller2001noise,samedof2000multitunneling}. 

The variations in the signal decay time required daily re-optimizing the pole-zero time ($\tau_\textrm{PZ}$), which is typically close to the decay time~\cite{jordanov1994digital}. $\tau_\textrm{PZ}$, energy resolution and pile-up are optimal and when the filtered waveforms have a trapezoidal shape with minimum slope in the central section and minimum over- or undershoot after the trapezoid. For each $\tau_\textrm{PZ}$ optimization, we used only the first 10-minute data file from each day since the signal decay time of each STJ is constant throughout an ADR cycle. 
We initially set  $\tau_\textrm{PZ}=200~\upmu$s to process the data and selected events without pile-up by rejecting all events whose standard deviations of the mean time, pre-signal baseline or post-signal baseline differed from their average by more than 2$\sigma$.
We then systematically varied $\tau_\textrm{PZ}$ from 32~$\upmu$s~(40~samples) to 560~$\upmu$s~(700 samples) to optimize the flatness of the central section of the trapezoid.

\begin{figure}[t]
    \centering
    \includegraphics[width=\columnwidth]{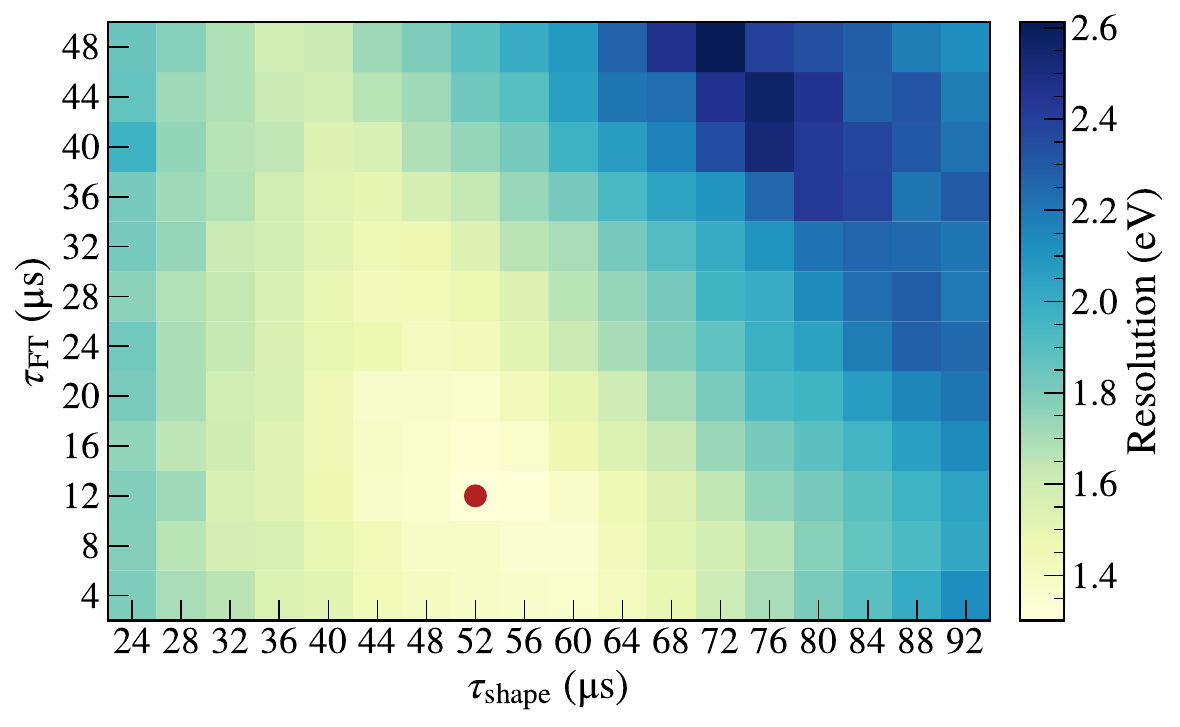}
    \caption{Heat map of the FWHM energy resolution at 105~eV for a typical STJ pixel as a function of $\tau_\mathrm{shape}$ and $\tau_\textrm{FT}$. The red dot at [52,~12]~$\upmu$s shows the optimal values of the filter parameters. }
    \label{fig:resolutionHeatMap}
\end{figure}

We then used the first 10-minute data file from the first unblinded date to investigate the impact of varying $\tau_\textrm{shape}$ and the flat-top time $\tau_\textrm{FT}$ on the energy resolution.
In this preliminary analysis, the 105~eV laser photopeak (30~photons) was selected and its resolution was calculated for various combinations of $\tau_\textrm{shape}$ and  $\tau_\textrm{FT}$. 
For each combination, the PZ correction was applied to minimize the gradient of the flat-top region. 
Figure~\ref{fig:resolutionHeatMap} shows the result of this investigation for a typical channel. 
We observed that the energy resolution is reasonably flat near $\tau_\textrm{shape} \approx50~\upmu$s and $\tau_\textrm{FT} \approx12~\upmu$s in all channels.
For ease of analysis, the filtering parameters for the long trapezoid was set to  $\tau_\textrm{shape}=52~\upmu$s~(65~samples) and  $\tau_\textrm{FT}=12~\upmu$s~(15 samples) in all channels.
The small variations in energy resolution did not affect our sensitivity, as the electron capture peaks in the BeEST spectra are much broader than the energy resolution of the STJ. 
For the short trapezoid, we used a 16~$\upmu$s $\tau_\textrm{shape}$~(20~samples) and a 4~$\upmu$s  $\tau_\textrm{FT}$~(5~samples) with no PZ correction to investigate the early part of the waveforms. 

\subsection{Triggering}\label{sec:trigger}

Pulses were triggered when the output of the fast trapezoidal filter exceeded the median baseline level by at least 5$\sigma$, where $\sigma$ describes the standard deviation of the baseline. 
The baseline level was calculated after an initial coarse cut that removed all data points that vary from the median signal by more than $\pm 3 \sigma$ to avoid the influence of any pulses in the data stream. 
The remaining baseline samples were then used to calculate a finer median and the standard deviation $\sigma$ for each 10-minute segment. 
To prevent multiple triggering from a single pulse, each trigger event was followed by a dead time set to the length of the long trapezoid plus twice the baseline lengths, which is 196~$\upmu$s for $\tau_\textrm{shape}=52~\upmu$s,  $\tau_\textrm{FT}=12~\upmu$s, and $\tau_\mathrm{base}=40~\upmu$s.

\begin{figure}
    \centering
    \includegraphics[width=\columnwidth]{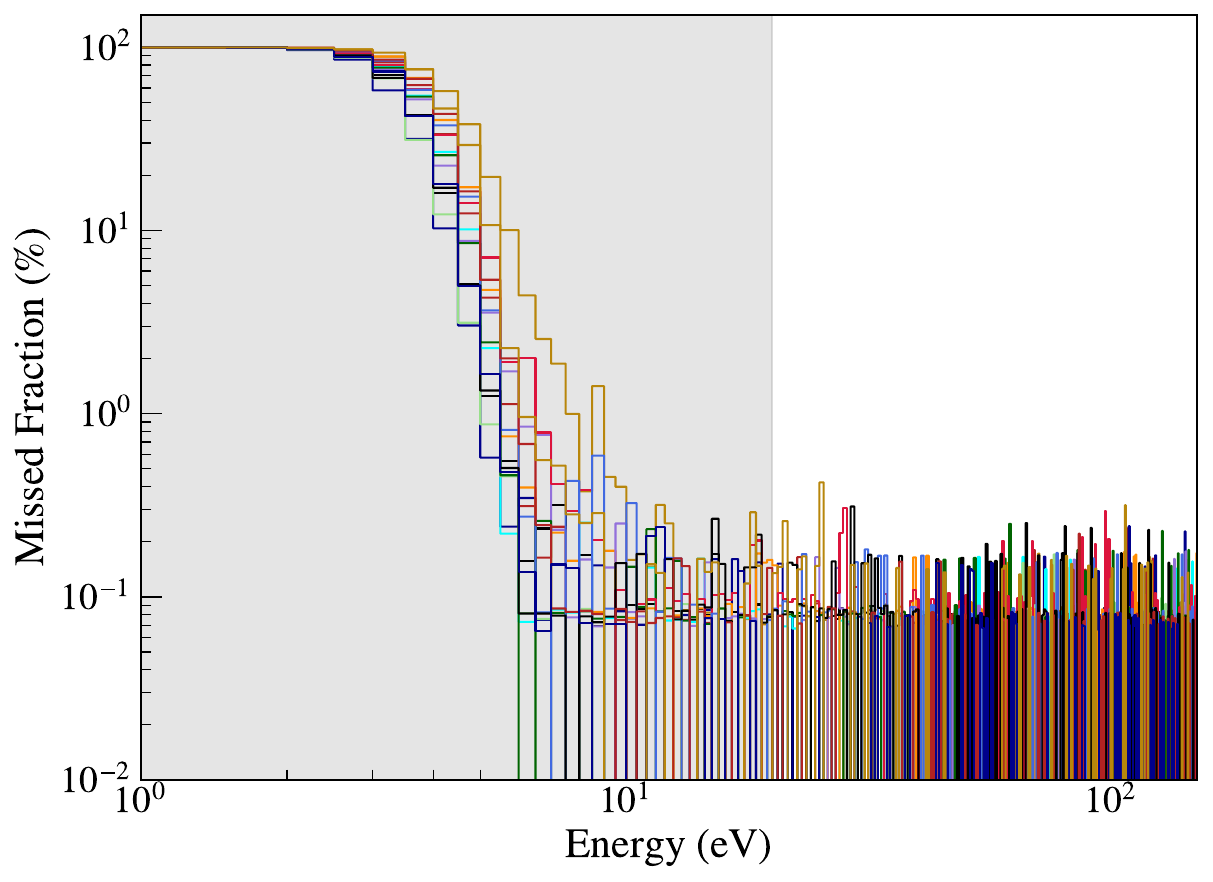}
    \caption{Percentage of known pulses from a waveform generator \textit{not} detected at different energies. 
    Colors represent 15~different pixels.
    Above $\approx$10~eV, the detection efficiency is $>99.9$\% for all pixels in the array.}
    \label{fig:efficiency}
\end{figure}

We have tested the triggering efficiency as a function of energy numerically by inserting a known number of artificial signal waveforms at specific locations into the saved data stream. 
The waveforms, originally taken from the average laser signal at 105~eV~(30~laser photons), were scaled to simulate various energies based on the energy reconstruction algorithm explained in Section~\ref{sec:energy}, and then randomly inserted at times when no other signals were present.
No difference was observed between the pulse shapes of the laser and the EC events, except in the one channel that was excluded from the analysis, and the simulated waveforms accurately represented signals with known amplitudes.
We added an average of 1000~signals per minute to each of the 15~channels over Dataset~1 and processed the new data stream with the same algorithm as before.
A signal was counted as ``undetected'' if the data processing failed to record an inserted signal within $\pm5.6$~$\upmu$s of its known time stamp and within $\pm$5$\sigma_0$ of its known energy, where $\sigma_0$ is the resolution of the 105~eV laser peak for that channel.

Figure~\ref{fig:efficiency} shows the percentage of numerically inserted events that were \textit{not} accurately detected within the expected acceptance intervals. 
For energies above $\approx$10~eV, the detection efficiency exceeded 99.9\% and no peak-like structure was seen. 
The few isolated events above 10~eV that were missed were due to noise or small background signals in the original data that moved the detected energy outside the acceptance window. 
Widening the acceptance window recovered these events. 
Below energies of $\approx$10~eV, the triggering efficiency started to drop until it approached zero at $\approx$2~eV, with the exact detection threshold depending on the noise level of the particular detector. 
This does not affect the search for a primary sterile neutrino signature that is expected to fall into the energy range between $\approx$55 and $\approx$105~eV. 

\begin{figure}
    \centering
    \includegraphics[width=\columnwidth]{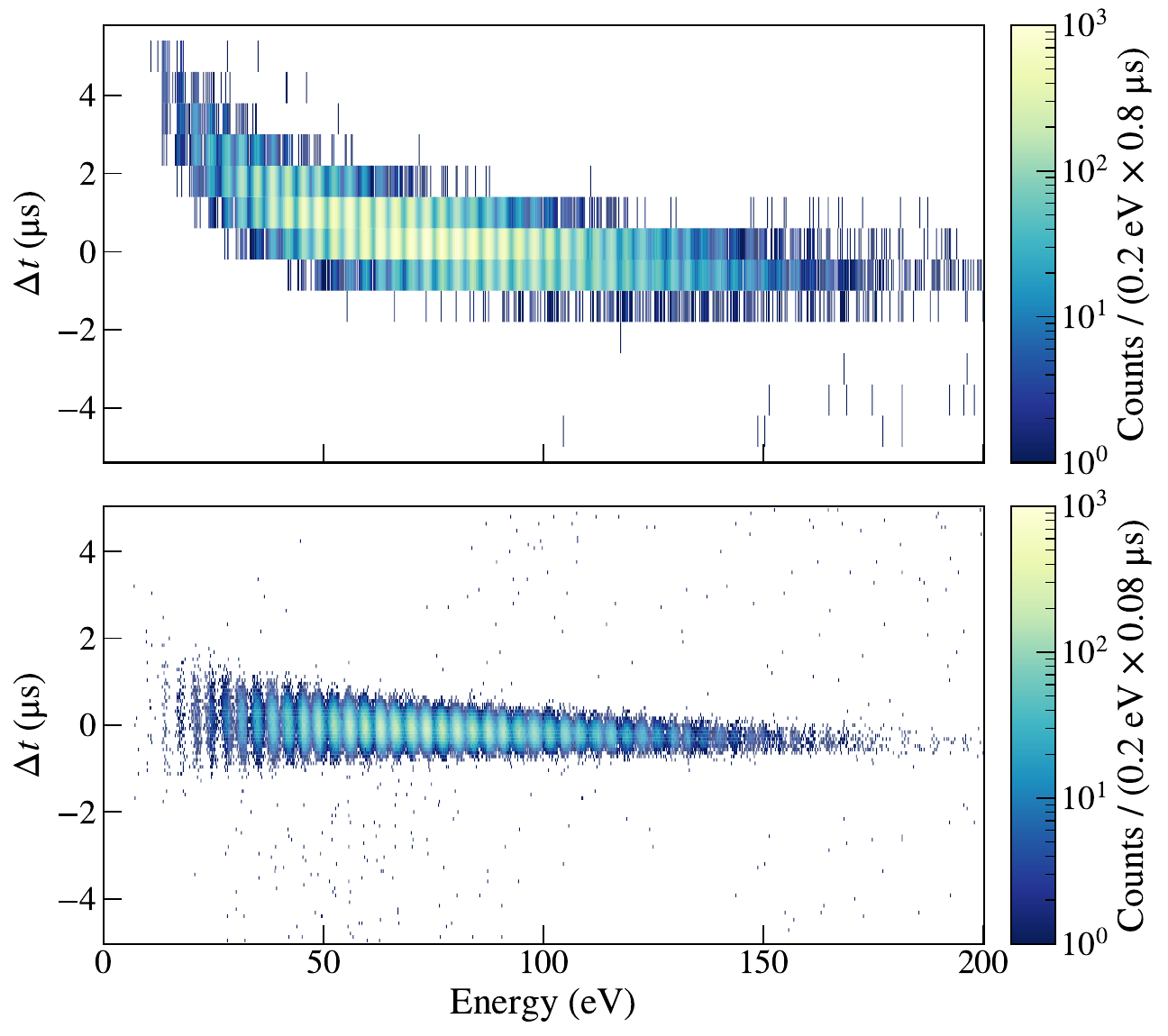}
    \caption{Time differences between laser pulses in two different pixels. (Top) Raw trigger time differences. (Bottom) Pulse arrival time differences after the timing correction. The 105~eV laser peak in one channel was selected as a reference, and the time differences were plotted against the energy in the other channel.}
    \label{fig:jitter}
\end{figure}

For a given threshold level, the raw trigger time, i.e., the time at which the fast trapezoidal filter exceeded the triggering threshold, varied relative to the pulse onset as a function of the pulse amplitude due to energy walk (Fig.~\ref{fig:jitter}, top). 
We therefore fit linear functions to the pre-pulse baseline and to the rising edge of the slow trapezoid and extracted the pulse arrival time as the intersection of the two fits~(Fig.~\ref{fig:pulse_shape}, inset). 
This improved the timing accuracy to $\approx\pm0.5~\upmu$s FWHM in the energy range of interest of [20,120]~eV for the arrival time differences of nominally simultaneous laser signals~(Fig.~\ref{fig:jitter}, bottom). 
The timing accuracy could be higher than the sampling rate of 0.8~$\upmu$s because the fit procedure used information from the entire rising edge of the pulse to average out noise. 
This accuracy enhanced the robustness for both pulse heights and pulse shape parameter extraction. 
It is also important to identify coincident laser signals and background events while keeping the coincidence window small.

\subsection{Energy reconstruction}\label{sec:energy}

Pulse amplitudes were extracted from the slow trapezoidal filter. We extract the peak from the ADC value in the center of the trapezoid’s flat top, which occurs 58~$\upmu$s after the pulse onset for shaping times of 52~$\upmu$s and flat-top times of 12~$\upmu$s. 
To minimize the effect of baseline fluctuations, the pulse base was determined from the average of the waveform in the interval [-40,-20]~$\upmu$s before the pulse onset. 

Similar to Phase-II of the experiment~\cite{beest2021phase2}, we used the externally triggered pulsed UV~laser for absolute energy calibration. The laser intensity was adjusted so that the peak of the laser envelope falls in the region between 20 and 120~eV for most of the pixels. 
The laser signals were separated from the recoil events by time-tagging the events coinciding with the laser triggers. 
Each 10-minute data file was calibrated separately to correct for variations in the laser intensity and any other slowly time-varying systematics. 
Since the calibration precision for a single 10-minute file was still limited by statistics, we added the signals from two files before and two files after the target file and generate a laser spectrum using the cumulative 50-minute data. 
We fit the 3- to 71-photon peaks ([10.5,248.5]~eV) in the laser spectrum with a superposition of Gaussians.
To increase the robustness of the fit, we fit each peak individually while including two adjacent laser peaks on each side of the target peak. 
This provided accurate values of peak centroids and widths, even in pixels with a somewhat reduced energy resolution.

\begin{figure}
    \centering
    \includegraphics[width=\columnwidth]{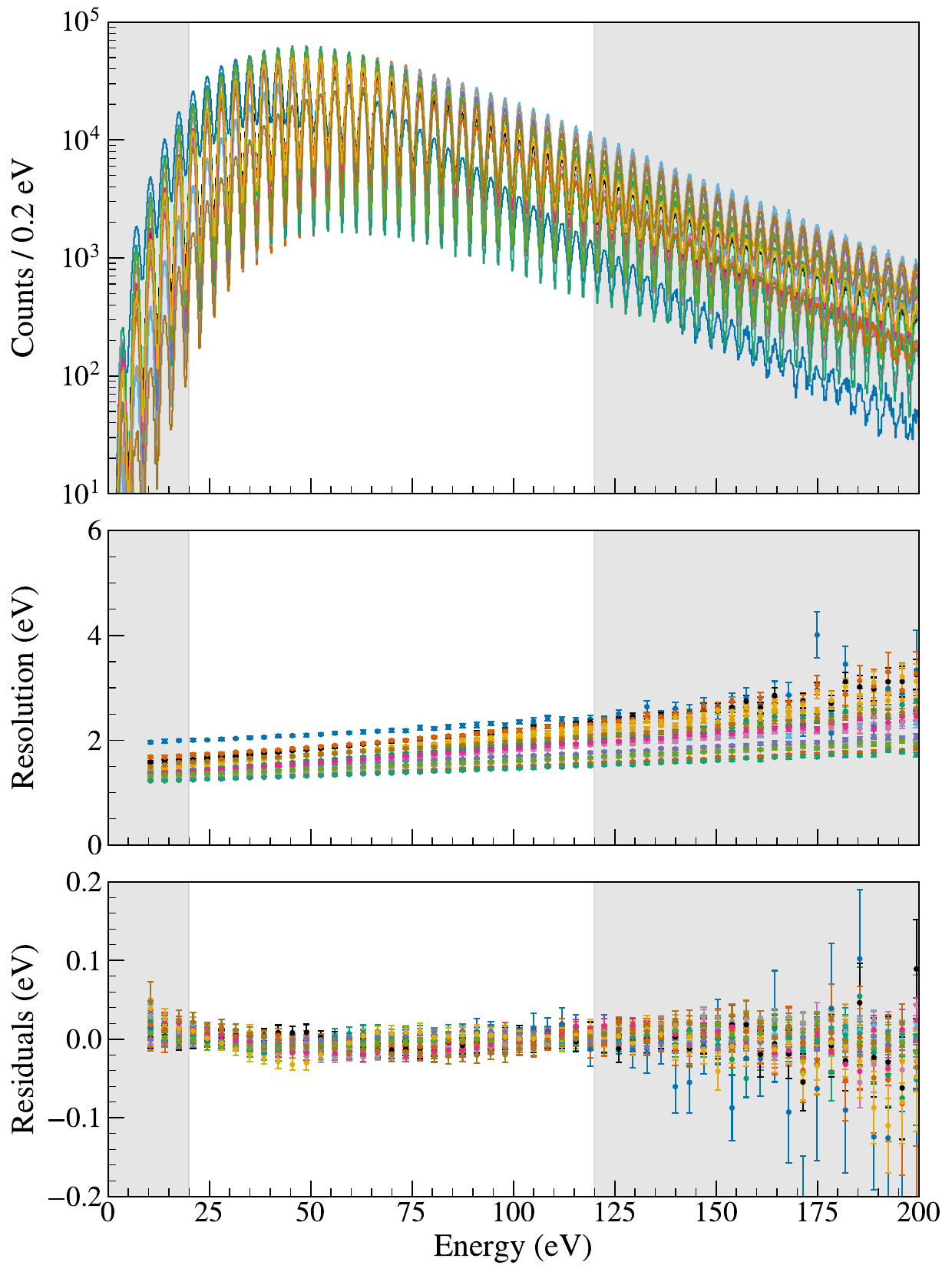}
    \caption{Calibrated laser spectra for $\approx20$~hours of BeEST Phase-III data. 
    Colors represent 15~different pixels.
    The ROI of [20,120]~eV is shown as the white background.
    (Top) Laser calibration spectra for all 15 channels for one day. 
    (Middle) FWHM energy resolution of individual laser peaks varies between $\approx$1.5~eV and $\approx$2.5~eV in the ROI.
    (Bottom) Residuals from a quadratic energy calibration. Residuals from a quadratic energy calibration are <0.1~eV throughout the ROI for all channels.  
    }
    \label{fig:laserSpectrum}
\end{figure}

During this process, we identified two subtle effects that affect laser signals differently than $^7$Be signals and that need to be corrected for accurate calibration. 
Both were due to the fact that the calibration laser produces simultaneous signals in all pixels, while the $^7$Be decays occurred randomly in time and pixel.
The first effect was caused by the shared ground wire for groups of 9 pixels (Fig.~\ref{fig:STJPhoto}), which was resistive between the detector chip and the preamplifier at room temperature. 
Any signal produced a small voltage drop across this wiring resistance, which was then amplified and added to all signals in pixels that share the same ground wire. 
This produced an offset that is proportional to the laser intensity and the wiring resistance.

The second effect was caused by absorption of scattered laser photons in the Si substrate between pixels. 
This generated high-energy phonons in the substrate that could propagate to the STJs and break additional Cooper pairs in them. 
For constant laser intensity, this produced a constant average offset that must be subtracted for consistent calibration~\cite{ponce2018u235}. 
Since the laser intensity varied significantly beyond statistical fluctuations in Phase-III~(Fig.~\ref{fig:laserSpectrum}, top), this effect produces an offset that is systematically related to the laser intensity.

\begin{figure}
    \centering
    \includegraphics[width=\columnwidth]{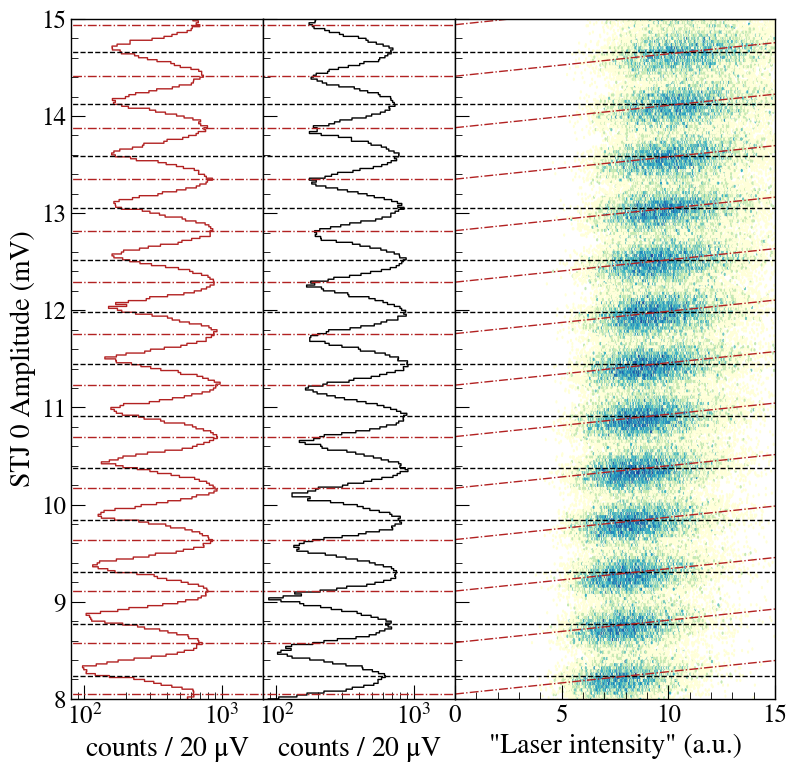}
    \caption{
    An illustration of how the laser intensity affects the laser signal amplitude in one channel.  
    In the right plot, the average of laser signals from 15 other pixels is taken as a proxy for the laser intensity. 
    The centroids of laser clusters are marked with horizontal black dashed lines.
    Each cluster, representing the events with the same number of photons absorbed in STJ-0, showed linear dependence on the laser intensity due to crosstalk and photon absorption in the Si substrate.
    A linear fit to the events with the same number of photons (red dashed lines) allowed correcting for this effect by extrapolating to zero laser intensity. 
    The two histograms compare the laser spectrum before (black) and after (red) this correction.
    }
    \label{fig:substrateCorrection}
\end{figure}

Therefore, resistive crosstalk and substrate events both produced gain changes and offsets that scaled with the intensity of the laser.
We illustrate this effect by plotting the signal amplitudes from one channel against the laser intensity of each event. 
For this, the average of coincident laser signals was used as a measure of the laser intensity. 
Figure~\ref{fig:substrateCorrection}~(right) shows that the signal for the \textit{same} laser peak in the \textit{same} pixel increases as the laser intensity increases. 

To correct for these changes, we applied a linear correction to the laser peaks~(red dash-dotted lines in Fig.~\ref{fig:substrateCorrection}) and extrapolated the lines to zero laser intensity, to obtain the laser peak positions without the effects of resistive crosstalk and substrate heating~(Fig.~\ref{fig:substrateCorrection} left).
At zero intensity, the corrected laser peaks deviated from the peak centroids before correction~(Fig.~\ref{fig:substrateCorrection} center), providing corrected calibration gain and offset terms. 
This correction produced the laser spectrum whose calibration was applied to the $^7$Be signal from the same pixel.

The centroid positions $V_n$ of the $n-$th laser peaks were then extracted via a weighted fit to a second-order polynomial:

\begin{equation}\label{eq:calibration_polynomial}
    V_n=c_0+c_1E_n+c_2E_n^2~,
\end{equation}
\noindent 
where $E_n=n\times3.49865$~eV is the energy of the $n$-th peak, and $c_0, c_1$and $c_2$ are the offset, gain, and non-linearity of the calibration, respectively.
Laser peaks whose centroid uncertainty or width uncertainty exceeds 20\% of their fit values are considered to be bad fits and excluded from the fit to ensure a robust calibration. 

Figure~\ref{fig:laserSpectrum} illustrates the calibrated laser spectra for all 15~channels from a single day of measurement, with their resolution values and residuals at the individual laser peaks. 
Using the calibration polynomial in Eq.~\ref{eq:calibration_polynomial}, the resulting spectra had fits with calibration residuals less than 0.05~eV across the entire ROI.
This uncertainty corresponds to a sterile neutrino mass of 26~keV and sets the low-mass limit for the experiment. 
Higher-order non-linearities in the setup were due to the ADC and do not directly limit the sensitivity of the search. 
The residuals were often small even outside the ROI from 20~to 120~eV, indicating that the ADC non-linearity of the NI PXIe-6356 digitizers is small, achieving an $\mathcal{O}(10)$~meV precision across a range of several hundred~eV~\cite{bray2024daq}.

\begin{figure}
    \centering
    \includegraphics[width=\columnwidth]{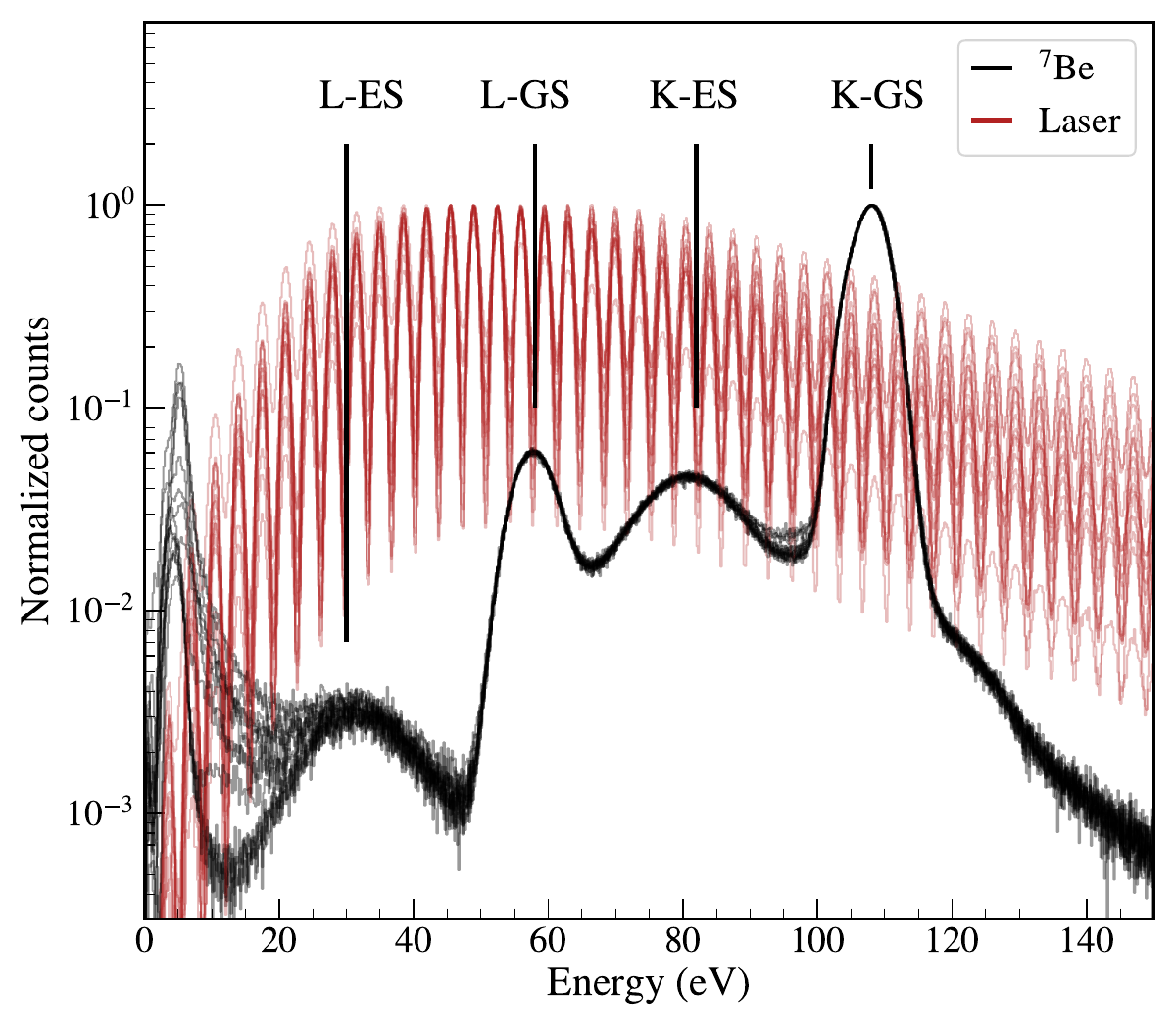}
    \caption{Normalized laser spectra and associated $^7$Be spectra for all 15~pixels from a single day of measurement.}
    \label{fig:energyShifts}
\end{figure}

Figure~\ref{fig:energyShifts} shows the calibrated laser and $^7$Be spectra for all pixels from a single day of measurement, normalized to account for different signal rates. 
As expected, the spectra show four primary peaks, two for K-capture into the ground state (K-GS) and into the excited nuclear state (K-ES) of $^7$Li, and two corresponding L-capture peaks (L-GS and L-ES)~\cite{beest2020lkratio}. 
The full description of their spectral shapes are explained in detail in Sec.~\ref{sec:bkg}.
The standard deviations of the two GS peak positions across all channels were $\pm0.12$~eV, and STJ detectors had energy resolutions between 1.5 and 2.7~eV FWHM at 105~eV~(30~photons) in Phase-III of the BeEST experiment~(Fig.~\ref{fig:laserSpectrum}). 
The narrow laser peaks illustrate that the contribution of the STJ detector resolution to the signal width is negligible.

\section{Data Cleaning}\label{sec:dc}

To achieve a sensitivity on the mixing fraction for a heavy neutrino mass state 
below $10^{-3}$ with the BeEST experiment, an accurate understanding of the background spectrum due to active-neutrino decays is required. 
This includes pile-up and interactions of  $^7$Be $\gamma$-rays in the Si substrate, as well as external sources like pick-up or cosmic background events. 
These background contributions can be identified and removed prior to spectral analysis. 
This is referred to as \textit{data cleaning}. 
It encompasses the removal of periods with high calibration uncertainty or high noise, of pick-up, pile-up and of events triggering multiple detectors simultaneously. 
It is desirable that the removal steps do not depend on the energy of the events to limit bias in the final spectra. 
This section details the data cleaning algorithms implemented for Phase-III of the BeEST experiment.

\subsection{Poor calibration}

If the distribution of the laser peaks fluctuates due to laser instabilities, the calibration uncertainty may become high enough that the spectral shape is distorted. 
This is particularly detrimental as inaccurate energy determination can produce a false signal mimicking a shifted sterile neutrino spectrum.

\begin{figure}
    \centering
    \includegraphics[width=\columnwidth]{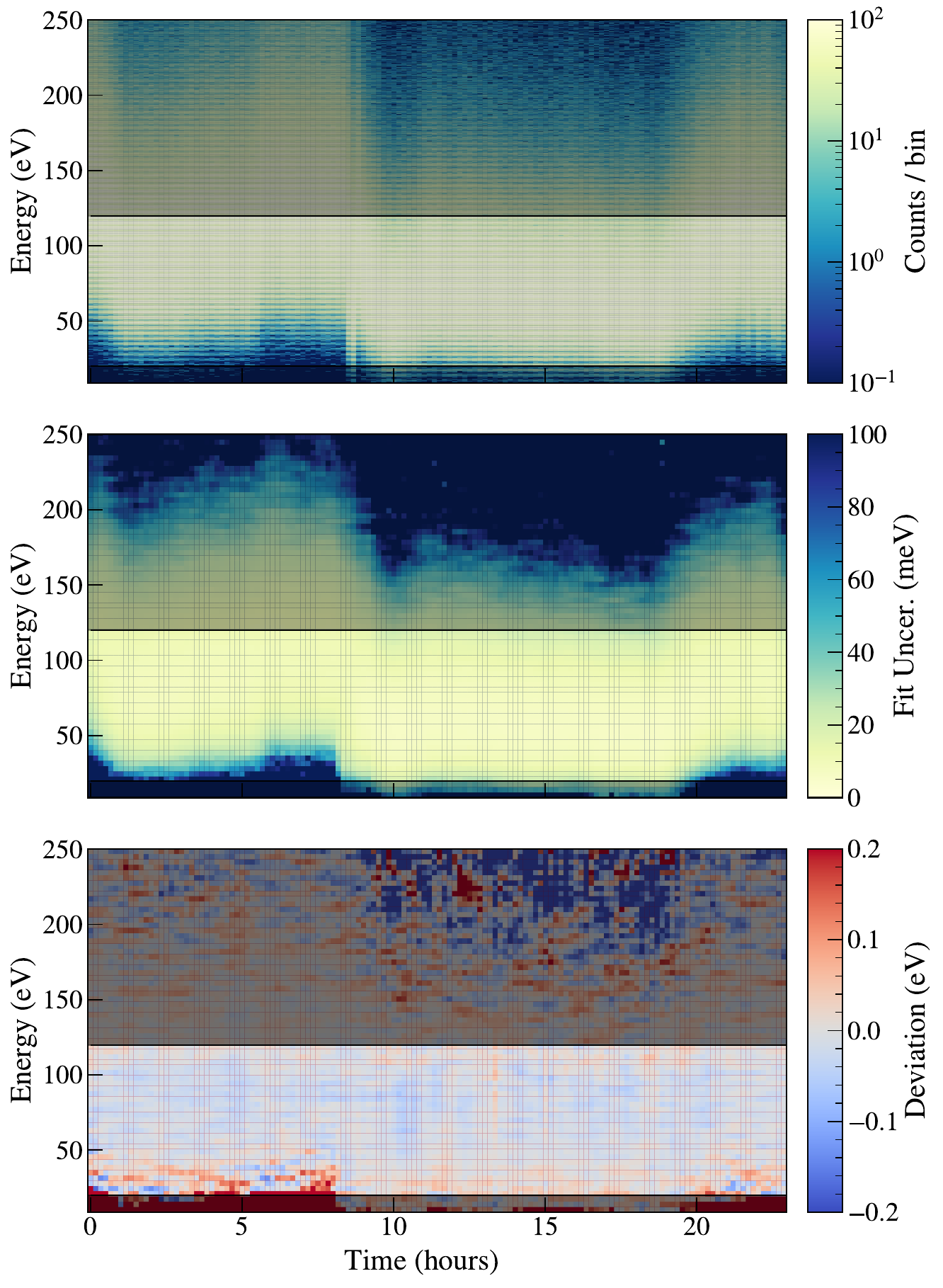}
    \caption{Illustration of the calibration laser stability over one day from a single detector. (Top) 2D histogram of laser calibration peak intensities. (Middle) Energy uncertainty of the centroid fit position at each laser peak. (Bottom) Deviation of measured laser peak positions from the true laser energies. The ROI is highlighted.}
    \label{fig:laserDrift}
\end{figure}

\begin{figure}[t]
    \centering
    \includegraphics[width=\columnwidth]{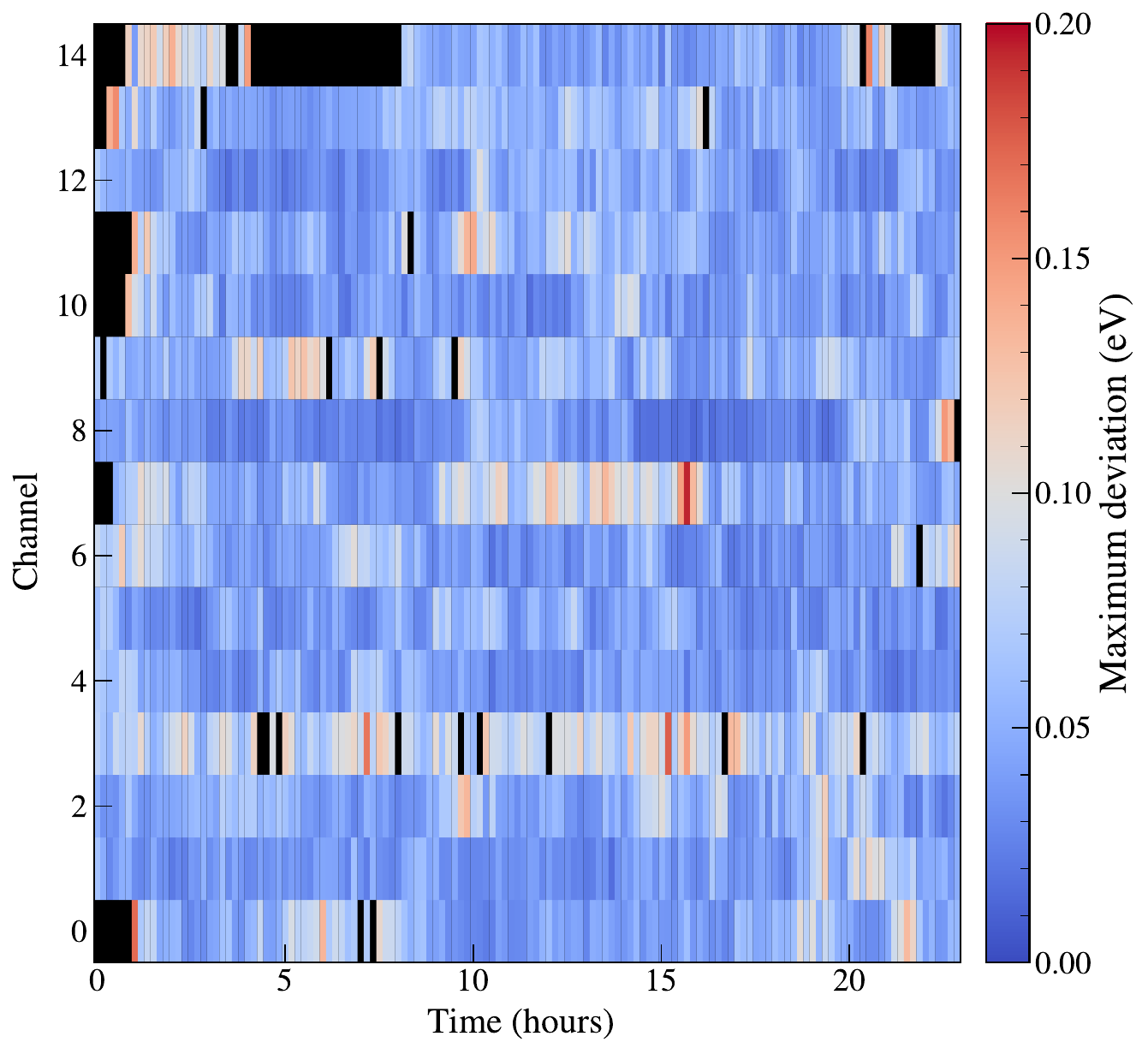}
    \caption{Energy fit residuals in the ROI for all 15 channels over one day. All rejected datasets are masked in black.}
    \label{fig:calibrationError}
\end{figure}

Figure~\ref{fig:laserDrift} (top) illustrates how the laser peak's envelope changed over time within one day of measurement. 
The reliability of the calibration is closely correlated to the envelope change.
For a single detector channel, we rejected 10-minute calibration periods if any of the laser peaks in the ROI~(6-34~photons) exhibits a centroid uncertainty higher than 0.2~eV~(corresponding to 51~keV sterile neutrino mass). 
We also rejected 10-minute periods if the $R^2$ value of the energy calibration was lower than 0.999~998, the value at which the $R^2$ distribution showed an increased count due to erratic datasets. 
The heat map of maximum residuals within the ROI from a single day measurement is shown in Fig.~\ref{fig:calibrationError}.

In the dataset we analyzed, 3.9\% of the data were rejected due to high calibration uncertainty. 
As the entire period is removed, this data cleaning cut does not have energy dependence, and does not bias the shapes of the accepted spectra.

\subsection{Pick-up}\label{sec:pickup}

\begin{figure}[t]
    \centering
    \includegraphics[width=\columnwidth]{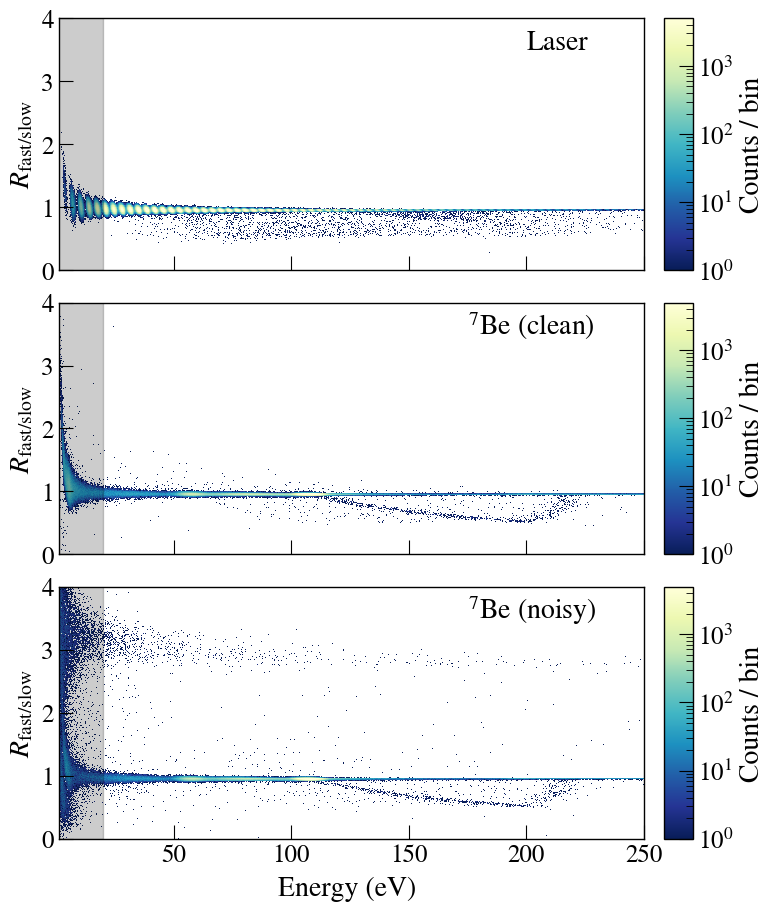}
\caption{Comparison of the rejection parameter $R_\textrm{fast/slow}$ with respect to energy for laser events~(top), electron capture events in a quiet channel~(middle) and electron capture events in a channel with high pick-up rate (bottom). Regions below the ROI of 20~eV are shaded in gray. Pile-up-induced structure is also shown at low $R_\textrm{fast/slow}$  values. }
    \label{fig:shotlong_zoomout}
\end{figure}

Periods with increased electronic pick-up were also rejected.  
While most detector pixels did not suffer from pick-up most of the time, certain pixels were more susceptible to pick-up than others. 
Pick-up signals had different waveforms than $^7$Be signals and could therefore be identified from the filtered signals. 
We defined the amplitude ratio of the fast and the slow filter signals as a rejection parameter $R_\textrm{fast/slow}$. 
For isolated ``good'' laser or $^7$Be signals, $R_\textrm{fast/slow}$ had a value around 0.9 (Fig.~\ref{fig:shotlong_zoomout}). 
Pick-up signals, on the other hand, tend to be faster and therefore tend to have an increased value of $R_\textrm{fast/slow}$. 
They also produced filtered signals mostly at low amplitudes outside the energy band of interest.
Bipolar pick-up signals also produced $R_\textrm{fast/slow}$ values below zero.

We applied a count-rate-based data cleaning step to eliminate periods of high pick-up rate from the data. 
We tagged events with $R_\textrm{fast/slow}>2$ or $R_\textrm{fast/slow}<0$ as pick-up events, and counted the number of pick-up-tagged events in each 10-minute file in each detector.
We then removed periods where the pick-up count in each 10-minute file was higher than  $m+5\sqrt{m}$ , where $m$ is the median of the pick-up counts for a single day of data taking. 
In the dataset used here, 2.9\% of the data were removed by this data cleaning step. 

\begin{figure}[t]
    \centering
    \includegraphics[width=\columnwidth]{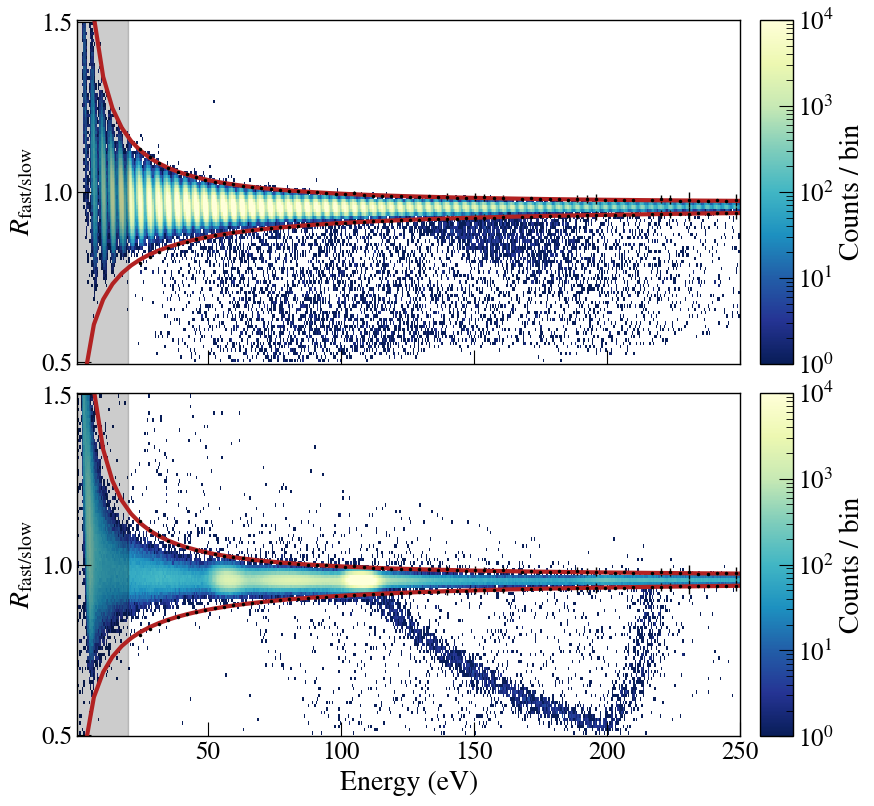}
    \caption{Rejection parameter $R_\textrm{fast/slow}$ of the amplitudes of the short and long trapezoids for (top) laser events and  (bottom) electron capture events. Regions below the ROI of 20~eV are shaded in gray. The red solid lines indicate the  $\pm 5\sigma_\textrm{rms}$ selection bands derived from the laser-tagged calibration events. The events with low $R_\textrm{fast/slow}$ are due to pile-up.}
    \label{fig:pileupscatter}
\end{figure}

\subsection{Pile-up}~\label{sec:pileupdc}

Pile-up events occur when two or more signals that are close in time overlap on each other and consist of two types: pile-up between two independent EC events and pile-up of EC events with laser pulses. 
While the pile-up discrimination efficiency for the former depends strongly on the time separation between the two events and the timing resolution of the detectors, we could remove almost all of the latter case with high confidence. 
This is thanks to the known arrival time of the laser pulses. 
We therefore applied a  $(\tau_\textrm{shape}+\tau_\textrm{FT})=\pm64~\upmu$s dead time at the time of laser injections. 
Given that the laser pulse rate was 100 Hz, we eliminated 1.28\% of the total data by applying this data cleaning step.
Since the entire period is removed, this data cleaning cut does not have energy dependence and does not bias the spectral shapes of the accepted spectra.

For pile-up between two EC events, we again made a use of the pulse shape parameter  $R_\textrm{fast/slow}$. 
For pulses separated by less than $\tau_\textrm{FT}=6~\upmu$s~(50\% of $\tau_\textrm{FT}$ at which we calculate the amplitude), the summed energy of both pulses was recorded as a single event. 
If time separation between two pulses is between 6~$\upmu$s and $\tau_\textrm{shape}+0.5\tau_\textrm{FT}=58~\upmu$s, the rising part of the second pulse overlaps with the flat top of first pulse. 
In this case, the recorded energy depended on the time separation and varies between the sum of the two events (when the time difference is close to 6~$\upmu$s) to the energy of the first pulse (when it is close to 58~$\upmu$s).
In both cases, the rejection parameter $R_\textrm{fast/slow}$ was smaller than for fully separated events~(Fig. ~\ref{fig:pileupscatter}).
The timing resolution for identifying the pile-up of two K-GS events using this method was approximately 4~$\upmu$s, although separations under 6~$\upmu$s are recorded as a single event.

\begin{figure}
    \centering
    \includegraphics[width=\columnwidth]{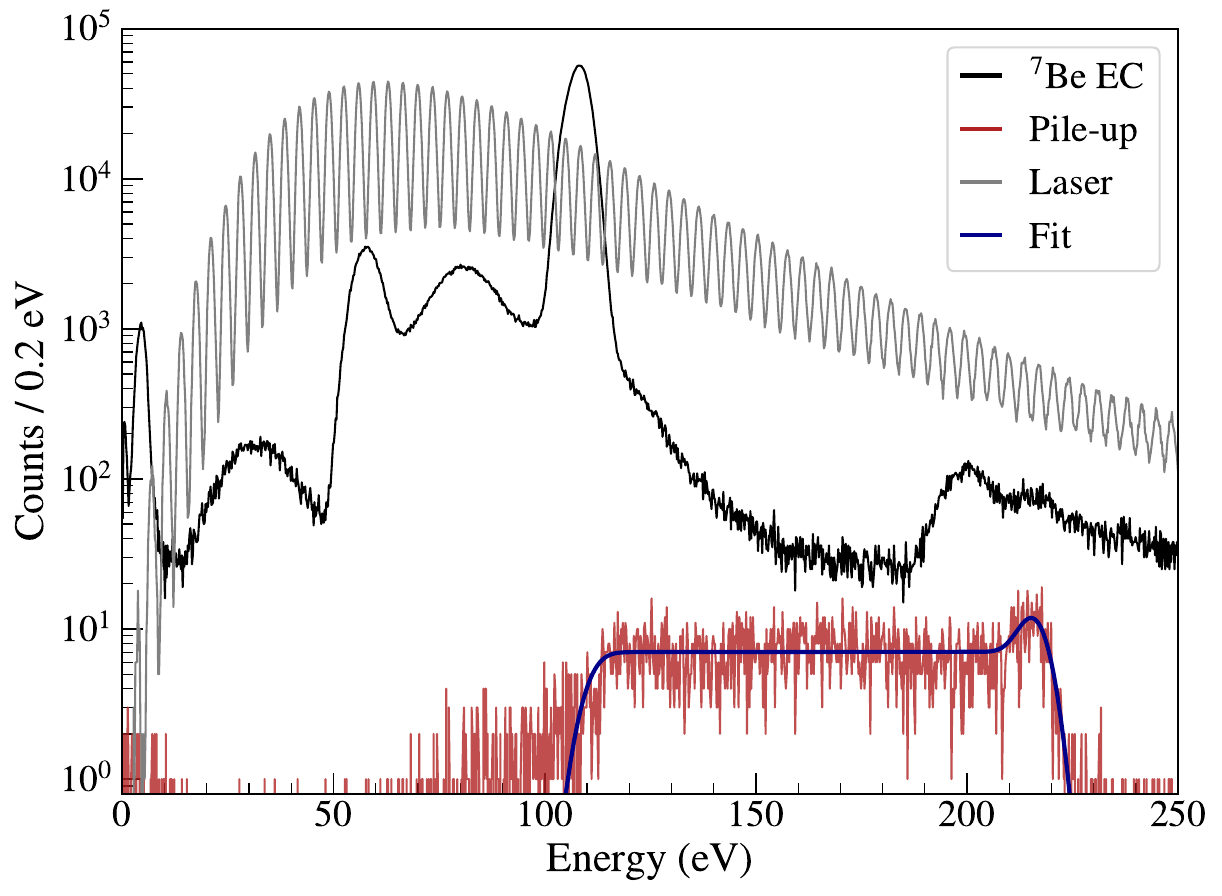}
    \caption{Spectrum for events with low $R_\textrm{short/long}$ values~(red), superimposed with pile-up and laser calibration spectra. The fit to pile-up spectrum is explained in Sec.~\ref{sec:pileup}. 
    The structure above 180~eV in the $^7$Be EC spectrum reflects the shake-up/shake-off effects, explained in Sec.~\ref{sec:shake}.}
    \label{fig:pileupspectrum}
\end{figure}

Figure~\ref{fig:pileupspectrum} shows the BeEST Phase-III spectrum for a one-day measurement from a single STJ pixel with the calibration laser events.  
The events characterized by low $R_\textrm{short/long}$ values are shown in the red spectrum, labeled as ``Pile-up''. 
The pile-up contribution was constant between the K-GS peak at $\approx$108 and $\approx$216~eV, with an additional peak at 216~eV due to two pulses arriving within 6~$\upmu$s.
The EC pile-up rejection efficiency dropped significantly when the two pulses are not separable with the detector timing resolution. 
Hence, instead of rejecting these events entirely, we included the spectral shape in our model fit~(Section~\ref{sec:pileup}).
The $^7$Be EC decay rates varied between 11~and~50~events/s in Dataset~1, with corresponding pile-up probabilities ranging from 0.6\% to 3\%, accounting for the 58$\upmu$s dead time. 
The pile-up probability decreased over time as the isotope decayed.

\subsection{Gamma-ray and muon induced background}~\label{sec:coincidence}

Another source of background was due to high-energy particles that interacted    in the Si substrate below the STJ detector array. 
These events were caused primarily by the 478~keV $\gamma-$rays that were emitted in 10.44(4)\% of the $^7$Be decays that populated the excited nuclear state of $^7$Li~\cite{nudat2002be7}. 
Cosmic muons and environmental radioactivity in the surrounding materials caused similar background events, but at a significantly lower rate. 
All these high-energy events in the Si substrate generated non-equilibrium phonons that could propagate to the STJ detector array, break Cooper pairs and cause a signal before thermalizing below the energy of the superconducting gap. 
Their distinguishing feature was that the phonons spread throughout the substrate and generated signals in several pixels simultaneously. 
Substrate events could therefore be identified by the simultaneous arrival of signals in at least 3~pixels within a coincidence window of 5.6~$\upmu$s. 
Considering the EC signal rates of $<100$~counts/s, the likelihood of a triple random coincidence is suppressed by $4\times 10^{-5}$ compared to the single hit event rate.

\begin{figure}
    \centering
    \includegraphics[width=\columnwidth]{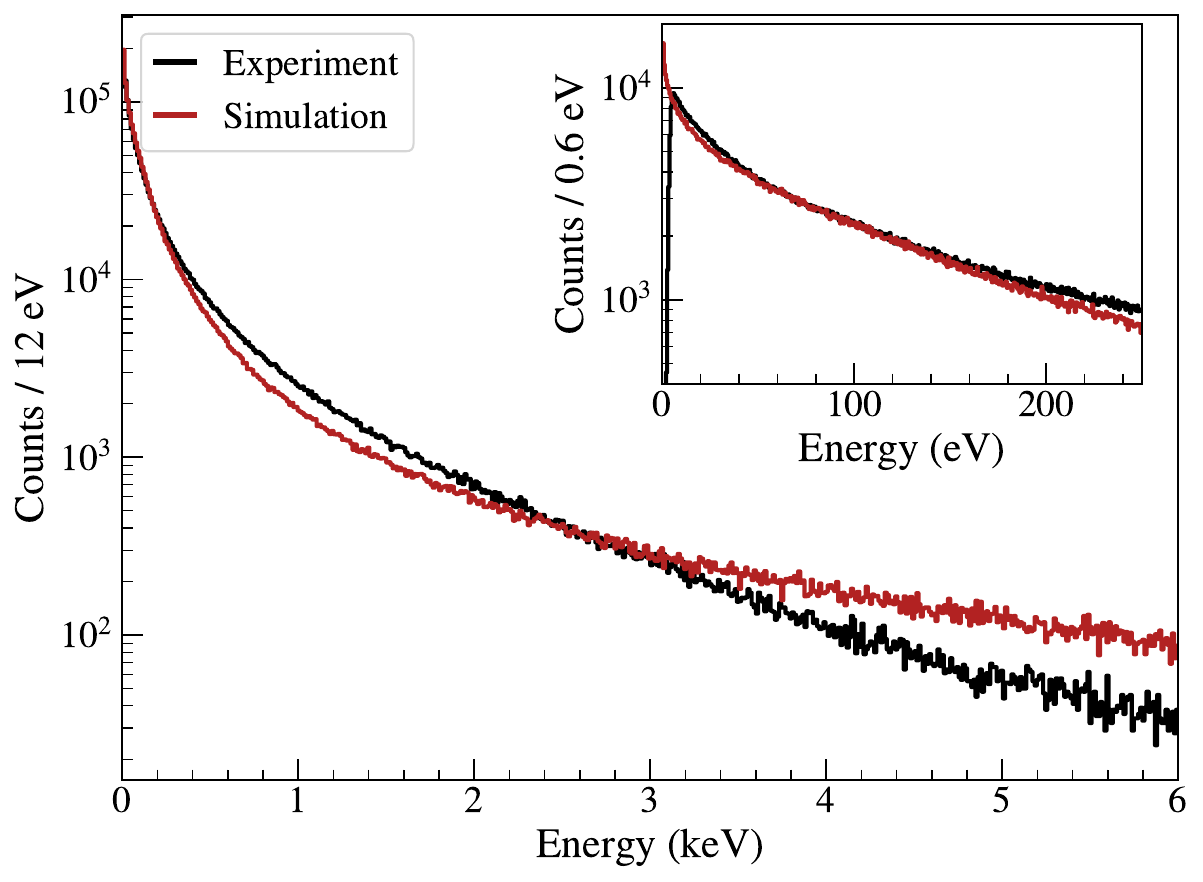}
    \caption{Background spectrum comparison of decay events with a detector multiplicity $\geq3$ from a single day measurement (black) to a Monte-Carlo model (red) that uses 478~keV $\gamma-$ray interactions in Si.}
    \label{fig:gammaSim}
\end{figure}

Figure~\ref{fig:gammaSim} shows the spectrum of the substrate events up to an energy of 6~keV. 
To first order, the distribution was given by the product of the energy deposited at a particular location in the Si substrate and the solid angle that an STJ detector subtends when seen from that location. 
We have simulated this distribution with a Monte-Carlo simulation, assuming a 478~keV gamma source uniformly distributed over the area of the detector array and a phonon transmission coefficient of 0.3 at the interface of the Ta base film to the oxidized Si substrate~\cite{kaplan1979acoustic}. 
The simulation with minimal assumptions exhibited a good agreement with the observed substrate spectra, confirming our assumption for the primary source of this background. 
The simulation showed some excess of events at low energies and fewer events at higher energies. 
This could be due to phonon thermalization below the gap energy during propagation in the Si substrate or by a detector non-linearity that increases at high energies. 
Phonon reflection at the substrate surface would also modify the spectrum. 
Including these effects can improve the fits, although the uncertainty about the relevant second-order effects makes it difficult to extract the underlying physics. 

\begin{figure}
    \centering
    \includegraphics[width=\columnwidth]{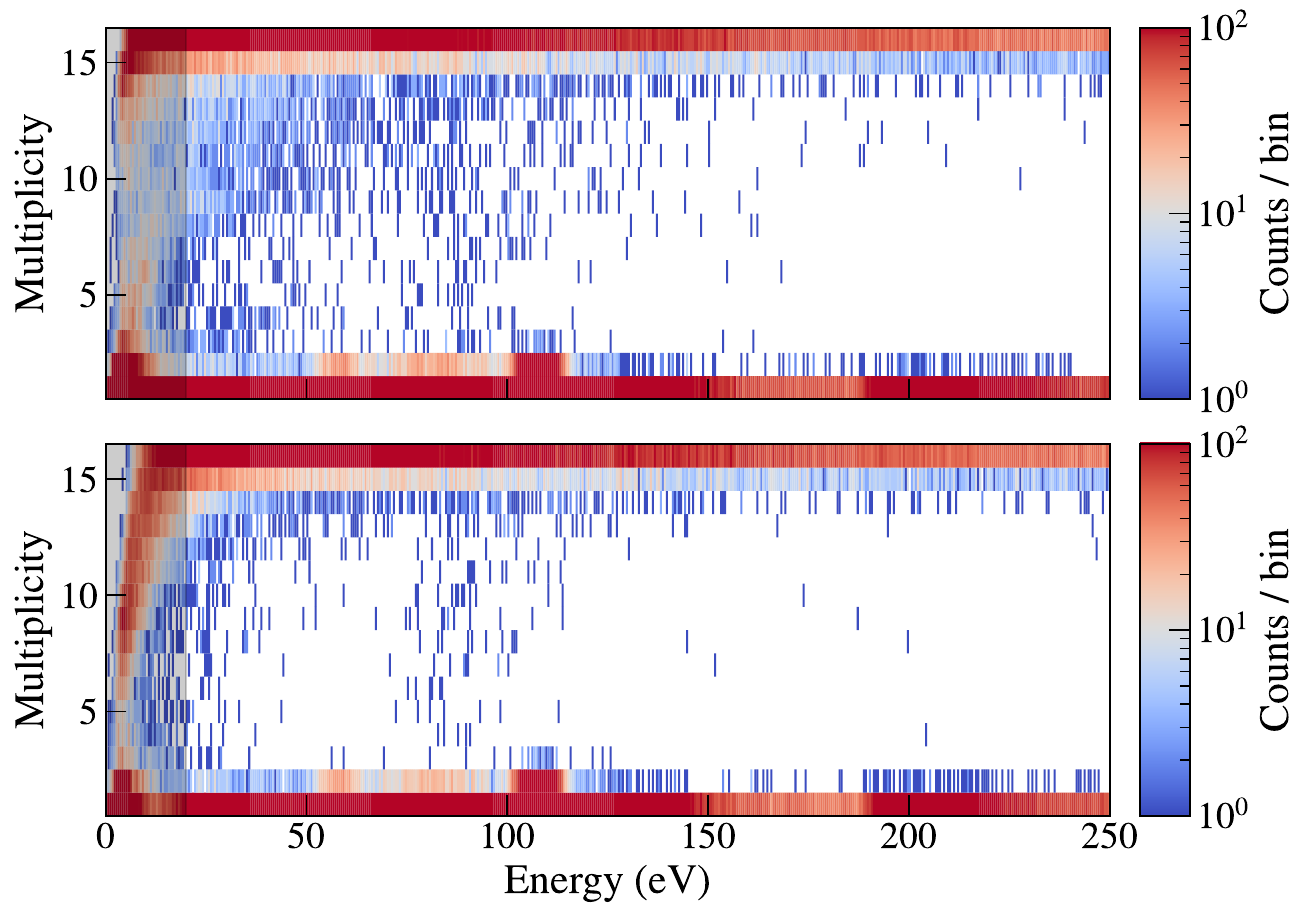}
    \caption{Multiplicity vs. energy for two channels. 
    High-multiplicity events were due to gamma and muon interactions in the substrate that generate simultaneous events in all pixels, except at low energies. Muliplicity-2 events at high energies were due to random coincidences and due to $^7$Be implanted between pixels at low energies. The gray shaded region is below the analysis threshold of 20~eV.}
    \label{fig:multiplicityVsE}
\end{figure}

The efficiency of our background rejection was expected to degrade at low energies, e.g. for substrate events caused by small-angle scattering close to one of the pixels that might trigger only one or two pixels in the immediate vicinity. 
 We have examined the magnitude of this effect by plotting the multiplicity distribution of signals as a function of energy (Fig.~\ref{fig:multiplicityVsE}). 
The reduced multiplicity is certainly observable, although it remains >5 even for energies significantly below the 20 eV analysis threshold. 
Certain channels exhibited additional low-multiplicity events at energies below 15~eV. 
However, these events form a distinctive continuum that was separated from the main 478 keV $\gamma-$distribution, indicating that the origin of these events was different.
We concluded that the tagging and removal efficiency of substrate events from the 478 keV $\gamma-$rays and muons is $\approx100$\% above 20~eV.

\subsection{$^7$Be implanted between pixels}~\label{sec:delayed}

Events with a multiplicity of 2 fall into two categories. 
One is due to random coincidences, the other is due to $^7$Be implanted between pixels. 
While the Si collimator in front of the detector array greatly reduced the number of $^7$Be ions implanted between detector pixels, its distance of $\approx$100~$\upmu$m above the STJ chip still allowed scattered ions to be implanted in the area that was nominally shielded. 
The Phase-III data showed that even scattered ions could be implanted deep enough so that they were not removed by repeated ethanol rinses after implantation.

\begin{figure}
    \centering
    \includegraphics[width=\columnwidth]{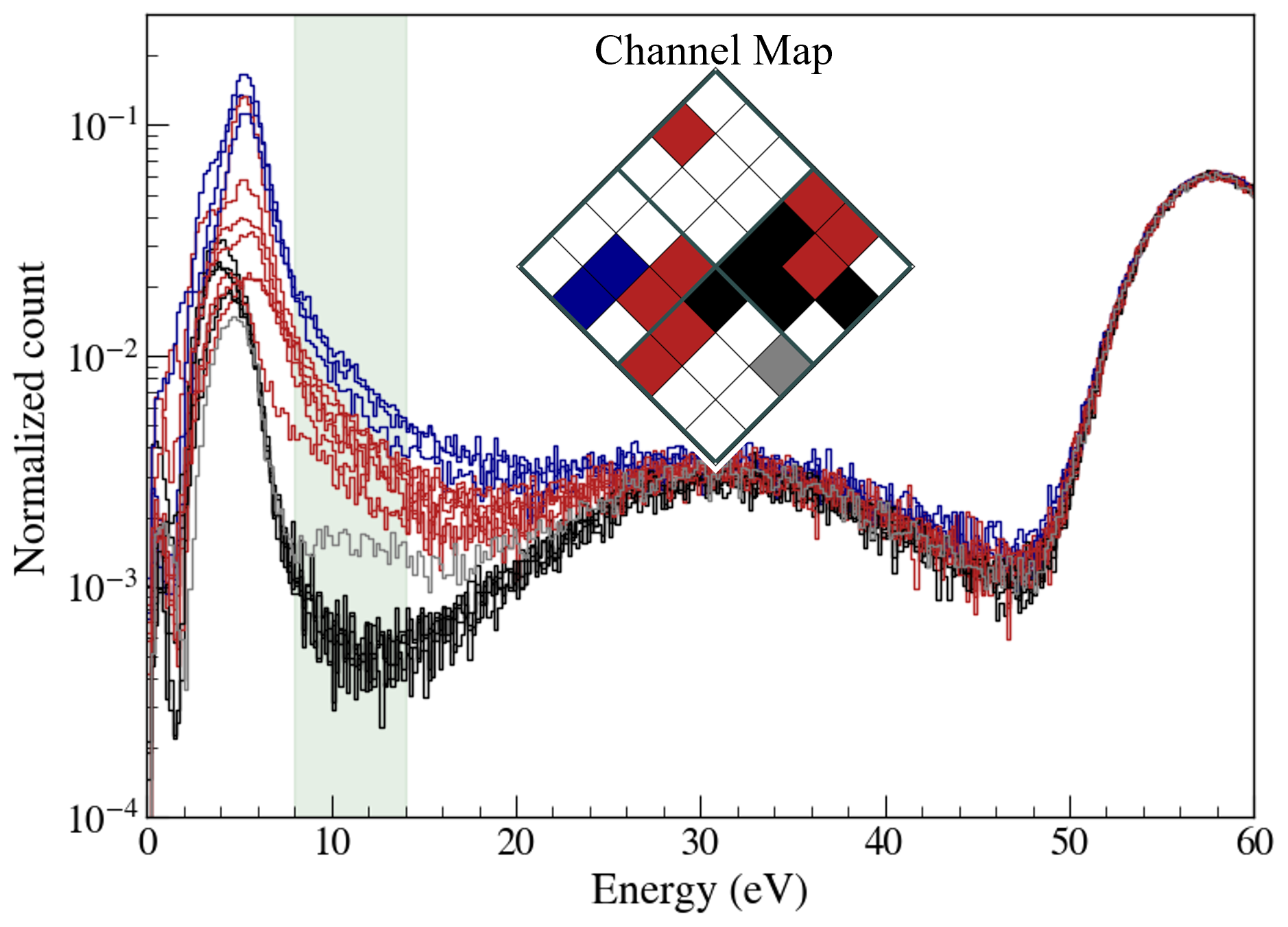}
     \caption{Comparison of low-energy events across the 15 pixels over a single day of measurement within the [0,60]~eV energy range. The spectra have been normalized to ensure that the K-GS peaks align with consistent amplitude. A position-dependent variance in the low-energy background, particularly within the [8,14]~eV region shaded in green, is evident.}
    \label{fig:lowESubstrate}
\end{figure}

Although the spectra from all STJ pixels exhibited almost identical patterns in the higher energy regions, there was a significant discrepancy in the spectral shapes of the low-energy features across different pixels.  Fig.~\ref{fig:lowESubstrate} illustrates that the low-energy background was the lowest in the area around the central pixel (black) and increased with distance from that area (red, blue).  
We attribute this to a misaligned Si collimator during $^7$Be implantation. 
The collimator could have been well-aligned near the center, accounting for the low continuum background in it and the neighbouring pixels. 
A slight rotation of the collimator could result in a higher $^7$Be implantation into the silicon substrate at distant pixels.
The misalignment may not be surprising given that  the Si collimator used for $^7$Be implantation in Phase-III was aligned and glued manually under a microscope.

\begin{figure}
    \centering
    \includegraphics[width=\columnwidth]{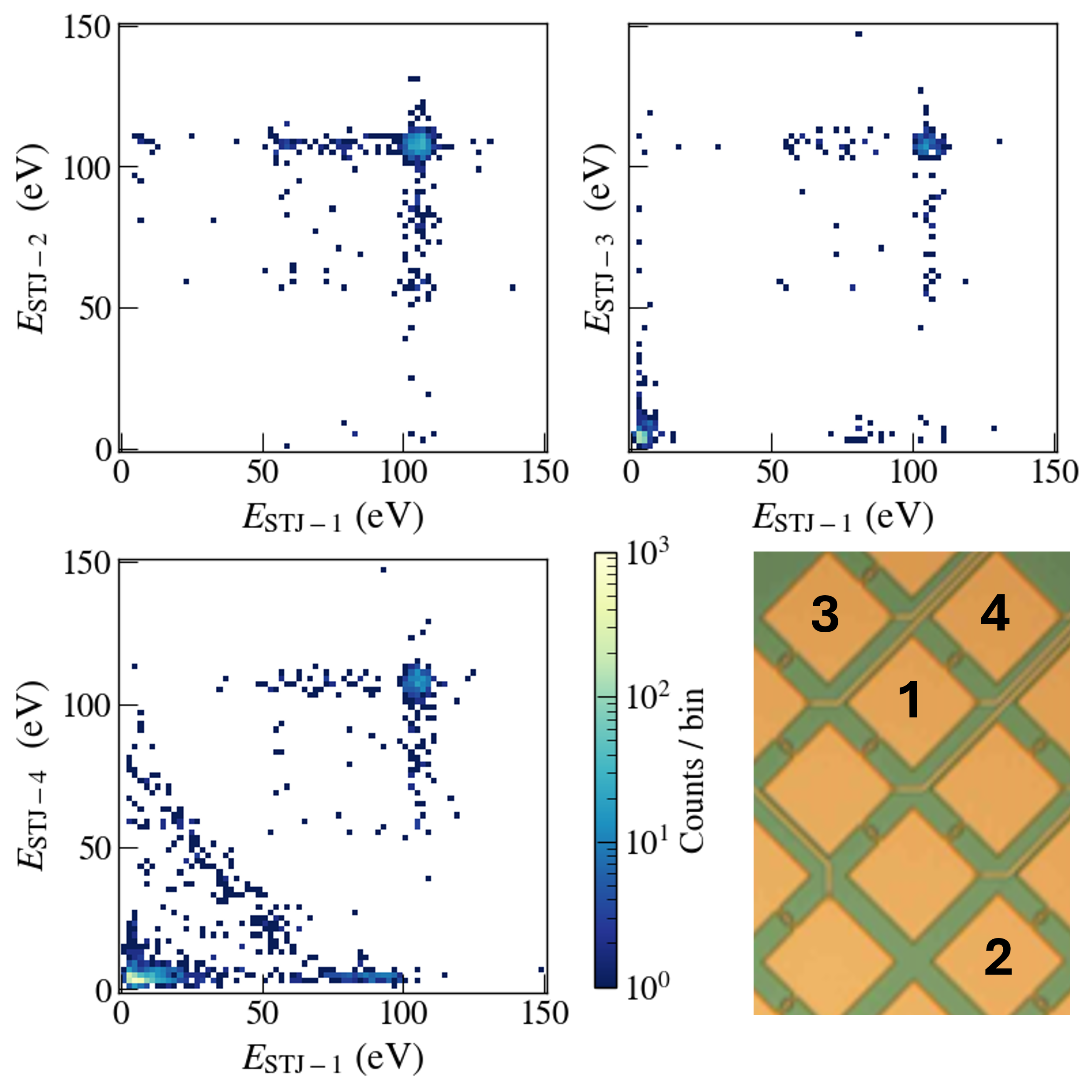}
    \caption{
    Energy correlation of multiplicity 2 events for three different combinations of detector pixels. Distant pixels~(top left) only showed random coincidences, dominated by the strong K-GS peak at 108~eV. 
    Neighboring pixels~(top right and bottom left) also showed coincidence at low energies due to $^7$Be implanted between pixels. 
    Neighboring pixels with a ground wire between them (bottom left) showed a band of events whose energies add up to $\approx$70~eV.}
    \label{fig:m2}
\end{figure}

Figure~\ref{fig:m2} illustrates the multiplicity-2 events in three different combinations of detector pixels. 
While distant pixels~(top~left) showed only random coincidences, adjacent pixels showed additional low energy clusters below 20~eV, indicating the presence of $^7$Be implanted in the Si substrate between them. 
Furthermore, pixels with a shared superconducting ground wire between them showed additional events whose energy adds up to $\approx$70~eV~(bottom~left).
Signal charges from these events were shared between the two pixels, with the relative fraction depending on the location of the decay along the ground connection. 

\begin{figure}
    \centering
    \includegraphics[width=\columnwidth]{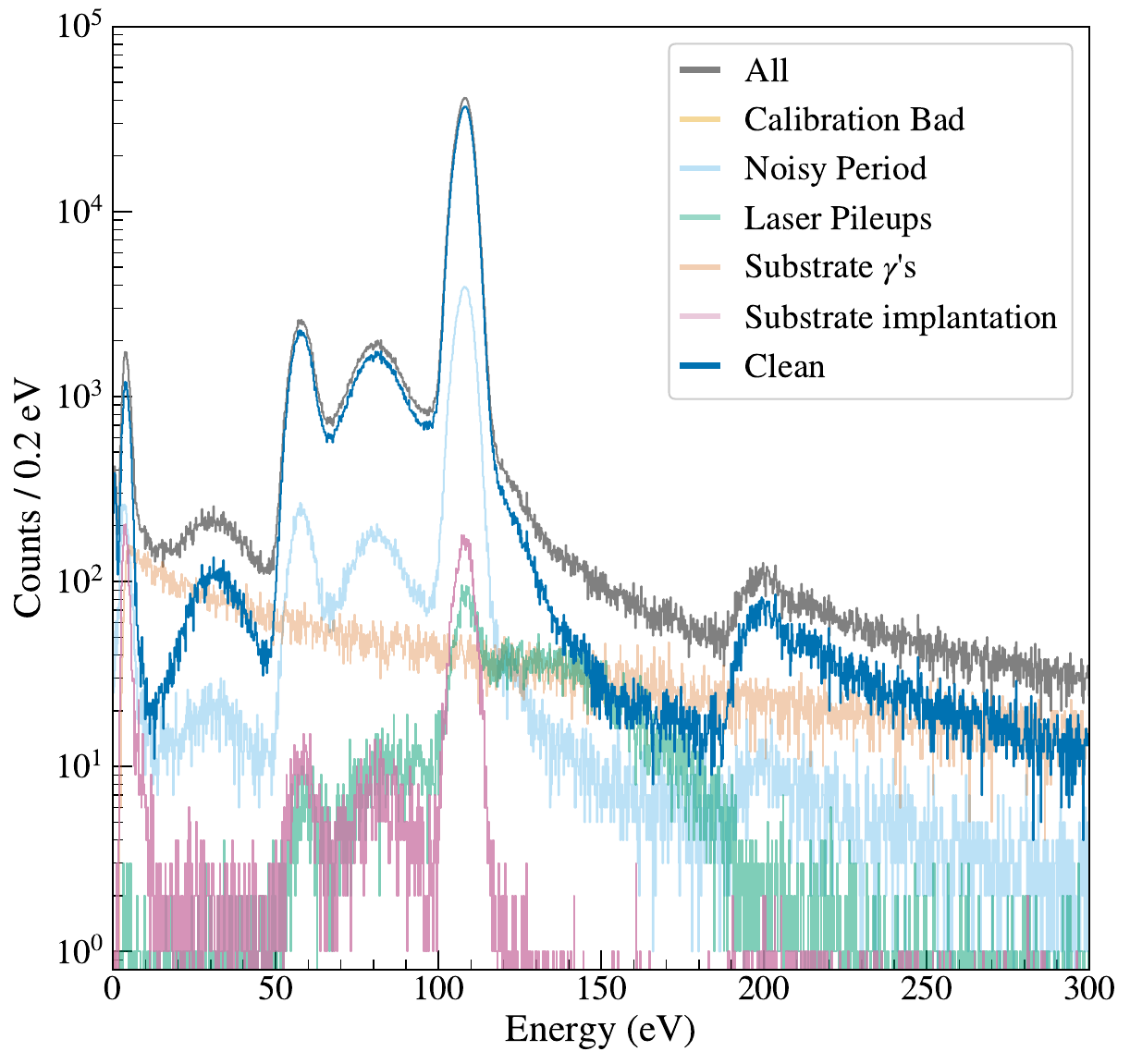}
    \caption{Various $^7$Be EC decay spectra of events that were rejected by each data cleaning step. The clean spectra were fit to the background model~(Section~\ref{sec:bkg}).}
    \label{fig:spectra_rejection}
\end{figure}

The implants in the Si substrate would also most commonly deposit $\approx$100 eV, but their energy was coupled to the detectors only through phonon transport. 
For decays immediately adjacent to a pixel, at most $\approx$50\% of the phonons would propagate towards the pixel, and some of them would propagate into the substrate without interacting with the detector interface. 
The remaining fraction has a probability of $\approx$30\% to propagate through the interface into the detector and produce a signal, while the other $\approx$70\% will be reflected into the substrate~\cite{kaplan1979acoustic}. 
The maximum energy deposited in the STJ detector was therefore around $\approx$15~eV for K-GS events in the substrate. 
Decays in the Si substrate at increasing distances from a detector would produce correspondingly smaller signals eventually failing to pass the trigger condition.
Therefore, we rejected all events with multiplicity 2.
Since the tagging efficiency of this background was less than 100\%, we include substrate events in our background modeling~(Section~\ref{sec:bkg}).  
Figure~\ref{fig:spectra_rejection} summarizes the impact of each data cleaning step on the spectra. 

\begin{figure}
    \centering
    \includegraphics[width=\columnwidth]{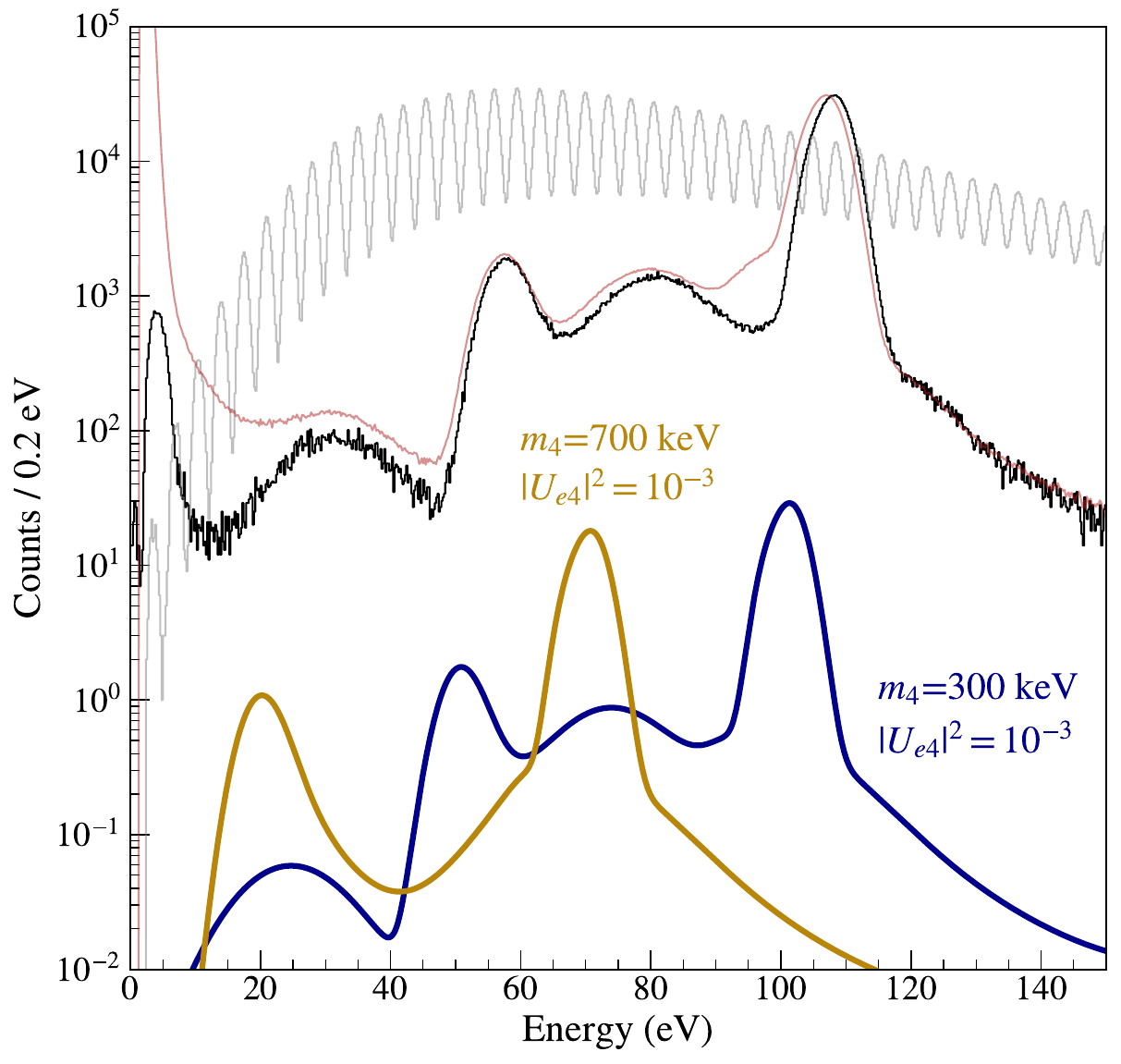}
    \caption{Comparison of the BeEST Phase-III spectrum to the Phase-II spectrum. The Phase-III EC spectrum after data cleaning (one channel, one day) is shown in black, the normalized Phase-II EC spectrum in red, and the laser spectrum in gray. The shift in the EC spectrum is due to the gain correction described in Sec.~\ref{sec:energy}.
    Hypothetical heavy neutrino signatures with $|U_\textrm{e4}|^2=0.001$ are overlaid in blue ($m_4=300$~keV) and yellow ($m_4=700$~keV). No excited state peaks are expected in the heavy neutrino signature above the 478~keV $Q$ value of the excited state decay.}
    \label{fig:comparisonToPhaseII}
\end{figure}

Figure~\ref{fig:comparisonToPhaseII} compares the BeEST Phase-III spectrum from a single pixel over one day of measurement~(black) after data cleaning to the Phase-II spectrum, normalized to the same K-GS peak amplitude~(red).  
The hypothetical heavy neutrino signatures of $m_4=300$~keV and $m_4=700$~keV with $|U_\textrm{e4}|^2=0.001$ are illustrated in blue and yellow, respectively. 
No excited-state peaks are anticipated in the heavy neutrino signature beyond the 478~keV $Q$ value of the excited-state decay.
This comparison clearly highlights the impact of the gain correction based on the laser intensity~(Sec.~\ref{sec:energy}), which appears as shifted peak positions. 
Improvements in background reduction are also evident, with a decrease in Auger-electron escape~(85-105~eV, Sec.~\ref{sec:escape}), rejection of $\gamma$ substrate events~(below 75~eV, Sec.~\ref{sec:coincidence}), and vetoing of events from $^7$Be implanted between pixels~(below 75~eV, Sec.~\ref{sec:delayed}).
Noise reduction below 20~eV, from enhanced offline triggering~(Sec.~\ref{sec:trigger}) and high-precision pulse shape analysis~(Sec.~\ref{sec:pickup}), also demonstrates the effectiveness of the new data cleaning procedure adapted in Phase-III.
The clean spectra could then be fit to an analytical model~(Section~\ref{sec:bkg}).

\section{Spectrum Modeling}\label{sec:bkg}

The background in our search for BSM physics is dominated by $^7$Be decay signals with the emission of an active neutrino. 
The neutrino does not interact in the sensor, and the measured signal reflects the recoil energy of the $^7$Li daughter nucleus plus the energy of the Li atomic recombination cascade to the ground state for the different decay channels.
Smaller contributions at low energies are due to substrate events that evaded the data cleaning procedures. 
Background from external sources is negligible because the STJ detectors have thickness of $<1~\upmu$m and muon interactions in the Si substrate are rejected along with $\gamma$-induced events. 

$^7$Be can capture a K or an L shell electron and decay either into the nuclear ground state or the first excited state of $^7$Li. 
The two-body nature of the EC final state and energy-momentum conservation allow for precise calculation of the nuclear recoil energy:

\begin{equation}
    E_\textrm{recoil}=\frac{Q_\textrm{EC}^2-m_\nu^2 c^4}{2(Q_\textrm{EC}+m_{\textrm{Li-7}}c^2)}~,
\end{equation}

\noindent
where $Q_\textrm{EC}$ is the $Q-$value of the EC decay, $m_{\textrm{Li-7}}c^2 = 6.5353663(5)$~GeV is the mass of the $^7$Li daughter nucleus~\cite{audi2012ame2012} and $m_\nu c^2$ is the mass of the active neutrino, taken as 0. 
For decays into the ground state of $^7$Li, the entire $Q_\textrm{EC}$ value of 861.963(23)~keV is released in a single transition that imparts a recoil energy of 56.836(3)~eV on the  $^7$Li daughter nucleus~\cite{bhandari2024q}. 
Decay into the excited state  $^7$Li$^*$  initially imparts a recoil energy of 11.30~eV, and  $^7$Li$^*$ subsequently decays to the ground state with a half-life of 72.8(20)~fs~\cite{nudat2002be7} emitting a 477.603(2)~keV $\gamma$-ray which adds another 17.45~eV recoil~\cite{beest2021phase2}.
Since the $\gamma$-emission occurs before the $^7$Li$^*$ nucleus has come to rest and its direction is independent of the $^7$Li$^*$ motion, decays into the excited state are Doppler broadened, with the exact line shape depending on the relaxation dynamics~\cite{bray2024doppler}.

K-capture creates a hole in the 1s level that relaxes through the emission of an Auger electron or an X-ray. 
This low-energy radiation is immediately re-absorbed and starts a relaxation cascade that deposits the  energy of the 1s core hole inside the detector within $\approx$1~ns~\cite{zehnder1995response}. 
If the entire relaxation energy is deposited in the sensor, the peak energy is equal to the sum of the $^7$Li recoil energy and the binding energy of the Li shell. The background spectrum therefore has four primary peaks: Two for capture of a K-electron and decay into the ground state~(K-GS) and excited state~(K-ES) of  $^7$Li, and two for the corresponding L capture peaks, (L-GS) and (L-ES)~(Section~\ref{sec:recoil}).
Auger electrons escaping from surface-implanted $^7$Be with partial energy deposition produce tails below the two K-capture peaks~(Section~\ref{sec:escape}).

For elemental Li, the 1s binding energy relative to the Fermi surface is 54.75~eV~\cite{bearden1967energylevel}, although the chemical environment and variations in the implant site symmetry in the Ta absorber can shift 1s binding energies by up to $\pm$2~eV~\cite{samanta2023material}. 
L-capture creates a hole in the n=2 shell, which is hybridized with the Ta 5d levels over a band of $\approx$5~eV width~\cite{samanta2023material}. 
The lifetime of the 1s hole adds a Lorentzian line width  $\gamma=0.03$~eV to the K-capture peaks~\cite{citrin1977broadening}. 
The lifetime of the 2s hole is sufficiently long to make lifetime broadening of the L-capture peaks negligible.

The vacancy created by the electron capture process, together with the accompanying nuclear charge modification, creates a change in the atomic potential felt by the $^7$Li electrons.
This sudden disturbance can result in a excitation in the remaining system.
The wave function of remaining outer-shell electrons is perturbed, and the electrons find themselves in a superposition of eigenstates, rather than a proper eigenstate of the atom. 
This perturbation is referred to as shake-up~(SU) if one or more electrons that did not participate in the EC decay are excited to a higher level within the atom in their final states. 
If the electrons are ejected to the continuum, the process is called shake-off~(SO). 
As the ``shaken'' electrons transition to higher energy states, they deposit extra energy in the detector. 
Shake-up~(SU) produces peaks at higher energy, and shake-off~(SO) produce the high energy tails above the main peaks~(Section~\ref{sec:shake}).

This section discusses the details of different physics contributions to the background spectrum and the model functions used in the spectral fits.

\subsection{Primary EC decay peaks}\label{sec:recoil}

In Phase-II, we found that a fit to the K-GS peak required the use of three Voigt functions $V(E;\mu_i,\sigma_i,A_i)$ with fixed lifetime broadening $\gamma=0.03$~eV. 
Here, $\mu_i,\sigma_i$ and $A_i$ are centroid, standard deviation and area of the peaks, and the subscript $i=1,2,3$ denotes individual component.
The central component of the peak contained $\approx$90\% of the total counts in the peak, and we associate it with the unperturbed position of the K-GS peak. 
The origin of the smaller component at lower energy is not fully understood yet, but might be associated with energy loss by secondary radiation~(Section~\ref{sec:secondary}) or the formation of lattice defects. 
An accurate fit to the high-energy side of the K-GS peak requires a third component, which is attributed to Li L electron shake-up~(Section~\ref{sec:shake}).  
For the L-capture peak, we observed no evidence of a lower-energy component or any correlation with secondary radiation~(Section~\ref{sec:secondary}). 
While the high-energy component due to electron shake-up is expected, it is entirely obscured by the broadened L-GS peak and the L-shake-off tail, which is anticipated to contain twice as many counts as the L shake-off tail for K capture~\cite{guerra2024shaking}.
Therefore, the L-capture peak into the ground state of $^7$Li~(L-GS) is fit with a single Gaussian function $G(E;\mu,\sigma,A)$, which reduces the risk of overfitting. 
The excited state peaks inherit the peak shape parameters from the ground state peaks, with their relative amplitudes initially set to 0.11657 to account for the \textsuperscript{7}Be branching ratio of 10.44(4)\% into the excited state~\cite{nudat2002be7}. 
They are convolved with a Gaussian function to model the Doppler broadening. 
The four primary peaks are therefore fit to the following functions:

\begin{equation}\label{eq:peaks}
    \begin{split}
    \small
        \mathrm{K-GS}&: \sum_{i=1}^3 V(E;\mu_{\mathrm{KGS},i},\sigma_{\mathrm{K}},A_{\mathrm{KGS},i}) \\
        \mathrm{K-ES}&: \Bigg[\sum_{i=1}^3 V(E;\mu_{\mathrm{KES},i},\sigma_{\mathrm{K}},0.11657A_{\mathrm{KGS},i})\Bigg]\circledast G(\sigma_{\mathrm{Doppler}}) \\
         \mathrm{L-GS}&: G(E;\mu_\mathrm{LGS},\sigma_\mathrm{L},A_{\mathrm{LGS}})\\
         \mathrm{L-ES}&: G(E;\mu_\mathrm{LES},\sigma_\mathrm{L},0.11657A_{\mathrm{LGS}})\circledast G(\sigma_{\mathrm{Doppler}})~,
    \end{split}
\end{equation}

\noindent
with common values of $\sigma_\mathrm{K}$ used for the three Voigt functions.
The symbol $\circledast G(\sigma_{\mathrm{Doppler}})$ denotes the Gaussian convolution to model the Doppler broadening, $\sigma_{\mathrm{Doppler}}$. 
Parameters determining the models are summarized in Table~\ref{tab:summary_main}.

\newpage

\subsection{Electron escape} \label{sec:escape}

The K-GS peaks exhibit a low-energy tail that extends into the region of the K-ES peak. 
This is the result of electron escape during the initial energy relaxation of the Auger electron for atoms that were deposited close to the surface of the STJ~\cite{bray2022monte}. 
The tails are a factor of 5 smaller in Phase-III than in Phase-II of the BeEST experiment~(Fig.~\ref{fig:comparisonToPhaseII}), because in Phase-III we repeatedly rinsed the detector chip with ethanol after \textsuperscript{7}Be implantation to remove $^7$Be at the surface. 
This tail was fit by an exponentially modified Gaussian that is offset from the K-GS peak by the minimum loss energy $E_\mathrm{Auger}$~\cite{beest2020lkratio,beest2021phase2}:

\begin{widetext}
\begin{equation}
\begin{split}
    \mathcal{P}_\mathrm{KGS-Auger}(E; &A_\mathrm{KGS-Auger},\mu_\mathrm{KGS},\sigma_\mathrm{K},E_\mathrm{Auger},k_\mathrm{Auger})\\
    & =\frac{A_\mathrm{KGS-Auger}}{2k_\mathrm{Auger}}\mathrm{exp}\Bigg(\frac{1}{k_\mathrm{Auger}}\Big((E-\mu_\mathrm{KGS}-E_\mathrm{Auger})+\frac{\lambda\sigma_\mathrm{K}^2}{2}\Big)\Bigg)\mathrm{erfc}\Bigg(\frac{(E-\mu_\mathrm{KGS}-E_\mathrm{Auger})}{\sqrt{2}\sigma_\mathrm{K}}+\frac{\sigma_\mathrm{K}}{\sqrt{2}k_\mathrm{Auger}}\Bigg)\\
    \mathcal{P}_\mathrm{KES-Auger}(E; &A_\mathrm{KGS-Auger},\mu_\mathrm{KES},\sigma_\mathrm{K},E_\mathrm{Auger},k_\mathrm{Auger})\\
    & =\frac{A_\mathrm{KGS-Auger}}{2k_\mathrm{Auger}}\mathrm{exp}\Bigg(\frac{1}{k_\mathrm{Auger}}\Big((E-\mu_\mathrm{KES}-E_\mathrm{Auger})+\frac{\lambda\sigma_\mathrm{K}^2}{2}\Big)\Bigg)\mathrm{erfc}\Bigg(\frac{(E-\mu_\mathrm{KES}-E_\mathrm{Auger})}{\sqrt{2}\sigma_\mathrm{K}}+\frac{\sigma_\mathrm{K}}{\sqrt{2}k_\mathrm{Auger}}\Bigg) \\ 
    &\circledast G(\sigma_{\mathrm{Doppler}})\times0.11657~.
\end{split}
\end{equation}
\end{widetext}

\noindent
Here, $A_\mathrm{KGS-Auger}$ represents the area of the K-GS Auger electron escape tail, $E_\mathrm{Auger}$ denotes the energy shift of the electron escape tail relative to the primary K-GS centroid  and $k_\mathrm{Auger}$ is the skewness parameter that sets the slope of the tail.

While this tail is not visible below the K-ES peak, the same escape mechanism will apply for K-capture decay into the excited state. 
We therefore add an exponentially modified Gaussian tail to the K-ES peak, constrained to the same fractional probability as for the K-GS peak but convolved with the Doppler broadening. 
We do not expect similar low-energy tails for the L-capture peaks, since the electron binding energies for L electrons are mostly below the work function in Ta~\cite{samanta2023material}.

\begin{figure}
    \centering
    \includegraphics[width=\columnwidth]{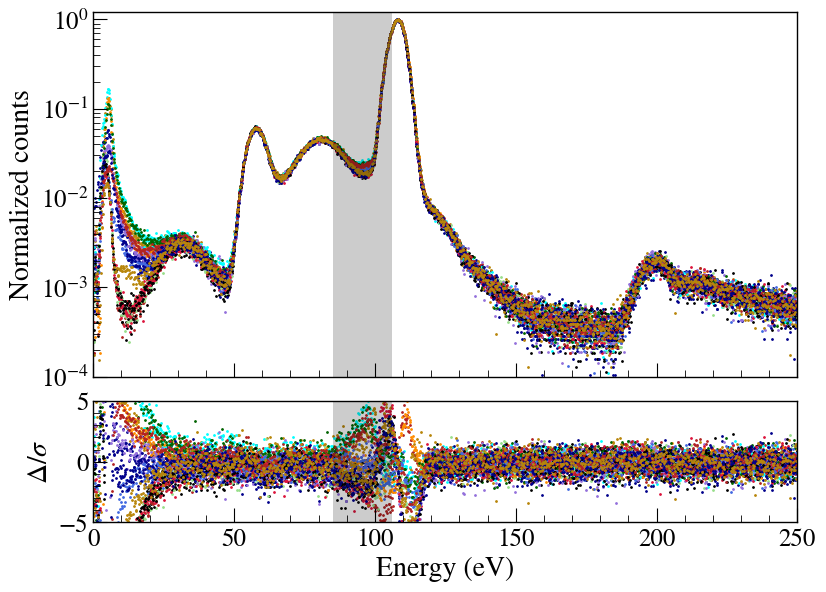}
    \caption{(Top) Normalized $^7$Be EC decay spectra from all 15 pixels from a single day of measurement. (Bottom) Residuals from the average spectrum.  
    Variations in the measured spectra are primarily due to variations in the $^7$Be implant distribution. This causes variations in the signals from implants into the Si substrate at low energies and in the electron escape tail between ~85~and~105~eV (gray band). The deviation at $\approx110$~eV is due to small difference in energy resolution between pixels~(Fig.~\ref{fig:laserSpectrum}).}
    \label{fig:histogramComparison}
\end{figure}

Figure~\ref{fig:histogramComparison} illustrates the normalized spectra from all 15 STJ channels. 
The spectra agree well with one another except at very low energies and in the [85,106]~eV region where the electron escape tail dominates. 
These variations are due to small differences in the $^7$Be implant distribution in each pixel.

\subsection{Shake-up and shake-off}~\label{sec:shake}

Shake-up and shake-off are possible for K- and L-electrons, after both K-capture and L-capture decays of $^7$Be. 
Each of the four electron capture peaks is therefore accompanied by two shake-up peaks (KSU and LSU) and two broad shake-off tails (KSO and LSO) at higher energies. 
To reduce the number of free parameters, we neglect the LSU peaks because they are low in energy and accounted for in the fit by broadening of the four primary peaks. 
We also neglect the KSU peaks for L-capture events because they overlap with the LSU peaks of the corresponding K-capture events. 
Finally, we neglect the KSO tails for the L-capture events because they are subdominant and overlap with the SO tails of the K-capture events. 
This leaves two shake-up peaks (KGS-KSU and KES-KSU) and six shake-off tails (KGS-LSO, KGS-KSO, KES-LSO, KES-KSO, LGS-LSO and LES-LSO) for the fit. 

We fit the SU peaks to Gaussian functions $G(E;\upmu,\sigma,A)$ parametrized by their centroids $\mu_\textrm{KGS-KSU}$ and $\mu_\textrm{KES-KSU}$; widths $\sigma_\textrm{KGS-KSU}$ and $\sigma_\textrm{KES-KSU}$; and areas $A_\textrm{KGS-KSU}$ and $A_\textrm{KES-KSU}$.
The K-ES K-SU peak is Doppler broadened and therefore  $\sigma_\textrm{KES-KSU}>\sigma_\textrm{KGS-KSU}$. 
The two areas are constrained to have the same ratio with the K-GS and K-ES areas, i.e., $A_\textrm{KGS-KSU}/A_\textrm{KGS}=A_\textrm{KES-KSU}/A_\textrm{KES}$, as the shake-up probability is not expected to depend on the nuclear state of $^7$Li that is populated in the decay. 
In Phase-II, we fit the SO tails to a function first introduced by Levinger in 1953~\cite{levinger1953electrons}, which has been used in previous precision decay studies~\cite{robertson2020shake} and which we have found to provide a reasonable description to the measured data despite the approximations used in its derivation~\cite{beest2021phase2}.
For Phase-III, we instead use a power function for the K-shake-off and a log-normal distribution for the L-shake-off probability density functions~(PDFs) respectively. 
This approach better aligns with new calculations of the atomic structure before and after electron capture, and will be discussed in detail in a future publication~\cite{guerra2024shaking}. 

\begin{widetext}
\begin{equation}
\begin{split}
    \mathrm{KGS-KSO}:&  \Bigg[A_\mathrm{KGS-KSO}~\Big(E-\mu_{\mathrm{KGS},1}\Big)^{p}\Bigg]\Bigg[\frac{-E_\mathrm{KGS-KSO}^{(p+1)}}{p+1} \Theta\Big(E-(\mu_{\mathrm{KGS},1}+E_\mathrm{KGS-KSO})\Big)\Bigg] \\
    \mathrm{LGS-KSO}:&  \Bigg[A_\mathrm{LGS-KSO}~\Big(E-\mu_{\mathrm{LGS}}\Big)^{p}\Bigg]\Bigg[\frac{-E_\mathrm{LGS-KSO}^{(p+1)}}{p+1} \Theta\Big(E-(\mu_{\mathrm{LGS}}+E_\mathrm{LGS-KSO})\Big)\Bigg] \\
    \mathrm{KGS-LSO}:& A_{\mathrm{KGS-LSO}}~\frac{1}{\sqrt{2\pi}s_{\mathrm{KGS-LSO}}\Big(E-E_\mathrm{KGS-LSO}\Big)} \mathrm{exp}\Bigg(-\frac{\mathrm{log}^2\big((E-E_\mathrm{KGS-LSO})/a_{\mathrm{KGS-LSO}}\big)}{2s_{\mathrm{KGS-LSO}}^2}\Bigg) \\ 
    \mathrm{LGS-LSO}:& A_{\mathrm{LGS-LSO}}~\frac{1}{\sqrt{2\pi}s_{\mathrm{LGS-LSO}}\Big(E-E_\mathrm{LGS-LSO}\Big)} \mathrm{exp}\Bigg(-\frac{\mathrm{log}^2\big((E-E_\mathrm{LGS-LSO})/a_{\mathrm{LGS-LSO}}\big)}{2s_{\mathrm{LGS-LSO}}^2}\Bigg) \\ 
\end{split}
\end{equation}
\end{widetext}

\noindent
Here, $p$ is the power law decay scale parameter; $E_\mathrm{KGS-KSO}, E_\mathrm{LGS-KSO}, E_\mathrm{KGS-LSO}$ and $E_\mathrm{LGS-LSO}$ are the onset energies of the shake-off tails. 
For the log-normal distributions, $s_{\mathrm{KGS-LSO}}$ and $s_{\mathrm{LGS-LSO}}$ are the standard distribution and $a_{\mathrm{KGS-LSO}}$ and $a_{\mathrm{LGS-LSO}}$ are the energy scale parameters.
$\Theta$ is the Heaviside step function.
These PDFs are convolved with a Gaussian function with the same rms width  $\sigma_\mathrm{K,L}$  as the peaks to account for the observed broadening.
We no longer constrain the area of the shake-off tail to the same fraction for L- and K-capture peaks since L captures are predicted to have stronger shake-off probabilities than K captures~\cite{guerra2024shaking}.
The fit results in the region from >220~eV, where the K-GS K-SO spectrum is dominant, again shows good agreement with the data.

\subsection{Pile-up}~\label{sec:pileup}

The background model for pile-up of two K-GS events~(Fig.~\ref{fig:pileupspectrum}, blue) consists of a flat section from $\mu_{\mathrm{KGS},1}$ to $2\mu_{\mathrm{KGS},1}$ and an exponentially modified Gaussian peak at $2\mu_{\mathrm{KGS},1}$ with the pile-up resolution $\sigma_\mathrm{PU}$ and the skewness parameter $k_\mathrm{PU}$.
This function models the pile-up events discussed in Section~\ref{sec:pileupdc}, and matches the observed data well. 
This pile-up model is included in our fit of the background.
Pile-up contributions from other combinations, such as (K-GS,~L-GS) or (K-GS,~K-ES), are an order of magnitude weaker and therefore neglected.

\subsection{ $^7$Be implanted into Si substrate}

As discussed in Sec.~\ref{sec:delayed}, the background from $^7$Be implantation in the Si substrate is heavily geometry-dependent, because the alignment of the Si collimator and the amount of $^7$Be scattering from it varies between pixels. 
In addition, this background cannot be fully rejected by coincidence vetoing because not all of these events produce a signal in multiple pixels. 
Instead, this background is included in the model as a single exponential in the ROI. 
Initial estimates for the amplitude of this background are extracted from the counts in the [12,14]~eV region~(Fig.~\ref{fig:lowESubstrate}). 
In some pixels, this approximation produces accurate fits down to the lowest energies~(Fig.~\ref{fig:4eVpeak}).
In others, a single exponential does not fit the data within the statistical accuracy for energies <20 eV (Fig.~\ref{fig:lowESubstrate}). 
We therefore restrict the analysis of the spectra to energies >20~eV and model the remaining background as a single exponentially decaying spectrum.
For future experiments, it is desirable to remove this background completely by replacing the Si collimator with a patterned photoresist collimator deposited directly onto the STJ chip.

\subsection{Secondary radiation}~\label{sec:secondary}

\begin{figure}
    \centering
    \includegraphics[width=\columnwidth]{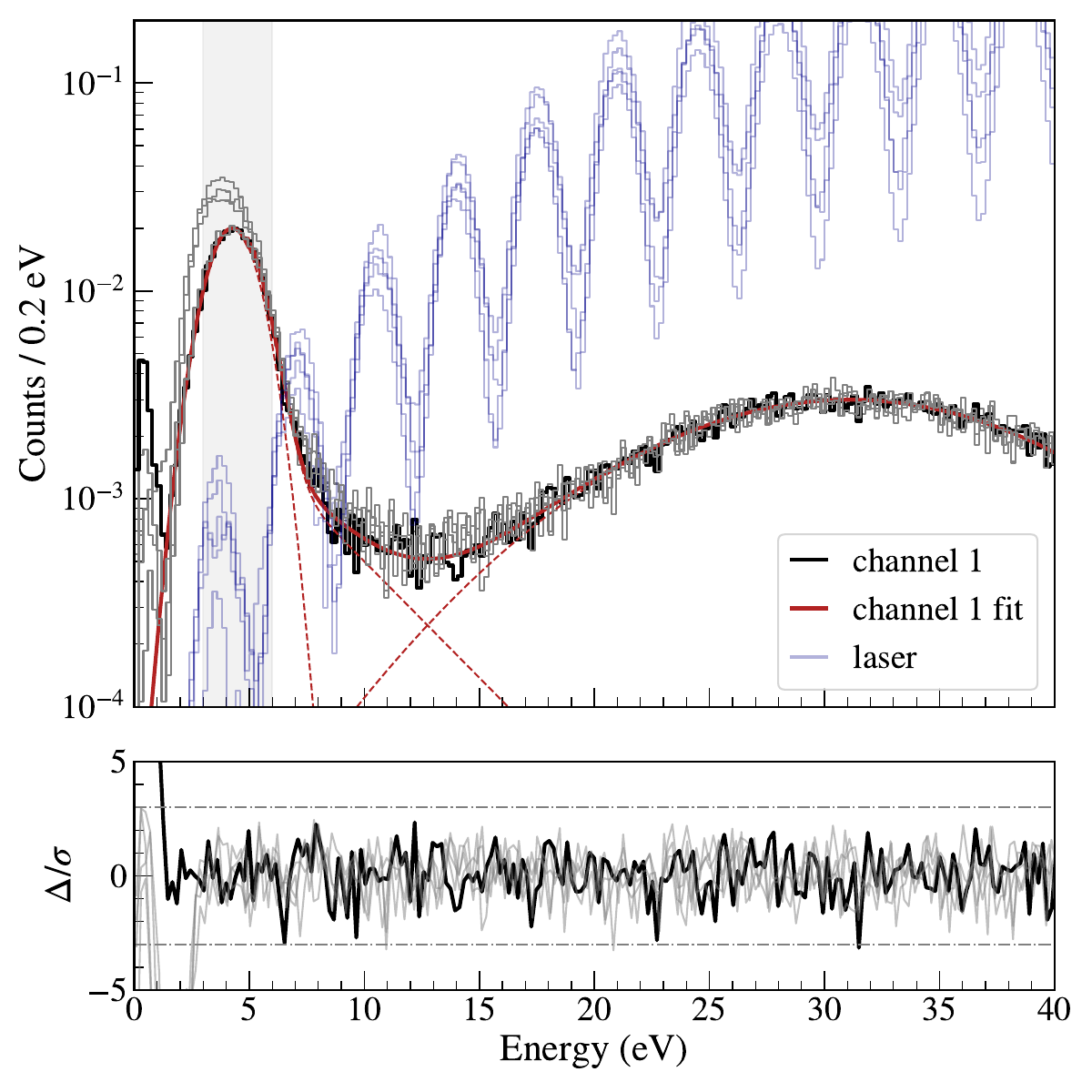}
    \caption{Spectral shapes of the low energy region in the group with the lowest background. The spectrum and the corresponding fit for one channel are presented as an example. The noise peak below 2~eV and the 3.5~eV peak from secondary radiations are illustrated.}
    \label{fig:4eVpeak}
\end{figure}

All channels exhibited a noise peak at the lowest energies below 2~eV, as expected for the low trigger level~(Fig.~\ref{fig:4eVpeak}).
These events could be efficiently identified by their high amplitude ratio $R_\textrm{short/long}$ of the short and long trapezoidal filters.
All spectra also exhibited a peak around 3.5~eV that -- in contrast to the noise peak at the lowest energies -- had the same pulse shapes as laser- and $^7$Be-induced signals. 
In addition, this peak decayed like the rest of the spectrum in proportion to the decay of $^7$Be.
Both observations suggest that these events are caused by the decay of $^7$Be.
The peak at 3.5~eV had a width of $\approx$3~eV FWHM that was significantly wider than the laser signals at the same energy~(Fig.~\ref{fig:4eVpeak}). 
This suggests that it was not due to a monochromatic source but a distribution of events with a width of order $\approx$2.5~eV.

\begin{figure}
    \centering
    \includegraphics[width=\columnwidth]{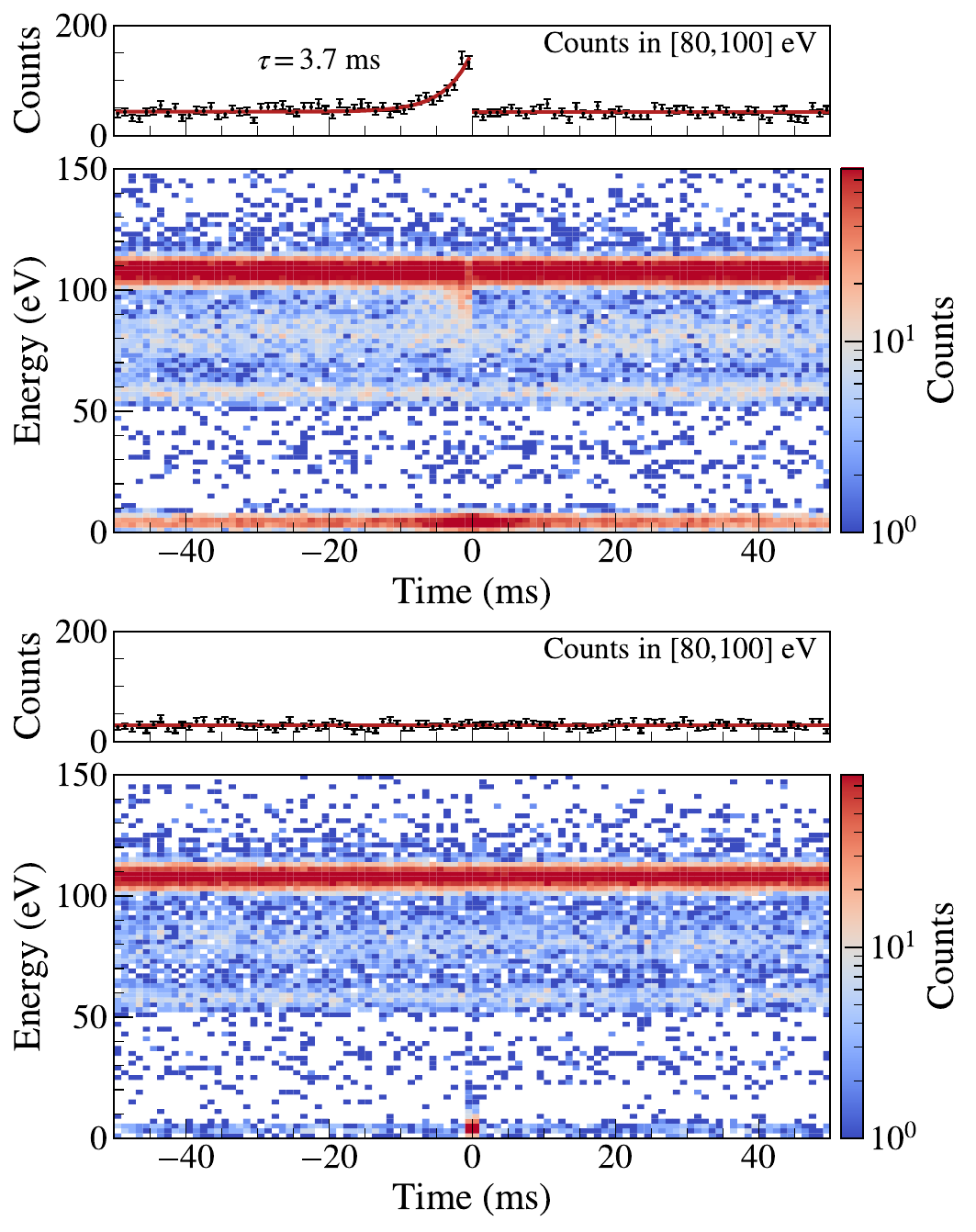}
    \caption{2D histograms of events in one channel that were time-tagged by the [2,10]~eV events. (Top): events tagged by the same channel. (Bottom) events tagged by an adjacent channel with shared ground wire. Histograms of events in [80,100] eV are also illustrated at the top of each 2D plots.  See text for details.}
    \label{fig:time2DHist}
\end{figure}

The origin of the spectrum at 3.5~eV can be studied from temporal correlations with other events. 
Figure~\ref{fig:time2DHist} (top) shows the distribution of time differences between signals below 10~eV at $t=0$ and other signals in the energy range between 0 and 150~eV. 
Most of the signals were random and uniformly distributed in time, dominated by the strong K-capture peak at $\approx$108~eV. 
However, the \textit{same} detector showed an increase of events in the  energy range from 80 to 100~eV immediately preceding a low-energy event, within a time interval of a few milliseconds. 
This energy range was dominated by events in which an energetic electron escaped from the detector during the initial relaxation of the Auger electron~(Sec.\ref{sec:escape}).
This delayed correlation was not seen between two separate pixels~(Fig.~\ref{fig:time2DHist}~bottom).

This suggests that the escaping electrons may excite metastable states during their escape that subsequently decay and re-deposit their energy in the detector. 
Specifically, oxygen vacancies in Ta$_2$O$_5$---which is present on the surface of STJ pixels---produce that an isolated defect state 1.5~eV below the edge of the conduction band~\cite{guo2014defect}. 
Irregular surface (hydr)oxides may produce metastable states at different energies which broadens the peak. 
Metastable states can also be located, e.g., in the oxide on the surface of the Si collimator in front of the detector. 
Not all of the metastable states are necessarily excited by electrons from K-capture events. 
It is also possible that some of them were excited by $\gamma$-rays that are produced in $\approx$10\% of all $^7$Be decays. 
The time distribution of the increased number of events in the energy range from 80 to 100~eV~(Fig.~\ref{fig:time2DHist}) corresponded to a lifetime of the metastable state of $3.7\pm0.6$~ms. 
This value was the same for all pixels, suggesting a common mechanism for the peak at 3.5~eV. 
We consider metastable states in the Ta$_2$O$_5$ surface oxide excited by escaping electrons to be the most likely explanation.

\subsection{Fit range}

While our sterile neutrino search focuses on finding a BSM signature below the K-GS peak at 108~eV, the robustness of the fit in the ROI can be enhanced by extending the fit to higher energies. 
Some SM features from the $^7$Be EC decay such as the pile-up of two K-GS events~(up to 220~eV) and the K-GS K-shake-up peak~($\approx$200~eV) can appear at energies above the ROI, and the spectrum is dominated by the shake-off continuum above 230~eV. 
We therefore set the upper bound of the fit to 250~eV. 
At low-energy, we used 20~eV as the lower bound of the fit~(Sec.~\ref{sec:secondary}).
Both detection efficiency~(Sec.~\ref{sec:daq}) and triggering efficiency~(Sec.~\ref{sec:trigger}) were high and flat above this bound, ensuring the robustness of our fit.

\subsection{Code implementation}

The software developed to carry out the analyses has been compiled into a Python-based toolkit, which is utilized to generate the final energy spectra. 
The toolkit was  developed using Python 3.10.8 and incorporates \texttt{iminuit}, a Python implementation of the MINUIT2 C++ library~\cite{iminuit}.
It was initially developed using Dataset~1, and its capabilities were verified with Datasets~2~and~3.   
This process was fully automated, eliminating the need for manual intervention, and successfully produced final spectra for all 15 channels. 
On a single node of the Borax high-performance computing cluster at Lawrence Livermore National Laboratory~\cite{borax},  the processing of 3~TB data from 16~pixels in a single day takes $\approx$7~hours.
This translates to 14~hours with 25~computing~nodes to process the entire BeEST Phase-III data.

\section{Results and Discussion}\label{sec:results}

\begin{table*}
  
    \caption{Summary of fit parameters of the BeEST Phase-III spectral model. The parameters for the pile-up spectra and values of the areas, marked with an asterisk~(*) vary among pixels and days. The last column shows the average and the rms variations for all 15~pixels from a single day.}
    \label{tab:summary_main}
    \begin{tabular}{|c|c|c |c |c |c |c|}
    \hline
         &Parameter & Note & unit & Initial & Bounds &Fit values\\
        \hline
        \hline
         \multirow{15}{*}{\makecell{Primary \\peaks}}  &$R$& $^7$Li$_\mathrm{ES}$/ $^7$Li$_\mathrm{GS}$ ratio & - & 0.11657& [0.11432,11882]&$0.11630\pm0.00011$\\
          &$\gamma$& Lorentzian life-time broadening & eV& 0.03 &  -  &Fixed\\
         &$A_\textrm{KGS,1}$& Area of the primary K-GS component& count &  $0.232M$\footnote{$M$ denotes the number of events in [106,110]~eV.\label{footnote_a}}& $[0.19M,0.28M]$&*\\
         &$\mu_\mathrm{KGS,1}$& Primary K-GS peak position & eV & 108& [107,109]&$108.50\pm0.11$\\
         &$A_\textrm{KGS,2}$& Area of the secondary K-GS component & count &  $0.11M$\footref{footnote_a}& $[0.08M,0.14M]$&*\\
         &$s_{K,2}$ & Shift of the secondary K-GS peak from $\mu_{\mathrm{KGS},1}$ & eV & -3.2& [-3.8,-2.6]&$-3.24\pm0.04$\\
         &$A_\textrm{KGS,3}$& Area of the tertiary K-GS component & count &  $0.0246M$\footref{footnote_a}& $[0.02M,0.03M]$&*\\
         &$s_{K,3}$ & Shift of the tertiary K-GS peak from $\mu_{\mathrm{KGS},1}$  & eV & 2.532& [2.1,2.9]&$2.44\pm0.10$\\
         &$\sigma_\mathrm{K}$ & rms width of all three K-GS peak components  & eV & 2.05& [1.7,2.8]&$2.10\pm0.04$\\ 
         &$\mu_\mathrm{KES,1}$& K-ES peak position & eV &  81& [79,83]&$81.33\pm0.11$\\
         &$\sigma_{\mathrm{Doppler}}$& rms Doppler broadening of the excited peaks & eV & 6.97&  [6,8]&$7.15\pm0.03$\\
         &$A_\textrm{L}$& Area of the L-GS peak & count & $0.0185M$\footref{footnote_a}& $[0.015,0.025]$&*\\
         &$\mu_\mathrm{LGS}$ & L-GS peak position & eV &  57& [55,59]&$56.93\pm0.12$\\
         &$\sigma_\mathrm{L}$ & L-GS peak width & eV & 2.75& [2.1,3.5]&$2.87\pm0.07$\\
         &$\mu_\mathrm{LES}$ & L-ES peak position & eV &  30.5& [29,33]&$30.6\pm0.4$\\
        \hline
        \hline
         \multirow{3}{*}{\makecell{Auger \\ electron \\ escape}} &$A_\mathrm{KGS-Auger}$ & Area of K-GS Auger electron escape tail & count & $M_A$\footnote{$M_A$ denotes the number of events in [97,99]~eV.}& $[0.3M_A,2M_A]$&*\\
         &$E_\mathrm{Auger}$ & Onset energy of the electron escape tail relative to $\mu_{\mathrm{KGS},1}$& eV & 7.5& [5.5,9.5]&$7.72\pm0.32$\\
         &$k_\mathrm{Auger}$ & Skewness parameter for  electron escape tail & eV &9.3& [4.3,13.3]&$7.2\pm0.9$\\
        \hline
        \hline
        \multirow{16}{*}{\makecell{Shake-up / \\Shake-off}}  &$A_\mathrm{KGS-KSU}$ & Area of the KGS-KSU peak & count & $0.0195M$\footref{footnote_a} & [0.01$M$,0.03$M$] &* \\
         &$E_\mathrm{KGS-KSU}$ & Energy of the KGS-KSU peak relative to $\mu_{\mathrm{KGS},1}$   & eV & 88 &  [87,92] &$89.37\pm0.27$\\
         &$\sigma_\mathrm{KGS-KSU}$ & rms width of the KGS-KSU peak  & eV & 4.7 &  [4,6] &$4.66\pm0.16$\\
         &$A_{\mathrm{KGS-KSO}}$& Area of the KGS-KSO tail & count & $0.00448M$\footref{footnote_a}& [0.003$M$,0.008$M$]&* \\
         &$E_\mathrm{KGS-KSO}$ & Offset energy of the KGS-KSO tail relative to $\mu_{\mathrm{KGS},1}$  & eV & 107 &  [80,160] &$93.1\pm1.0$\\ %%%%%%%%%%%%
         &$p$& Decay scale of the K-SO tails & eV$^{-1}$ & -2.5&  [-5,-2]&$-2.01\pm0.07$\\
         &$A_{\mathrm{LGS-KSO}}$& Area of the LGS-KSO tail & count & $0.00124M$\footref{footnote_a}& [0.0001$M$,0.004$M$]&* \\
         &$E_\mathrm{LGS-KSO}$ & Offset energy of the LGS-KSO tail relative to $\mu_{\mathrm{LGS}}$  & eV & 62&  [54.5,69.5]&$64.28\pm0.54$\\
         &$A_\mathrm{KGS-LSO}$ & Area of the KGS-LSO tail & count & $0.007M$\footref{footnote_a}&  [0.001$M$,0.012$M$]&* \\
         &$E_\mathrm{KGS-LSO}$ & Offset energy the KGS-LSO tail relative to $\mu_{\mathrm{KGS},1}$   & eV & 1&  [0,2.5]&$0.17\pm0.25$\\
         &$s_\mathrm{KGS-LSO}$ & Shape parameter of the KGS-LSO tail & - & 0.9&  [0.7,3.0]&$0.689\pm0.032$\\
         &$a_\mathrm{KGS-LSO}$ & Scale parameter of the KGS-LSO tail & eV$^{-1}$ & 9.32&  [6.8,11.8]&$9.90\pm0.23$\\
         &$A_\mathrm{LGS-LSO}$ & Area of the LGS-LSO tail & count & $0.006M$\footref{footnote_a}&  [0.001$M$,0.02$M$]&* \\
         &$E_\mathrm{LGS-LSO}$ & Offset energy of the LGS-LSO tail relative to $\mu_{\mathrm{LGS}}$    & eV & 0.01 &  [0,2.51] & $1.16\pm0.54$ \\
         &$s_\mathrm{LGS-LSO}$ & Shape parameter of the LGS-LSO tail & - & 0.9&  [0.7,3.0]&$1.49\pm0.22$\\
         &$a_\mathrm{LGS-LSO}$ & Scale parameter of the LGS-LSO tail & eV$^{-1}$ & 3&  [1,6]&$2.60\pm0.47$\\
        \hline
        \hline
         \multirow{4}{*}{\makecell{Pile-up}} &$A_\mathrm{flat} $ & Area of flat region in the pileup spectrum& count & *\footnote{Initial values are independently determined from the fit to low short/long events spectra.\label{footnote_c}} & * &* \\
         &$A_\mathrm{peak} $ & Area of peak region in the pileup spectrum & count & *\footref{footnote_c} & * &* \\
         &$\sigma_\mathrm{PU} $ & rms width of pile-up spectrum & eV & *\footref{footnote_c}  & * &* \\
         &$k_\mathrm{PU} $ & Skewness of the pile-up peak & eV & *\footref{footnote_c}  & * &* \\
        \hline
        \hline
    \end{tabular}
\end{table*}

\begin{figure*}
    \centering
    \includegraphics[width=\textwidth]{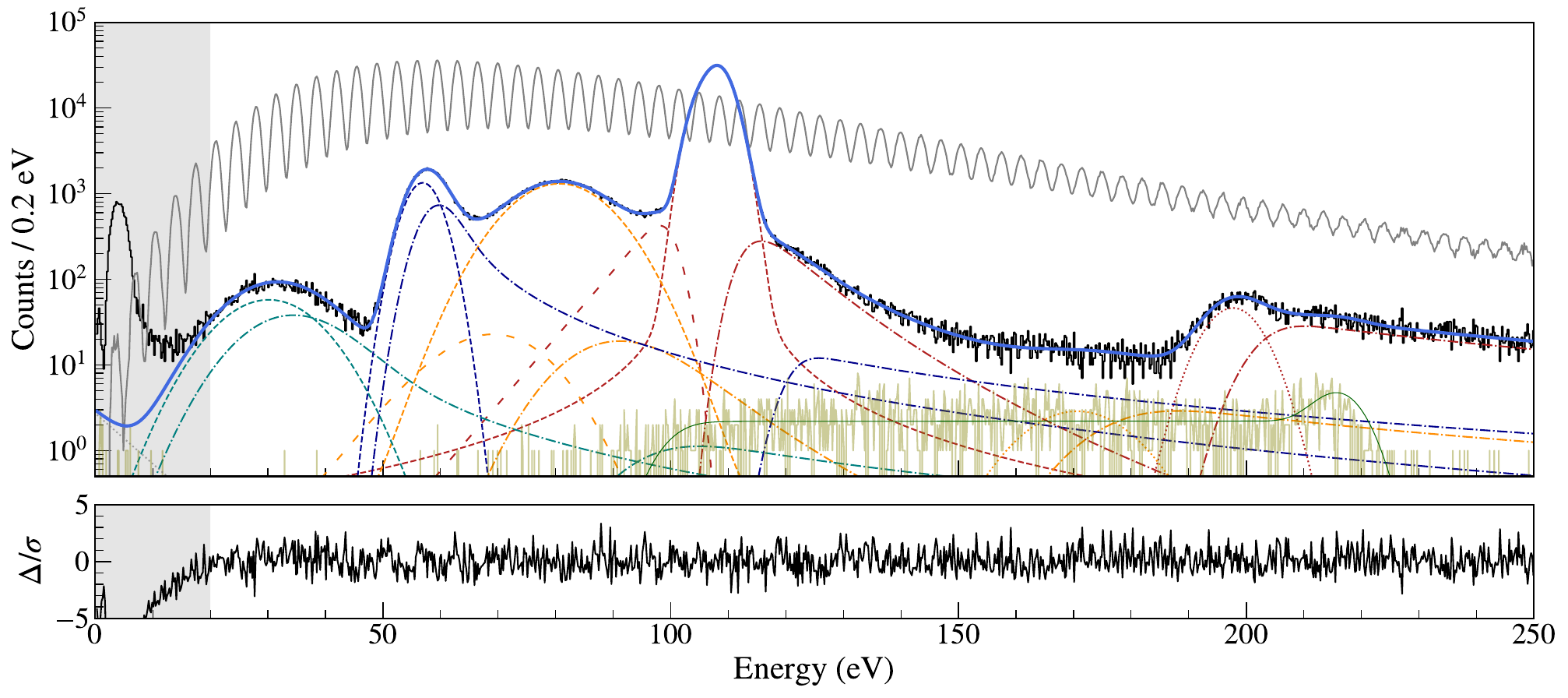}
    \caption{The BeEST Phase-III spectrum modeling. 
    Clean electron capture spectrum~(black), laser calibration spectrum~(gray solid), pile-up spectrum~(olive), K-capture spectra to ground and excited states~(red and orange), L-capture spectra to ground and excited states~(dark blue and teal), the low-energy background from secondary radiation and substrate implantation~(gray dotted), the final fit spectrum~(blue). Capture peaks are indicated with dashed lines, while shake-up/shake-off spectra are shown with dash-dotted lines. The residuals are small in the fit range above 20~eV, with $\chi^2_\textrm{red}=$1.05 and a $p$-value of 0.10. 
    }
    \label{fig:fit}
\end{figure*}

Figure~\ref{fig:fit} shows a full fit of the $^7$Be decay spectrum for a single STJ pixel from one day of released data. 
All data cleaning steps from Section~\ref{sec:dc} were applied to remove intervals of poor calibration accuracy and high pick-up. 
In addition, the spectrum no longer contained the broad background due to gamma interactions in the Si substrate and only part of the $^7$Be decay between pixels.
Pile-up of two $^7$Be decay signals in the same pixel could not be fully removed and was therefore modeled. 
The fit used the model functions introduced in Section~\ref{sec:bkg}, and the fit parameters extracted from all pixels on the same day are summarized in Table~\ref{tab:summary_main}. 

Figure~\ref{fig:fit} confirmed our earlier assumptions for the basic physics from Phase-II of the BeEST experiment. 
The spectrum quality was significantly improved over Phase-II, primarily because of better background rejection and data cleaning. 
The analog DAQ in Phase-II did not allow for pulse shape analysis, and we were thus limited in our analysis.
Also, the use of a single STJ pixel in Phase-II did not allow coincidence measurements to identify and reject the gamma-induced background, nor the background from $^7$Be implantation between pixels. 
For Phase-II, we had therefore assumed that these backgrounds could both be approximated by exponentially decaying functions. 
While this produced a high-quality fit with $\chi^2_\textrm{red}\approx 1.05$~\cite{beest2020lkratio}, the approximation did not match the new measured data in detail (Fig.~\ref{fig:gammaSim}). 
Removing this background completely based on coincidence vetoing greatly  reduced the background index and improved the visibility of the peaks for the different $^7$Be decay channels, especially at low energy.

The quality of the fits and the consistency of the fit parameters across different pixels (Table~\ref{tab:summary_main}), however, also illustrates that some spectral details are not yet fully understood. 
The centroid energy of 108.50(11)~eV for the primary K-GS peak was consistently lower than expected for a recoil energy of 56.826(9)~eV~\cite{beest2021phase2} plus the Li atomic relaxation energy of about 54.75~eV~\cite{bearden1967energylevel}.
The discrepancy of $\approx$3~eV could not be explained by chemical shifts of the Li K binding energy in Ta alone~\cite{samanta2023material}. 
This could mean that there was a systematic error in the energy calibration that we have not yet identified, although the consistency of the data (Table~\ref{tab:summary_main}) makes this appear unlikely. 
It could also suggest that nuclear signals were slightly quenched in STJ detectors~\cite{zehnder1995response}. 
Both hypotheses are consistent with a measured centroid energy of 56.93(12)~eV for the L-GS peak, which includes the same recoil energy of 56.826(9)~eV plus an average energy around $\approx$3~eV for the Li 2s hole in a Ta matrix~\cite{samanta2023material}. 
For now, we leave this question open, since it does not affect the sensitivity of the search for sterile neutrinos. 

We again found that an accurate fit of the primary peak K-GS peak required the sum of three components listed in Table~\ref{tab:summary_main}. 
The energy of the central primary component corresponds to the bare $^7$Li recoil plus the relaxation of the Li~1s hole. 
The higher-energy component was to account for the shake-up of a Li L electron into a bound state~(K-GS L-SU) which was not otherwise accounted for in the fit. 
We speculate that the lower-energy component of the K-GS peak could be due to excitation of metastable states in surface oxides~(Sec.~\ref{sec:secondary}) or events that involve energy loss due to lattice damage by the \textsuperscript{7}Li recoil. 
Typical energies of Frenkel pairs are a few eV, and the centroid shift of -3.24(4)~eV is consistent with the formation of one or two Frenkel pairs for a subset of events. 
All three components of the K-GS peak could be fit to a Voigt function with the same width, although this width of 4.95(9)~eV FWHM significantly exceeded the detector resolution for photon detection of $\approx$2~eV FWHM. 
Only part of this excess can be explained by chemical shifts of the Li 1s level in different sites of the Ta matrix~\cite{samanta2023material}. 
We speculate that the broadening may be caused by shake-up of multiple 5d electrons from the Ta matrix in which $^7$Be is embedded. 
But currently, the sources of the multiple peaks and their broadening are still being investigated.

Interestingly, the peak from L-capture decay to the $^7$Li ground state (L-GS) could be fit by a single Gaussian function, albeit one whose FWHM of 6.76(16)~eV was even wider than that of the K-GS peak. 
Note that this value was affected by the choice of the function used to model the L shake-off tail. 
Here, we used a log-normal function to model the L shake-off tail, whose peak accounts for some of the observed L-GS width~(Fig.~\ref{fig:fit}). 
If a different function without such a peak were chosen to describe the shake-off, the width of the Gaussian component of the L-GS peak could have increased. 
Initially, we had considered the increased L-GS width to be due to the variations in Li 2s energy levels, which fall into the Ta 5d band and are hybridized with it. 
In fact, density functional theory simulations suggest that the Li 2s levels do vary more widely than the Li 1s levels. 
However, the calculated Li 2s distribution of ~4~eV FWHM is still significantly less than observations~\cite{samanta2023material}. 
It may be possible to explain the observed width by the distribution of Li 2s levels in combination with L shake-up. 
Most of the Li 2s levels are located within ~4 eV of the Fermi energy~\cite{samanta2023material}, and the work function in Ta is 4.4~eV. 
Shake-up into bound level can therefore produce states that have up to ~8.4~eV in energy, and shake-up effects are stronger for L capture than for K capture. 
This hypothesis is currently being investigated.

As expected, the K-capture peak into the excited nuclear state $^7$Li$^*$ is Doppler broadened due to the in-flight emission of a 478~keV gamma ray. 
However, it is somewhat surprising that a simple Gaussian function matches the K-ES peak shape well. 
The peak shape depends on the ratio of the $^7$Li$^*$ lifetime and the time of the $^7$Li$^*$ ion to slow-down in the Ta matrix. 
Different models can be assumed for this slow-down, but none of them predict a Gaussian shape for the K-ES and L-ES peaks~\cite{voytas2001lkratio}. Accurate measurements of the excited-state peak shapes and detailed models of the ion slow-down at low energies might be able to shed a light on this question~\cite{bray2024doppler}.

The shake-off tails are currently fit to ad-hoc functions that are chosen because they provide consistent fits to the data with  $\chi^2_\textrm{red}\approx1$. 
Interestingly, none of the model functions from the literature~\cite{levinger1953electrons,echo2018shake,robertson2020shake} can match the observations as well, although this may not be surprising given the approximations used in their derivation. 
Our parametrization matched the shake-off tails above the K-capture peaks, where the background from other effects is small after gamma-induced signals have been removed by coincidence vetoing (Sec.~\ref{sec:coincidence}). 
It is more problematic for the L shake-off tails of the L-capture peaks, which overlap with the K-capture peaks and fall into the energy range of the expected sterile neutrino signal. 
Fortunately, we do not observe any structure in the shake-off tails, in the sense that the Phase-III spectra can be fit with $\chi^2_\textrm{red}\approx1$ using simple functions that vary smoothly over energy scales larger than the expected width of a sterile neutrino signal (Fig.~\ref{fig:fit}). 
Still, uncertainties in the L shake-off tails cause systematic uncertainties that will have to be modeled more accurately.

The electron escape tails below the K-capture peaks are greatly reduced in Phase-III compared to earlier phases because the detector chip has been rinsed in ethanol repeatedly after implantation to remove $^7$Be from the detector surface. 
These surface deposits were likely caused by  $^7$Be scattering at the Si collimator that had been placed $\approx$100~$\upmu$m in front of the detector chip during implantation in an attempt to restrict the  $^7$Be implants to the active detector area. 
 Scattering likely also caused the  $^7$Be distribution inside the STJs to differ slightly pixel-to-pixel, which in turn causes small variations in the magnitude of electron escape tails (Fig.~\ref{fig:histogramComparison}). 
 These tails can be modeled adequately by an exponentially modified Gaussian function, even though initial Monte-Carlo simulations suggest that details of the tail may differ from that shape~\cite{bray2022monte}. 
 Future experiments could reduce the variations in the $^7$Be implantation profile by replacing the Si collimator with a photoresist mask directly patterned onto the STJ detector chip. 
Ultimately, the escape tails could possibly be removed completely by depositing a Ta cap layer onto the STJs after $^7$Be implantation.

\begin{figure*}
    \centering
    \includegraphics[width=\textwidth]{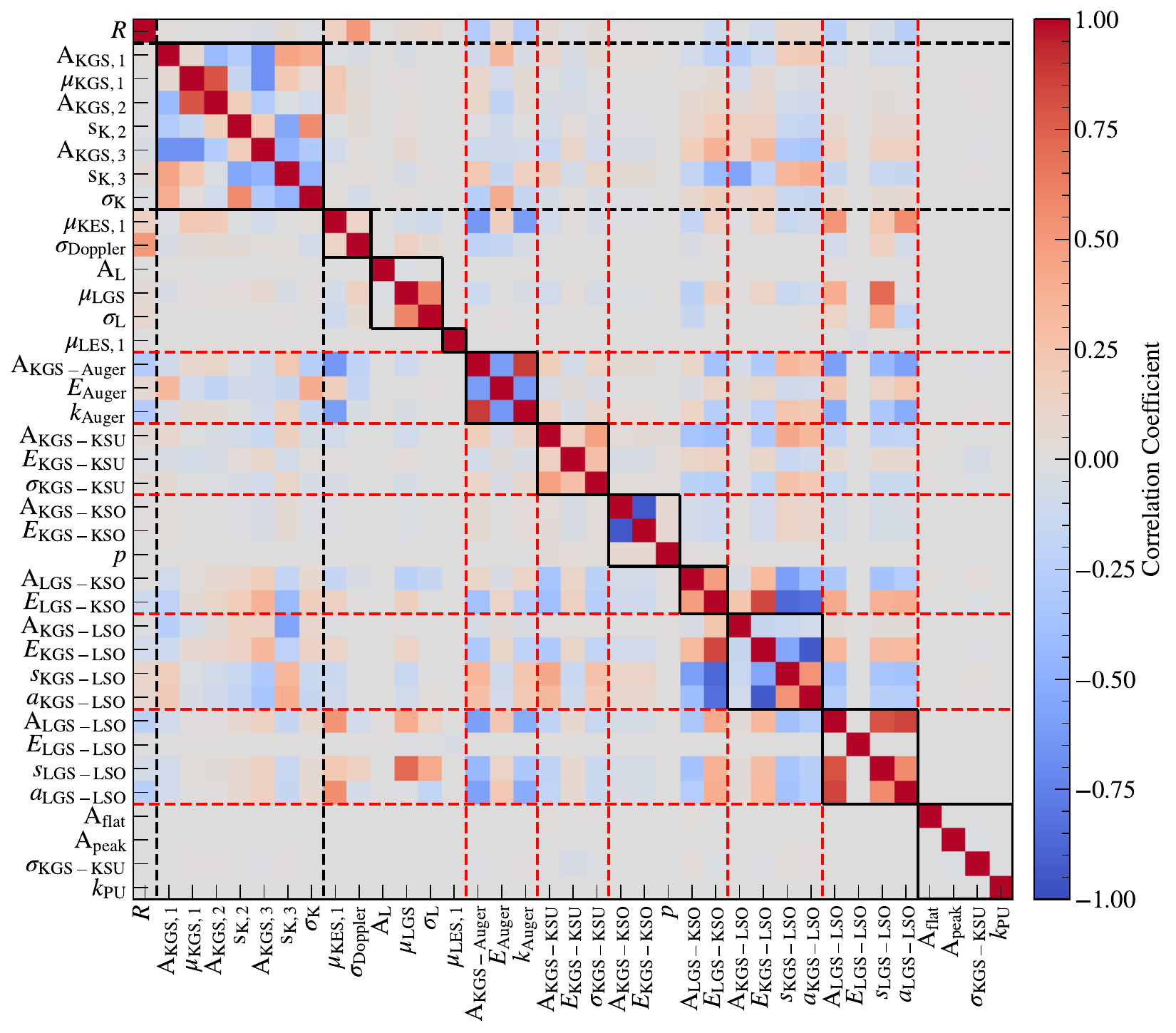}
    \caption{Correlation map showing the interdependencies between parameters in the BeEST Phase-III spectral fit~(Table~\ref{tab:summary_main}). Darker regions indicate stronger (anti)correlations, highlighting parameter pairs that impact the model's systematic stability and precision. Black squares enclose parameters within the same PDFs. Black dashed lines mark the most important K-GS peaks, while the red dashed lines indicate the positions of K-ES and L-capture peaks, the Auger-electron PDF, K-shakeups, K-shake-offs, L-shake-offs, and pile-up PDF.}
    \label{fig:covariance}
\end{figure*}

Figure~\ref{fig:covariance} illustrates the correlations between the parameters in the BeEST Phase-III spectral fit, highlighting their interdependencies. 
Darker colors indicate stronger correlations, either positive or negative. 
Black squares group parameters from the same PDFs, where the parameters are often strongly correlated.
This is expected, as parameters from the same PDF typically work together to describe the spectral component.

More interestingly, there are some strongly correlated parameters across different PDFs, outside the black squares. 
For example, the K-ES~(60-100~eV) and L-GS~(50-70~eV) peaks are moderately correlated with both the Auger-electron escape PDF~(60-100~eV) and the LGS-LSO PDF~(50-100~eV). 
The Auger-electron PDF is also correlated with the nearby K-GS peak, which is the most important feature for modeling the heavy neutrino signature. 
The K-GS peak is also mildly correlated with the KGS-LSO PDF, which forms the high-energy tail of the KGS peak. 
In contrast, the KGS-KSU and KGS-KSO PDFs have minimal cross-PDF correlations, as their values are well constrained by the high-energy part of the spectrum. 

The cross-PDF correlations of the K-GS peak parameters, especially with the Auger-electron escape and the L-shake-off PDFs, indicate that the BeEST Phase-III heavy neutrino search may be systematically influenced in the energy regions above 90~eV and below 70~eV in the spectrum, where these PDFs dominate. 
These regions correspond to the sterile neutrino mass ranges below 500~keV and above 700~keV, respectively. 
Performing a simultaneous fit to all 15~channels may help reducing the systematic uncertainties from the Auger-electron escape, as its rate varies among pixels. 
In future phases, we plan to fabricate a tantalum capping layer over the STJs after the $^7$Be implantation, which is expected to  remove the Auger-electron escape. 
In addition, a more detailed study on atomic calculations of the shake-up/shake-off effects in solid-state tantalum will be conducted.
These improvements will contribute to reducing the systematic uncertainties in the experiment.

The high detector resolution, the statistical accuracy and the good reproducibility of the data currently raise more new questions about the underlying physics than they answer. 
Most of them appear to be caused by the electron capture decay of $^7$Be occurring inside the STJ detector and the associated interactions between the electrons involved in the decay and the detector materials. 
These effects have not received a lot of attention in the literature so far, although we expect them to become increasingly important as the energy resolution of quantum sensors continues to improve.

\section{Projected Sensitivity}

\begin{figure}
    \centering
    \includegraphics[width=\columnwidth]{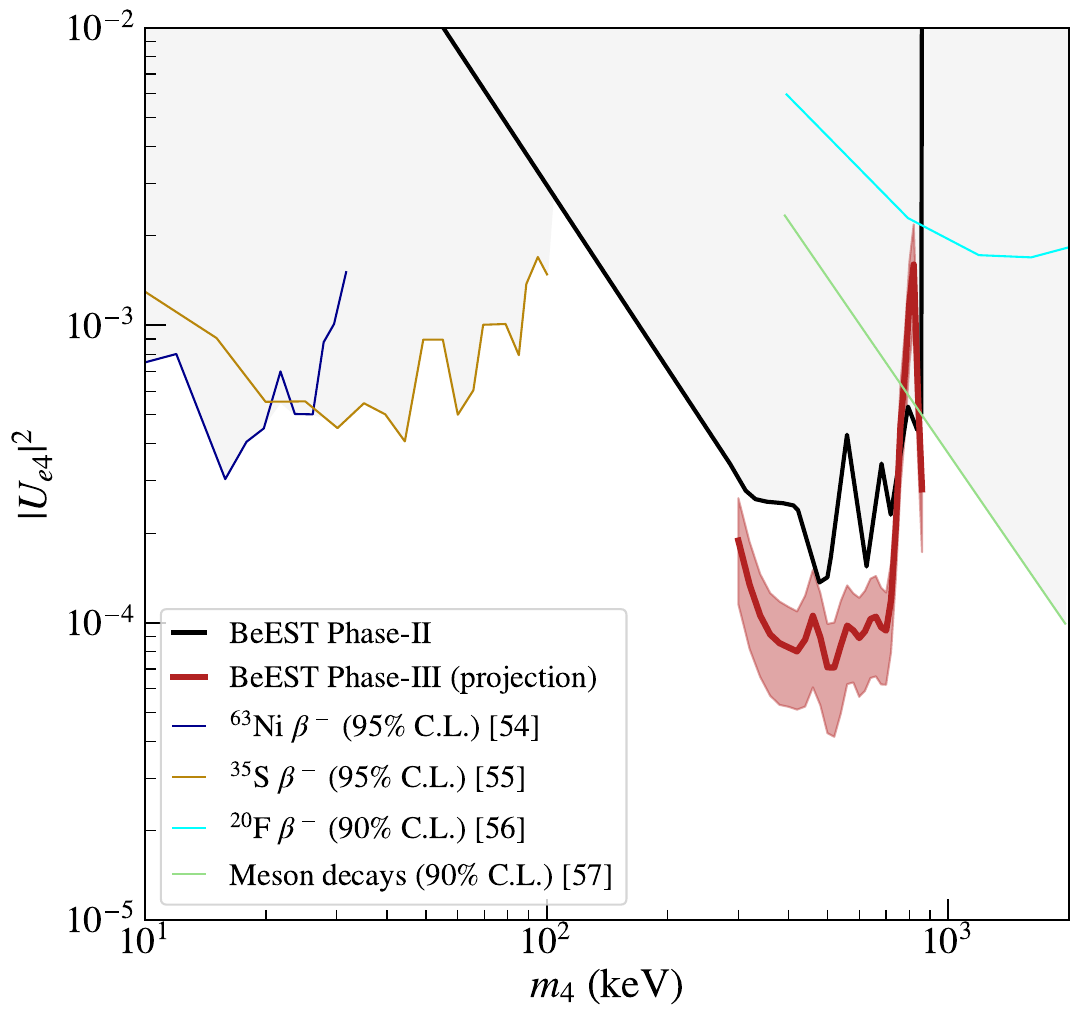}
    \caption{Projected sensitivity for the BeEST Phase-III with $1\sigma$ uncertainty band. Only the statistical considerations were made. Previous limits are shown in gray shaded regions~\cite{Ni1999limit,S2000limit,Cu1983limit,F1990limit,meson2019limit}.}
    \label{fig:projection}
\end{figure}

We estimated the projected sensitivity of the BeEST Phase-III data set for a heavy neutrino mass search using the spectral fit in Fig.~\ref{fig:fit}, accounting for the statistical contributions only. 
The fit spectra from a single channel and one day were scaled to match the Phase-III exposure of 15~channels over 50~days, accounting for the 53.3~days half-life of $^7$Be. 
We then created 100 Monte-Carlo simulations to create a sensitivity projection. 
The results are presented in Fig.~\ref{fig:projection}, along with previous limits from $^{63}$Ni~\cite{Ni1999limit}, $^{35}$S~\cite{S2000limit}, $^{64}$Cu~\cite{Cu1983limit} and $^{20}$F~\cite{F1990limit} beta decays, as well as meson decays~\cite{meson2019limit}. 
Limits are correctly shifted to higher energies compared to Phase-II due to the calibration correction with laser intensity~(Sec.~\ref{sec:energy}).

The sensitivity is improved in the region below 450~keV due to the reduction of Auger-electron escape. 
The improvement is more modest in the high-mass region above 700~keV, where uncertainties in the spectral shapes of the shake-off tails limit the sensitivity. 
Overall, the BeEST Phase-III sensitivity achieves roughly by a factor of 3 improvement over Phase-II, while also correcting the shifts in the limit curve caused by previous calibration inaccuracies.
This sensitivity estimate accounts only for statistical uncertainties, assuming that the background models are accurate. 
With increased exposure, previously unobserved features may emerge, which could change the projected limit.

\section{Conclusion}\label{sec:conclusion}

We have developed an analysis procedure for Phase-III of the BeEST experiment. 
Phase-III utilized a new data acquisition system that continuously captured the output of 15~STJ detector pixels with implanted $^7$Be at 16-bit and 0.8~$\upmu$s sampling. 
This allowed testing different algorithms on the continuous data stream to optimize STJ energy resolution, correct for drifts and reject periods of reduced calibration accuracy. 
We have used pulse shape analysis based on two different trapezoidal filters to identify pile-up and reject occasional pick-up. 
In addition, gamma-induced substrate events were rejected by vetoing coincident signals in different detector pixels. 
These data cleaning steps significantly reduced systematic uncertainties, brought out the peaks from the different $^7$Be decay channels for accurate analysis and improved the quality of the subsequent fit.

The clean spectra were fit to a sum of Voigt and Gaussian functions with shake-up and shake-off tails above and electron escape tails below the peaks. 
We have applied the optimized cleaning and fitting procedure to one day of unblinded data from a two-month run in late 2022. 
Fit parameters were mostly consistent pixel-to-pixel, except for minor variations in parameters affected by details of the $^7$Be implant distribution in a detector pixel. 
On the other hand, several spectral details, most importantly the centroids and widths of the spectral peaks, consistently differed from our current understanding of the underlying physics. 
And while our mathematical description of the shake-off and escape tails provided good fits to the observed spectra, they were based on ad-hoc assumptions rather than first-principles calculations. 
These observations motivate the development of improved models of the electron capture decay, particularly modifications required for decays inside detector materials. 
Assuming that the current background model is accurate, Phase-III will increase the sensitivity of the BeEST experiment by roughly a factor of 3, although this number may be altered if unexpected spectral features arise as additional data are unblinded. 
In either case, the new data cleaning algorithms and spectral fits were of high quality and sufficiently robust to be executed on large data sets without human intervention. 
They pave the way for accurate and reliable limits on sterile neutrinos and other physics beyond the standard model once the full Phase-III data set is unblinded.

\section{Acknowledgment}

The BeEST experiment is funded by the U.S. Department of Energy, Office of Nuclear Physics under Award Numbers SCW1758 and DE-SC0021245, the LLNL Laboratory Directed Research and Development program through Grants No. 19-FS-027 and No. 20-LW-006, the Gordon and Betty Moore Foundation (10.37807/GBMF11571), the European Metrology Programme for Innovation and Research (EMPIR) Projects No. 17FUN02 MetroMMC and No. 20FUN09 PrimA-LTD, and the FCT—Funda\c{c}\~{a}o para a Ci\^encia e a Tecnologia~(Portugal) through national funds in the framework of the projects UID/04559/2020 (LIBPhys). TRIUMF receives federal funding via a contribution agreement with the National Research Council of Canada. Francisco Ponce is supported by Pacific Northwest National Laboratory, which is managed for the US Department of Energy by Battelle under contract DE-AC05-76RL01830. This work was performed under the auspices of the U.S. Department of Energy by Lawrence Livermore National Laboratory under Contract No. DE-AC52-07NA27344. 

\bibliography{__biblio}% Produces the bibliography via BibTeX.

\begin{thebibliography}{58}%
\makeatletter
\providecommand \@ifxundefined [1]{%
 \@ifx{#1\undefined}
}%
\providecommand \@ifnum [1]{%
 \ifnum #1\expandafter \@firstoftwo
 \else \expandafter \@secondoftwo
 \fi
}%
\providecommand \@ifx [1]{%
 \ifx #1\expandafter \@firstoftwo
 \else \expandafter \@secondoftwo
 \fi
}%
\providecommand \natexlab [1]{#1}%
\providecommand \enquote  [1]{``#1''}%
\providecommand \bibnamefont  [1]{#1}%
\providecommand \bibfnamefont [1]{#1}%
\providecommand \citenamefont [1]{#1}%
\providecommand \href@noop [0]{\@secondoftwo}%
\providecommand \href [0]{\begingroup \@sanitize@url \@href}%
\providecommand \@href[1]{\@@startlink{#1}\@@href}%
\providecommand \@@href[1]{\endgroup#1\@@endlink}%
\providecommand \@sanitize@url [0]{\catcode `\\12\catcode `\$12\catcode `\&12\catcode `\#12\catcode `\^12\catcode `\_12\catcode `\%12\relax}%
\providecommand \@@startlink[1]{}%
\providecommand \@@endlink[0]{}%
\providecommand \url  [0]{\begingroup\@sanitize@url \@url }%
\providecommand \@url [1]{\endgroup\@href {#1}{\urlprefix }}%
\providecommand \urlprefix  [0]{URL }%
\providecommand \Eprint [0]{\href }%
\providecommand \doibase [0]{https://doi.org/}%
\providecommand \selectlanguage [0]{\@gobble}%
\providecommand \bibinfo  [0]{\@secondoftwo}%
\providecommand \bibfield  [0]{\@secondoftwo}%
\providecommand \translation [1]{[#1]}%
\providecommand \BibitemOpen [0]{}%
\providecommand \bibitemStop [0]{}%
\providecommand \bibitemNoStop [0]{.\EOS\space}%
\providecommand \EOS [0]{\spacefactor3000\relax}%
\providecommand \BibitemShut  [1]{\csname bibitem#1\endcsname}%
\let\auto@bib@innerbib\@empty
%</preamble>
\bibitem [{\citenamefont {Patrignani}\ \emph {et~al.}(2016)\citenamefont {Patrignani} \emph {et~al.}}]{patrignani2016darkmatter}%
  \BibitemOpen
  \bibfield  {author} {\bibinfo {author} {\bibfnamefont {C.}~\bibnamefont {Patrignani}} \emph {et~al.} (\bibinfo {collaboration} {Particle Data Group}),\ }\bibfield  {title} {\bibinfo {title} {{Review of Particle Physics}},\ }\href {https://doi.org/10.1088/1674-1137/40/10/100001} {\bibfield  {journal} {\bibinfo  {journal} {Chin. Phys. C}\ }\textbf {\bibinfo {volume} {40}},\ \bibinfo {pages} {100001} (\bibinfo {year} {2016})}\BibitemShut {NoStop}%
\bibitem [{\citenamefont {Fukuda}\ \emph {et~al.}(1998)\citenamefont {Fukuda} \emph {et~al.}}]{skk1998evidence}%
  \BibitemOpen
  \bibfield  {author} {\bibinfo {author} {\bibfnamefont {Y.}~\bibnamefont {Fukuda}} \emph {et~al.} (\bibinfo {collaboration} {Super-Kamiokande Collaboration}),\ }\bibfield  {title} {\bibinfo {title} {{Evidence for Oscillation of Atmospheric Neutrinos}},\ }\href {https://doi.org/10.1103/PhysRevLett.81.1562} {\bibfield  {journal} {\bibinfo  {journal} {Phys. Rev. Lett.}\ }\textbf {\bibinfo {volume} {81}},\ \bibinfo {pages} {1562} (\bibinfo {year} {1998})}\BibitemShut {NoStop}%
\bibitem [{\citenamefont {Ahmad}\ \emph {et~al.}(2001)\citenamefont {Ahmad} \emph {et~al.}}]{sno2001oscillation}%
  \BibitemOpen
  \bibfield  {author} {\bibinfo {author} {\bibfnamefont {Q.~R.}\ \bibnamefont {Ahmad}} \emph {et~al.} (\bibinfo {collaboration} {SNO Collaboration}),\ }\bibfield  {title} {\bibinfo {title} {{Measurement of the Rate of ${\ensuremath{\nu}}_{e}+\mathit{d}\ensuremath{\rightarrow}\mathit{p}+\mathit{p}+{\mathit{e}}^{\ensuremath{-}}$ Interactions Produced by {$^{8}B$} Solar Neutrinos at the {Sudbury Neutrino Observatory}}},\ }\href {https://doi.org/10.1103/PhysRevLett.87.071301} {\bibfield  {journal} {\bibinfo  {journal} {Phys. Rev. Lett.}\ }\textbf {\bibinfo {volume} {87}},\ \bibinfo {pages} {071301} (\bibinfo {year} {2001})}\BibitemShut {NoStop}%
\bibitem [{\citenamefont {Bilenky}(2015)}]{bilenky2015neutrino}%
  \BibitemOpen
  \bibfield  {author} {\bibinfo {author} {\bibfnamefont {S.~M.}\ \bibnamefont {Bilenky}},\ }\bibfield  {title} {\bibinfo {title} {{Neutrino in standard model and beyond}},\ }\href {https://doi.org/10.1134/S1063779615040024} {\bibfield  {journal} {\bibinfo  {journal} {Phys. Part. Nuclei}\ }\textbf {\bibinfo {volume} {46}},\ \bibinfo {pages} {475} (\bibinfo {year} {2015})}\BibitemShut {NoStop}%
\bibitem [{\citenamefont {Wu}\ \emph {et~al.}(1957)\citenamefont {Wu}, \citenamefont {Ambler}, \citenamefont {Hayward}, \citenamefont {Hoppes},\ and\ \citenamefont {Hudson}}]{wu1957parity}%
  \BibitemOpen
  \bibfield  {author} {\bibinfo {author} {\bibfnamefont {C.~S.}\ \bibnamefont {Wu}}, \bibinfo {author} {\bibfnamefont {E.}~\bibnamefont {Ambler}}, \bibinfo {author} {\bibfnamefont {R.~W.}\ \bibnamefont {Hayward}}, \bibinfo {author} {\bibfnamefont {D.~D.}\ \bibnamefont {Hoppes}},\ and\ \bibinfo {author} {\bibfnamefont {R.~P.}\ \bibnamefont {Hudson}},\ }\bibfield  {title} {\bibinfo {title} {Experimental test of parity conservation in beta decay},\ }\href {https://doi.org/10.1103/PhysRev.105.1413} {\bibfield  {journal} {\bibinfo  {journal} {Phys. Rev.}\ }\textbf {\bibinfo {volume} {105}},\ \bibinfo {pages} {1413} (\bibinfo {year} {1957})}\BibitemShut {NoStop}%
\bibitem [{\citenamefont {Pontecorvo}(1968)}]{pontecorvo1968neutrino}%
  \BibitemOpen
  \bibfield  {author} {\bibinfo {author} {\bibfnamefont {B.}~\bibnamefont {Pontecorvo}},\ }\bibfield  {title} {\bibinfo {title} {Neutrino experiments and the problem of conservation of leptonic charge},\ }\href@noop {} {\bibfield  {journal} {\bibinfo  {journal} {Sov. Phys. JETP}\ }\textbf {\bibinfo {volume} {26}},\ \bibinfo {pages} {165} (\bibinfo {year} {1968})}\BibitemShut {NoStop}%
\bibitem [{\citenamefont {Gariazzo}\ \emph {et~al.}(2015)\citenamefont {Gariazzo}, \citenamefont {Giunti}, \citenamefont {Laveder}, \citenamefont {Li},\ and\ \citenamefont {Zavanin}}]{gariazzo2015light}%
  \BibitemOpen
  \bibfield  {author} {\bibinfo {author} {\bibfnamefont {S.}~\bibnamefont {Gariazzo}}, \bibinfo {author} {\bibfnamefont {C.}~\bibnamefont {Giunti}}, \bibinfo {author} {\bibfnamefont {M.}~\bibnamefont {Laveder}}, \bibinfo {author} {\bibfnamefont {Y.~F.}\ \bibnamefont {Li}},\ and\ \bibinfo {author} {\bibfnamefont {E.~M.}\ \bibnamefont {Zavanin}},\ }\bibfield  {title} {\bibinfo {title} {Light sterile neutrinos},\ }\href {https://doi.org/10.1088/0954-3899/43/3/033001} {\bibfield  {journal} {\bibinfo  {journal} {J. Phys. G: Nucl. Part. Phys.}\ }\textbf {\bibinfo {volume} {43}},\ \bibinfo {pages} {033001} (\bibinfo {year} {2015})}\BibitemShut {NoStop}%
\bibitem [{\citenamefont {Giunti}\ and\ \citenamefont {Lasserre}(2019)}]{giunti2019annual}%
  \BibitemOpen
  \bibfield  {author} {\bibinfo {author} {\bibfnamefont {C.}~\bibnamefont {Giunti}}\ and\ \bibinfo {author} {\bibfnamefont {T.}~\bibnamefont {Lasserre}},\ }\bibfield  {title} {\bibinfo {title} {{eV-Scale Sterile Neutrinos}},\ }\href {https://doi.org/10.1146/annurev-nucl-101918-023755} {\bibfield  {journal} {\bibinfo  {journal} {Annu. Rev. Nucl. Part. Sci.}\ }\textbf {\bibinfo {volume} {69}},\ \bibinfo {pages} {163} (\bibinfo {year} {2019})}\BibitemShut {NoStop}%
\bibitem [{\citenamefont {Böser}\ \emph {et~al.}(2020)\citenamefont {Böser}, \citenamefont {Buck}, \citenamefont {Giunti}, \citenamefont {Lesgourgues}, \citenamefont {Ludhova}, \citenamefont {Mertens}, \citenamefont {Schukraft},\ and\ \citenamefont {Wurm}}]{boser2020status}%
  \BibitemOpen
  \bibfield  {author} {\bibinfo {author} {\bibfnamefont {S.}~\bibnamefont {Böser}}, \bibinfo {author} {\bibfnamefont {C.}~\bibnamefont {Buck}}, \bibinfo {author} {\bibfnamefont {C.}~\bibnamefont {Giunti}}, \bibinfo {author} {\bibfnamefont {J.}~\bibnamefont {Lesgourgues}}, \bibinfo {author} {\bibfnamefont {L.}~\bibnamefont {Ludhova}}, \bibinfo {author} {\bibfnamefont {S.}~\bibnamefont {Mertens}}, \bibinfo {author} {\bibfnamefont {A.}~\bibnamefont {Schukraft}},\ and\ \bibinfo {author} {\bibfnamefont {M.}~\bibnamefont {Wurm}},\ }\bibfield  {title} {\bibinfo {title} {Status of light sterile neutrino searches},\ }\href {https://doi.org/https://doi.org/10.1016/j.ppnp.2019.103736} {\bibfield  {journal} {\bibinfo  {journal} {Prog. Part. Nucl. Phys.}\ }\textbf {\bibinfo {volume} {111}},\ \bibinfo {pages} {103736} (\bibinfo {year} {2020})}\BibitemShut {NoStop}%
\bibitem [{\citenamefont {Diaz}\ \emph {et~al.}(2020)\citenamefont {Diaz}, \citenamefont {Argüelles}, \citenamefont {Collin}, \citenamefont {Conrad},\ and\ \citenamefont {Shaevitz}}]{diaz2020where}%
  \BibitemOpen
  \bibfield  {author} {\bibinfo {author} {\bibfnamefont {A.}~\bibnamefont {Diaz}}, \bibinfo {author} {\bibfnamefont {C.}~\bibnamefont {Argüelles}}, \bibinfo {author} {\bibfnamefont {G.}~\bibnamefont {Collin}}, \bibinfo {author} {\bibfnamefont {J.}~\bibnamefont {Conrad}},\ and\ \bibinfo {author} {\bibfnamefont {M.}~\bibnamefont {Shaevitz}},\ }\bibfield  {title} {\bibinfo {title} {Where are we with light sterile neutrinos?},\ }\href {https://doi.org/https://doi.org/10.1016/j.physrep.2020.08.005} {\bibfield  {journal} {\bibinfo  {journal} {Phys. Rep.}\ }\textbf {\bibinfo {volume} {884}},\ \bibinfo {pages} {1} (\bibinfo {year} {2020})}\BibitemShut {NoStop}%
\bibitem [{\citenamefont {Seo}(2021)}]{seo2021review}%
  \BibitemOpen
  \bibfield  {author} {\bibinfo {author} {\bibfnamefont {S.-H.}\ \bibnamefont {Seo}},\ }\bibfield  {title} {\bibinfo {title} {Review of sterile neutrino experiments},\ }in\ \href@noop {} {\emph {\bibinfo {booktitle} {PARTICLE PHYSICS at the Year of 150th Anniversary of the Mendeleev’s Periodic Table of Chemical Elements: Proceedings of the Nineteenth Lomonosov Conference on Elementary Particle Physics}}}\ (\bibinfo {organization} {World Scientific},\ \bibinfo {year} {2021})\ pp.\ \bibinfo {pages} {10--16}\BibitemShut {NoStop}%
\bibitem [{\citenamefont {Dasgupta}\ and\ \citenamefont {Kopp}(2021)}]{dasgupta2021sterile}%
  \BibitemOpen
  \bibfield  {author} {\bibinfo {author} {\bibfnamefont {B.}~\bibnamefont {Dasgupta}}\ and\ \bibinfo {author} {\bibfnamefont {J.}~\bibnamefont {Kopp}},\ }\bibfield  {title} {\bibinfo {title} {Sterile neutrinos},\ }\href {https://doi.org/https://doi.org/10.1016/j.physrep.2021.06.002} {\bibfield  {journal} {\bibinfo  {journal} {Phys. Rep.}\ }\textbf {\bibinfo {volume} {928}},\ \bibinfo {pages} {1} (\bibinfo {year} {2021})},\ \bibinfo {note} {sterile neutrinos}\BibitemShut {NoStop}%
\bibitem [{\citenamefont {Janot}\ and\ \citenamefont {Jadach}(2020)}]{janot2019improved}%
  \BibitemOpen
  \bibfield  {author} {\bibinfo {author} {\bibfnamefont {P.}~\bibnamefont {Janot}}\ and\ \bibinfo {author} {\bibfnamefont {S.}~\bibnamefont {Jadach}},\ }\bibfield  {title} {\bibinfo {title} {{Improved Bhabha cross section at LEP and the number of light neutrino species}},\ }\href {https://doi.org/https://doi.org/10.1016/j.physletb.2020.135319} {\bibfield  {journal} {\bibinfo  {journal} {Phys. Lett. B}\ }\textbf {\bibinfo {volume} {803}},\ \bibinfo {pages} {135319} (\bibinfo {year} {2020})}\BibitemShut {NoStop}%
\bibitem [{\citenamefont {Abdallah}\ \emph {et~al.}(2005)\citenamefont {Abdallah} \emph {et~al.}}]{DELPHI2003photon}%
  \BibitemOpen
  \bibfield  {author} {\bibinfo {author} {\bibfnamefont {J.}~\bibnamefont {Abdallah}} \emph {et~al.} (\bibinfo {collaboration} {{DELPHI Collaboration}}),\ }\bibfield  {title} {\bibinfo {title} {{Photon events with missing energy in $e^+-e^-$ collisions at $s^{1/2}$ = 130~GeV to 209~GeV}},\ }\href {https://doi.org/10.1140/epjc/s2004-02051-8} {\bibfield  {journal} {\bibinfo  {journal} {Eur. Phys. J. C}\ }\textbf {\bibinfo {volume} {38}},\ \bibinfo {pages} {395} (\bibinfo {year} {2005})}\BibitemShut {NoStop}%
\bibitem [{\citenamefont {Achard}\ \emph {et~al.}(2004)\citenamefont {Achard} \emph {et~al.}}]{L32003single}%
  \BibitemOpen
  \bibfield  {author} {\bibinfo {author} {\bibfnamefont {P.}~\bibnamefont {Achard}} \emph {et~al.} (\bibinfo {collaboration} {L3 Collaboration}),\ }\bibfield  {title} {\bibinfo {title} {{Single photon and multiphoton events with missing energy in $e^{+} e^{-}$ collisions at LEP}},\ }\href {https://doi.org/https://doi.org/10.1016/j.physletb.2004.01.010} {\bibfield  {journal} {\bibinfo  {journal} {Phys. Lett. B}\ }\textbf {\bibinfo {volume} {587}},\ \bibinfo {pages} {16} (\bibinfo {year} {2004})}\BibitemShut {NoStop}%
\bibitem [{\citenamefont {{ALEPH Collaboration, DELPHI Collaboration, L3 Collaboration, OPAL Collaboration, SLD Collaboration, LEP Electroweak Working Group, SLD Electroweak and Heavy Flavour Groups}}(2006)}]{aleph2006precision}%
  \BibitemOpen
  \bibfield  {author} {\bibinfo {author} {\bibnamefont {{ALEPH Collaboration, DELPHI Collaboration, L3 Collaboration, OPAL Collaboration, SLD Collaboration, LEP Electroweak Working Group, SLD Electroweak and Heavy Flavour Groups}}},\ }\bibfield  {title} {\bibinfo {title} {Precision electroweak measurements on the {Z} resonance},\ }\href {https://doi.org/https://doi.org/10.1016/j.physrep.2005.12.006} {\bibfield  {journal} {\bibinfo  {journal} {Phys. Rep.}\ }\textbf {\bibinfo {volume} {427}},\ \bibinfo {pages} {257} (\bibinfo {year} {2006})}\BibitemShut {NoStop}%
\bibitem [{\citenamefont {Aghanim}\ \emph {et~al.}(2020)\citenamefont {Aghanim} \emph {et~al.}}]{Planck2018results}%
  \BibitemOpen
  \bibfield  {author} {\bibinfo {author} {\bibfnamefont {N.}~\bibnamefont {Aghanim}} \emph {et~al.} (\bibinfo {collaboration} {{Planck Collaboration}}),\ }\bibfield  {title} {\bibinfo {title} {{{Planck 2018 results. VI. Cosmological parameters}}},\ }\href {https://doi.org/10.1051/0004-6361/201833910} {\bibfield  {journal} {\bibinfo  {journal} {Astron. Astrophys.}\ }\textbf {\bibinfo {volume} {641}},\ \bibinfo {pages} {A6} (\bibinfo {year} {2020})}\BibitemShut {NoStop}%
\bibitem [{\citenamefont {Fields}\ \emph {et~al.}(2020)\citenamefont {Fields}, \citenamefont {Olive}, \citenamefont {Yeh},\ and\ \citenamefont {Young}}]{Fields2019bigbang}%
  \BibitemOpen
  \bibfield  {author} {\bibinfo {author} {\bibfnamefont {B.~D.}\ \bibnamefont {Fields}}, \bibinfo {author} {\bibfnamefont {K.~A.}\ \bibnamefont {Olive}}, \bibinfo {author} {\bibfnamefont {T.-H.}\ \bibnamefont {Yeh}},\ and\ \bibinfo {author} {\bibfnamefont {C.}~\bibnamefont {Young}},\ }\bibfield  {title} {\bibinfo {title} {{{Big-Bang Nucleosynthesis after Planck}}},\ }\href {https://doi.org/10.1088/1475-7516/2020/03/010} {\bibfield  {journal} {\bibinfo  {journal} {JCAP}\ }\textbf {\bibinfo {volume} {03}},\ \bibinfo {pages} {010}}\BibitemShut {NoStop}%
\bibitem [{\citenamefont {Ivanov}\ \emph {et~al.}(2020)\citenamefont {Ivanov}, \citenamefont {Simonovi\'c},\ and\ \citenamefont {Zaldarriaga}}]{Ivanov2019cosmological}%
  \BibitemOpen
  \bibfield  {author} {\bibinfo {author} {\bibfnamefont {M.~M.}\ \bibnamefont {Ivanov}}, \bibinfo {author} {\bibfnamefont {M.}~\bibnamefont {Simonovi\'c}},\ and\ \bibinfo {author} {\bibfnamefont {M.}~\bibnamefont {Zaldarriaga}},\ }\bibfield  {title} {\bibinfo {title} {Cosmological parameters and neutrino masses from the final \textit{Planck} and full-shape {BOSS} data},\ }\href {https://doi.org/10.1103/PhysRevD.101.083504} {\bibfield  {journal} {\bibinfo  {journal} {Phys. Rev. D}\ }\textbf {\bibinfo {volume} {101}},\ \bibinfo {pages} {083504} (\bibinfo {year} {2020})}\BibitemShut {NoStop}%
\bibitem [{\citenamefont {Friedrich}\ \emph {et~al.}(2021)\citenamefont {Friedrich} \emph {et~al.}}]{beest2021phase2}%
  \BibitemOpen
  \bibfield  {author} {\bibinfo {author} {\bibfnamefont {S.}~\bibnamefont {Friedrich}} \emph {et~al.} (\bibinfo {collaboration} {BeEST Collaboration}),\ }\bibfield  {title} {\bibinfo {title} {{Limits on the Existence of sub-MeV Sterile Neutrinos from the Decay of $^{7}\mathrm{Be}$ in Superconducting Quantum Sensors}},\ }\href {https://doi.org/10.1103/PhysRevLett.126.021803} {\bibfield  {journal} {\bibinfo  {journal} {Phys. Rev. Lett.}\ }\textbf {\bibinfo {volume} {126}},\ \bibinfo {pages} {021803} (\bibinfo {year} {2021})}\BibitemShut {NoStop}%
\bibitem [{\citenamefont {Guerra}\ \emph {et~al.}()\citenamefont {Guerra} \emph {et~al.}}]{guerra2024shaking}%
  \BibitemOpen
  \bibfield  {author} {\bibinfo {author} {\bibfnamefont {M.}~\bibnamefont {Guerra}} \emph {et~al.} (\bibinfo {collaboration} {BeEST Collaboration}),\ }\href@noop {} {\bibinfo {title} {{Shake-up and Shake-off effects in the electron capture decay of $^7$Be (to be published)}}}\BibitemShut {NoStop}%
\bibitem [{\citenamefont {{Star Cryoelectronics}}()}]{starcryo}%
  \BibitemOpen
  \bibfield  {author} {\bibinfo {author} {\bibnamefont {{Star Cryoelectronics}}},\ }\href@noop {} {}\bibinfo {note} {\url{https://starcryo.com/}}\BibitemShut {NoStop}%
\bibitem [{\citenamefont {Fretwell}\ \emph {et~al.}(2020)\citenamefont {Fretwell} \emph {et~al.}}]{beest2020lkratio}%
  \BibitemOpen
  \bibfield  {author} {\bibinfo {author} {\bibfnamefont {S.}~\bibnamefont {Fretwell}} \emph {et~al.} (\bibinfo {collaboration} {BeEST Collaboration}),\ }\bibfield  {title} {\bibinfo {title} {{Direct Measurement of the $^{7}\mathrm{Be}$ $L/K$ Capture Ratio in Ta-Based Superconducting Tunnel Junctions}},\ }\href {https://doi.org/10.1103/PhysRevLett.125.032701} {\bibfield  {journal} {\bibinfo  {journal} {Phys. Rev. Lett.}\ }\textbf {\bibinfo {volume} {125}},\ \bibinfo {pages} {032701} (\bibinfo {year} {2020})}\BibitemShut {NoStop}%
\bibitem [{\citenamefont {Dilling}\ \emph {et~al.}(2014)\citenamefont {Dilling}, \citenamefont {Kr{\"u}cken},\ and\ \citenamefont {Merminga}}]{dilling2014isac}%
  \BibitemOpen
  \bibfield  {author} {\bibinfo {author} {\bibfnamefont {J.}~\bibnamefont {Dilling}}, \bibinfo {author} {\bibfnamefont {R.}~\bibnamefont {Kr{\"u}cken}},\ and\ \bibinfo {author} {\bibfnamefont {L.}~\bibnamefont {Merminga}},\ }\href@noop {} {\emph {\bibinfo {title} {ISAC and ARIEL: the TRIUMF radioactive beam facilities and the scientific program}}}\ (\bibinfo  {publisher} {Springer},\ \bibinfo {year} {2014})\BibitemShut {NoStop}%
\bibitem [{\citenamefont {Raeder}\ \emph {et~al.}(2014)\citenamefont {Raeder}, \citenamefont {Heggen}, \citenamefont {Lassen}, \citenamefont {Ames}, \citenamefont {Bishop}, \citenamefont {Bricault}, \citenamefont {Kunz}, \citenamefont {Mjøs},\ and\ \citenamefont {Teigelhöfer}}]{raeder2014iglis}%
  \BibitemOpen
  \bibfield  {author} {\bibinfo {author} {\bibfnamefont {S.}~\bibnamefont {Raeder}}, \bibinfo {author} {\bibfnamefont {H.}~\bibnamefont {Heggen}}, \bibinfo {author} {\bibfnamefont {J.}~\bibnamefont {Lassen}}, \bibinfo {author} {\bibfnamefont {F.}~\bibnamefont {Ames}}, \bibinfo {author} {\bibfnamefont {D.}~\bibnamefont {Bishop}}, \bibinfo {author} {\bibfnamefont {P.}~\bibnamefont {Bricault}}, \bibinfo {author} {\bibfnamefont {P.}~\bibnamefont {Kunz}}, \bibinfo {author} {\bibfnamefont {A.}~\bibnamefont {Mjøs}},\ and\ \bibinfo {author} {\bibfnamefont {A.}~\bibnamefont {Teigelhöfer}},\ }\bibfield  {title} {\bibinfo {title} {{An ion guide laser ion source for isobar-suppressed rare isotope beams}},\ }\href {https://doi.org/10.1063/1.4868496} {\bibfield  {journal} {\bibinfo  {journal} {Rev. Sci. Instrum.}\ }\textbf {\bibinfo {volume} {85}},\ \bibinfo {pages} {033309} (\bibinfo {year} {2014})}\BibitemShut {NoStop}%
\bibitem [{\citenamefont {Mostamand}\ \emph {et~al.}(2020)\citenamefont {Mostamand}, \citenamefont {Li}, \citenamefont {Romans}, \citenamefont {Ames}, \citenamefont {Kunz}, \citenamefont {Mj$\phi$s},\ and\ \citenamefont {Lassen}}]{mostamand2020production}%
  \BibitemOpen
  \bibfield  {author} {\bibinfo {author} {\bibfnamefont {M.}~\bibnamefont {Mostamand}}, \bibinfo {author} {\bibfnamefont {R.}~\bibnamefont {Li}}, \bibinfo {author} {\bibfnamefont {J.}~\bibnamefont {Romans}}, \bibinfo {author} {\bibfnamefont {F.}~\bibnamefont {Ames}}, \bibinfo {author} {\bibfnamefont {P.}~\bibnamefont {Kunz}}, \bibinfo {author} {\bibfnamefont {A.}~\bibnamefont {Mj$\phi$s}},\ and\ \bibinfo {author} {\bibfnamefont {J.}~\bibnamefont {Lassen}},\ }\bibfield  {title} {\bibinfo {title} {Production of clean rare isotope beams at {TRIUMF} ion guide laser ion source},\ }\href {https://doi.org/10.1007/s10751-020-1704-6} {\bibfield  {journal} {\bibinfo  {journal} {Hyperfine Interact.}\ }\textbf {\bibinfo {volume} {241}},\ \bibinfo {pages} {36} (\bibinfo {year} {2020})}\BibitemShut {NoStop}%
\bibitem [{\citenamefont {Lennarz}\ \emph {et~al.}()\citenamefont {Lennarz} \emph {et~al.}}]{lennarz2023implantation}%
  \BibitemOpen
  \bibfield  {author} {\bibinfo {author} {\bibfnamefont {A.}~\bibnamefont {Lennarz}} \emph {et~al.} (\bibinfo {collaboration} {BeEST Collaboration}),\ }\href@noop {} {\bibinfo {title} {{BeEST implantation (to be published)}}}\BibitemShut {NoStop}%
\bibitem [{\citenamefont {Ziegler}\ \emph {et~al.}(2010)\citenamefont {Ziegler}, \citenamefont {Ziegler},\ and\ \citenamefont {Biersack}}]{srim2010}%
  \BibitemOpen
  \bibfield  {author} {\bibinfo {author} {\bibfnamefont {J.~F.}\ \bibnamefont {Ziegler}}, \bibinfo {author} {\bibfnamefont {M.}~\bibnamefont {Ziegler}},\ and\ \bibinfo {author} {\bibfnamefont {J.}~\bibnamefont {Biersack}},\ }\bibfield  {title} {\bibinfo {title} {{SRIM} – the stopping and range of ions in matter (2010)},\ }\href {https://doi.org/https://doi.org/10.1016/j.nimb.2010.02.091} {\bibfield  {journal} {\bibinfo  {journal} {Nucl. Instrum. Methods Phys. Res. B.}\ }\textbf {\bibinfo {volume} {268}},\ \bibinfo {pages} {1818} (\bibinfo {year} {2010})},\ \bibinfo {note} {19th International Conference on Ion Beam Analysis}\BibitemShut {NoStop}%
\bibitem [{\citenamefont {Warburton}\ \emph {et~al.}(2015)\citenamefont {Warburton}, \citenamefont {Harris},\ and\ \citenamefont {Friedrich}}]{warburton2015xia}%
  \BibitemOpen
  \bibfield  {author} {\bibinfo {author} {\bibfnamefont {W.}~\bibnamefont {Warburton}}, \bibinfo {author} {\bibfnamefont {J.}~\bibnamefont {Harris}},\ and\ \bibinfo {author} {\bibfnamefont {S.}~\bibnamefont {Friedrich}},\ }\bibfield  {title} {\bibinfo {title} {High density processing electronics for superconducting tunnel junction x-ray detector arrays},\ }\href {https://doi.org/https://doi.org/10.1016/j.nima.2015.02.004} {\bibfield  {journal} {\bibinfo  {journal} {Nucl. Instrum. Methods Phys. Res. A.}\ }\textbf {\bibinfo {volume} {784}},\ \bibinfo {pages} {236} (\bibinfo {year} {2015})},\ \bibinfo {note} {{Symposium on Radiation Measurements and Applications 2014 (SORMA XV)}}\BibitemShut {NoStop}%
\bibitem [{\citenamefont {Kurakado}(1982)}]{kurakado1982stj}%
  \BibitemOpen
  \bibfield  {author} {\bibinfo {author} {\bibfnamefont {M.}~\bibnamefont {Kurakado}},\ }\bibfield  {title} {\bibinfo {title} {Possibility of high resolution detectors using superconducting tunnel junctions},\ }\href {https://doi.org/https://doi.org/10.1016/0029-554X(82)90654-1} {\bibfield  {journal} {\bibinfo  {journal} {Nucl. Instrum. Methods Phys. Res.}\ }\textbf {\bibinfo {volume} {196}},\ \bibinfo {pages} {275} (\bibinfo {year} {1982})}\BibitemShut {NoStop}%
\bibitem [{\citenamefont {Ponce}\ \emph {et~al.}(2018)\citenamefont {Ponce}, \citenamefont {Swanberg}, \citenamefont {Burke}, \citenamefont {Henderson},\ and\ \citenamefont {Friedrich}}]{ponce2018u235}%
  \BibitemOpen
  \bibfield  {author} {\bibinfo {author} {\bibfnamefont {F.}~\bibnamefont {Ponce}}, \bibinfo {author} {\bibfnamefont {E.}~\bibnamefont {Swanberg}}, \bibinfo {author} {\bibfnamefont {J.}~\bibnamefont {Burke}}, \bibinfo {author} {\bibfnamefont {R.}~\bibnamefont {Henderson}},\ and\ \bibinfo {author} {\bibfnamefont {S.}~\bibnamefont {Friedrich}},\ }\bibfield  {title} {\bibinfo {title} {Accurate measurement of the first excited nuclear state in $^{235}\mathrm{U}$},\ }\href {https://doi.org/10.1103/PhysRevC.97.054310} {\bibfield  {journal} {\bibinfo  {journal} {Phys. Rev. C}\ }\textbf {\bibinfo {volume} {97}},\ \bibinfo {pages} {054310} (\bibinfo {year} {2018})}\BibitemShut {NoStop}%
\bibitem [{\citenamefont {Bray}\ \emph {et~al.}(2024{\natexlab{a}})\citenamefont {Bray} \emph {et~al.}}]{bray2024daq}%
  \BibitemOpen
  \bibfield  {author} {\bibinfo {author} {\bibfnamefont {C.}~\bibnamefont {Bray}} \emph {et~al.} (\bibinfo {collaboration} {BeEST Collaboration}),\ }\bibfield  {title} {\bibinfo {title} {{The Data Acquisition System for Phase-III of the BeEST Experiment}},\ }\bibfield  {journal} {\bibinfo  {journal} {J. Low Temp. Phys.}\ }\href {https://doi.org/10.1007/s10909-024-03242-7} {10.1007/s10909-024-03242-7} (\bibinfo {year} {2024}{\natexlab{a}})\BibitemShut {NoStop}%
\bibitem [{\citenamefont {Wigmore}\ \emph {et~al.}(2004)\citenamefont {Wigmore}, \citenamefont {Boyd}, \citenamefont {Steele}, \citenamefont {Kozorezov},\ and\ \citenamefont {Bradley}}]{wigmore2004fluxtrapping}%
  \BibitemOpen
  \bibfield  {author} {\bibinfo {author} {\bibfnamefont {J.}~\bibnamefont {Wigmore}}, \bibinfo {author} {\bibfnamefont {P.}~\bibnamefont {Boyd}}, \bibinfo {author} {\bibfnamefont {A.}~\bibnamefont {Steele}}, \bibinfo {author} {\bibfnamefont {A.}~\bibnamefont {Kozorezov}},\ and\ \bibinfo {author} {\bibfnamefont {D.}~\bibnamefont {Bradley}},\ }\bibfield  {title} {\bibinfo {title} {Phonon excitation of quasiparticles in niobium and tantalum superconducting tunnel junction photon detectors},\ }\href {https://doi.org/https://doi.org/10.1016/j.nima.2003.11.340} {\bibfield  {journal} {\bibinfo  {journal} {Nucl. Instrum. Methods Phys. Res. A.}\ }\textbf {\bibinfo {volume} {520}},\ \bibinfo {pages} {263} (\bibinfo {year} {2004})},\ \bibinfo {note} {{Proceedings of the 10th International Workshop on Low Temperature Detectors}}\BibitemShut {NoStop}%
\bibitem [{\citenamefont {Hiller}\ \emph {et~al.}(2001)\citenamefont {Hiller}, \citenamefont {van~den Berg},\ and\ \citenamefont {Labov}}]{hiller2001noise}%
  \BibitemOpen
  \bibfield  {author} {\bibinfo {author} {\bibfnamefont {L.~J.}\ \bibnamefont {Hiller}}, \bibinfo {author} {\bibfnamefont {M.~L.}\ \bibnamefont {van~den Berg}},\ and\ \bibinfo {author} {\bibfnamefont {S.~E.}\ \bibnamefont {Labov}},\ }\bibfield  {title} {\bibinfo {title} {{Multiple-tunneling noise in superconducting tunnel junctions from partial current integration}},\ }\href {https://doi.org/10.1063/1.1428618} {\bibfield  {journal} {\bibinfo  {journal} {Appl. Phys. Lett.}\ }\textbf {\bibinfo {volume} {79}},\ \bibinfo {pages} {4441} (\bibinfo {year} {2001})}\BibitemShut {NoStop}%
\bibitem [{\citenamefont {Samedov}(2000)}]{samedof2000multitunneling}%
  \BibitemOpen
  \bibfield  {author} {\bibinfo {author} {\bibfnamefont {V.~V.}\ \bibnamefont {Samedov}},\ }\bibfield  {title} {\bibinfo {title} {Superconducting tunnel junction signal caused by quasielectron multitunneling},\ }\href {https://doi.org/https://doi.org/10.1016/S0168-9002(99)01328-5} {\bibfield  {journal} {\bibinfo  {journal} {Nuclear Instruments and Methods in Physics Research Section A: Accelerators, Spectrometers, Detectors and Associated Equipment}\ }\textbf {\bibinfo {volume} {444}},\ \bibinfo {pages} {59} (\bibinfo {year} {2000})}\BibitemShut {NoStop}%
\bibitem [{\citenamefont {Jordanov}\ \emph {et~al.}(1994)\citenamefont {Jordanov}, \citenamefont {Knoll}, \citenamefont {Huber},\ and\ \citenamefont {Pantazis}}]{jordanov1994digital}%
  \BibitemOpen
  \bibfield  {author} {\bibinfo {author} {\bibfnamefont {V.~T.}\ \bibnamefont {Jordanov}}, \bibinfo {author} {\bibfnamefont {G.~F.}\ \bibnamefont {Knoll}}, \bibinfo {author} {\bibfnamefont {A.~C.}\ \bibnamefont {Huber}},\ and\ \bibinfo {author} {\bibfnamefont {J.~A.}\ \bibnamefont {Pantazis}},\ }\bibfield  {title} {\bibinfo {title} {Digital techniques for real-time pulse shaping in radiation measurements},\ }\href {https://doi.org/https://doi.org/10.1016/0168-9002(94)91652-7} {\bibfield  {journal} {\bibinfo  {journal} {Nucl. Instrum. Methods Phys. Res. A.}\ }\textbf {\bibinfo {volume} {353}},\ \bibinfo {pages} {261} (\bibinfo {year} {1994})}\BibitemShut {NoStop}%
\bibitem [{\citenamefont {Tilley}\ \emph {et~al.}(2002)\citenamefont {Tilley}, \citenamefont {Cheves}, \citenamefont {Godwin}, \citenamefont {Hale}, \citenamefont {Hofmann}, \citenamefont {Kelley}, \citenamefont {Sheu},\ and\ \citenamefont {Weller}}]{nudat2002be7}%
  \BibitemOpen
  \bibfield  {author} {\bibinfo {author} {\bibfnamefont {D.}~\bibnamefont {Tilley}}, \bibinfo {author} {\bibfnamefont {C.}~\bibnamefont {Cheves}}, \bibinfo {author} {\bibfnamefont {J.}~\bibnamefont {Godwin}}, \bibinfo {author} {\bibfnamefont {G.}~\bibnamefont {Hale}}, \bibinfo {author} {\bibfnamefont {H.}~\bibnamefont {Hofmann}}, \bibinfo {author} {\bibfnamefont {J.}~\bibnamefont {Kelley}}, \bibinfo {author} {\bibfnamefont {C.}~\bibnamefont {Sheu}},\ and\ \bibinfo {author} {\bibfnamefont {H.}~\bibnamefont {Weller}},\ }\bibfield  {title} {\bibinfo {title} {Energy levels of light nuclei {A}=5, 6, 7},\ }\href {https://doi.org/https://doi.org/10.1016/S0375-9474(02)00597-3} {\bibfield  {journal} {\bibinfo  {journal} {Nucl. Phys. A}\ }\textbf {\bibinfo {volume} {708}},\ \bibinfo {pages} {3} (\bibinfo {year} {2002})}\BibitemShut {NoStop}%
\bibitem [{\citenamefont {Kaplan}(1979)}]{kaplan1979acoustic}%
  \BibitemOpen
  \bibfield  {author} {\bibinfo {author} {\bibfnamefont {S.~B.}\ \bibnamefont {Kaplan}},\ }\bibfield  {title} {\bibinfo {title} {Acoustic matching of superconducting films to substrates},\ }\href {https://doi.org/10.1007/BF00119193} {\bibfield  {journal} {\bibinfo  {journal} {J. Low Temp. Phys.}\ }\textbf {\bibinfo {volume} {37}},\ \bibinfo {pages} {343} (\bibinfo {year} {1979})}\BibitemShut {NoStop}%
\bibitem [{\citenamefont {Audi}\ \emph {et~al.}(2012)\citenamefont {Audi}, \citenamefont {Wapstra}, \citenamefont {Kondev}, \citenamefont {MacCormick}, \citenamefont {Xu},\ and\ \citenamefont {Pfeiffer}}]{audi2012ame2012}%
  \BibitemOpen
  \bibfield  {author} {\bibinfo {author} {\bibfnamefont {G.}~\bibnamefont {Audi}}, \bibinfo {author} {\bibfnamefont {A.}~\bibnamefont {Wapstra}}, \bibinfo {author} {\bibfnamefont {F.}~\bibnamefont {Kondev}}, \bibinfo {author} {\bibfnamefont {M.}~\bibnamefont {MacCormick}}, \bibinfo {author} {\bibfnamefont {X.}~\bibnamefont {Xu}},\ and\ \bibinfo {author} {\bibfnamefont {B.}~\bibnamefont {Pfeiffer}},\ }\bibfield  {title} {\bibinfo {title} {{The AME2012 atomic mass evaluation (II). Tables, graphs and references}},\ }\href {https://doi.org/10.1088/1674-1137/36/12/003} {\bibfield  {journal} {\bibinfo  {journal} {CPC}\ }\textbf {\bibinfo {volume} {36}},\ \bibinfo {pages} {1603} (\bibinfo {year} {2012})}\BibitemShut {NoStop}%
\bibitem [{\citenamefont {Bhandari}\ \emph {et~al.}(2024)\citenamefont {Bhandari}, \citenamefont {Bollen}, \citenamefont {Brunner}, \citenamefont {Gamage}, \citenamefont {Hamaker}, \citenamefont {Hockenbery}, \citenamefont {Gamage}, \citenamefont {Keblbeck}, \citenamefont {Leach}, \citenamefont {Puentes} \emph {et~al.}}]{bhandari2024q}%
  \BibitemOpen
  \bibfield  {author} {\bibinfo {author} {\bibfnamefont {R.}~\bibnamefont {Bhandari}}, \bibinfo {author} {\bibfnamefont {G.}~\bibnamefont {Bollen}}, \bibinfo {author} {\bibfnamefont {T.}~\bibnamefont {Brunner}}, \bibinfo {author} {\bibfnamefont {N.~D.}\ \bibnamefont {Gamage}}, \bibinfo {author} {\bibfnamefont {A.}~\bibnamefont {Hamaker}}, \bibinfo {author} {\bibfnamefont {Z.}~\bibnamefont {Hockenbery}}, \bibinfo {author} {\bibfnamefont {M.~H.}\ \bibnamefont {Gamage}}, \bibinfo {author} {\bibfnamefont {D.~K.}\ \bibnamefont {Keblbeck}}, \bibinfo {author} {\bibfnamefont {K.~G.}\ \bibnamefont {Leach}}, \bibinfo {author} {\bibfnamefont {D.}~\bibnamefont {Puentes}}, \emph {et~al.},\ }\bibfield  {title} {\bibinfo {title} {First direct $^{7}\mathrm{Be}$ electron-capture $q$-value measurement toward high-precision searches for neutrino physics beyond the standard model},\ }\href {https://doi.org/10.1103/PhysRevC.109.L022501} {\bibfield  {journal} {\bibinfo  {journal} {Phys. Rev. C}\ }\textbf {\bibinfo {volume}
  {109}},\ \bibinfo {pages} {L022501} (\bibinfo {year} {2024})}\BibitemShut {NoStop}%
\bibitem [{\citenamefont {Bray}\ \emph {et~al.}(2024{\natexlab{b}})\citenamefont {Bray} \emph {et~al.}}]{bray2024doppler}%
  \BibitemOpen
  \bibfield  {author} {\bibinfo {author} {\bibfnamefont {C.}~\bibnamefont {Bray}} \emph {et~al.} (\bibinfo {collaboration} {BeEST Collaboration}),\ }\bibfield  {title} {\bibinfo {title} {High-precision excited-state nuclear recoil spectroscopy with superconducting sensors},\ }\href {https://doi.org/10.48550/arXiv.2411.08076} {\bibfield  {journal} {\bibinfo  {journal} {arXiv:2411.08076}\ } (\bibinfo {year} {2024}{\natexlab{b}})}\BibitemShut {NoStop}%
\bibitem [{\citenamefont {Zehnder}(1995)}]{zehnder1995response}%
  \BibitemOpen
  \bibfield  {author} {\bibinfo {author} {\bibfnamefont {A.}~\bibnamefont {Zehnder}},\ }\bibfield  {title} {\bibinfo {title} {Response of superconductive films to localized energy deposition},\ }\href {https://doi.org/10.1103/PhysRevB.52.12858} {\bibfield  {journal} {\bibinfo  {journal} {Phys. Rev. B}\ }\textbf {\bibinfo {volume} {52}},\ \bibinfo {pages} {12858} (\bibinfo {year} {1995})}\BibitemShut {NoStop}%
\bibitem [{\citenamefont {Bearden}\ and\ \citenamefont {Burr}(1967)}]{bearden1967energylevel}%
  \BibitemOpen
  \bibfield  {author} {\bibinfo {author} {\bibfnamefont {J.~A.}\ \bibnamefont {Bearden}}\ and\ \bibinfo {author} {\bibfnamefont {A.~F.}\ \bibnamefont {Burr}},\ }\bibfield  {title} {\bibinfo {title} {Reevaluation of x-ray atomic energy levels},\ }\href {https://doi.org/10.1103/RevModPhys.39.125} {\bibfield  {journal} {\bibinfo  {journal} {Rev. Mod. Phys.}\ }\textbf {\bibinfo {volume} {39}},\ \bibinfo {pages} {125} (\bibinfo {year} {1967})}\BibitemShut {NoStop}%
\bibitem [{\citenamefont {Samanta}\ \emph {et~al.}(2023)\citenamefont {Samanta}, \citenamefont {Friedrich}, \citenamefont {Leach},\ and\ \citenamefont {Lordi}}]{samanta2023material}%
  \BibitemOpen
  \bibfield  {author} {\bibinfo {author} {\bibfnamefont {A.}~\bibnamefont {Samanta}}, \bibinfo {author} {\bibfnamefont {S.}~\bibnamefont {Friedrich}}, \bibinfo {author} {\bibfnamefont {K.~G.}\ \bibnamefont {Leach}},\ and\ \bibinfo {author} {\bibfnamefont {V.}~\bibnamefont {Lordi}},\ }\bibfield  {title} {\bibinfo {title} {Material effects on electron-capture decay in cryogenic sensors},\ }\href {https://doi.org/10.1103/PhysRevApplied.19.014032} {\bibfield  {journal} {\bibinfo  {journal} {Phys. Rev. Appl.}\ }\textbf {\bibinfo {volume} {19}},\ \bibinfo {pages} {014032} (\bibinfo {year} {2023})}\BibitemShut {NoStop}%
\bibitem [{\citenamefont {Citrin}\ and\ \citenamefont {Hamann}(1977)}]{citrin1977broadening}%
  \BibitemOpen
  \bibfield  {author} {\bibinfo {author} {\bibfnamefont {P.~H.}\ \bibnamefont {Citrin}}\ and\ \bibinfo {author} {\bibfnamefont {D.~R.}\ \bibnamefont {Hamann}},\ }\bibfield  {title} {\bibinfo {title} {Phonon broadening of x-ray photoemission line shapes in solids and its independence of hole state lifetimes},\ }\href {https://doi.org/10.1103/PhysRevB.15.2923} {\bibfield  {journal} {\bibinfo  {journal} {Phys. Rev. B}\ }\textbf {\bibinfo {volume} {15}},\ \bibinfo {pages} {2923} (\bibinfo {year} {1977})}\BibitemShut {NoStop}%
\bibitem [{\citenamefont {Bray}\ \emph {et~al.}(2022)\citenamefont {Bray}, \citenamefont {Hiller}, \citenamefont {Leach},\ and\ \citenamefont {Friedrich}}]{bray2022monte}%
  \BibitemOpen
  \bibfield  {author} {\bibinfo {author} {\bibfnamefont {C.~E.}\ \bibnamefont {Bray}}, \bibinfo {author} {\bibfnamefont {L.~J.}\ \bibnamefont {Hiller}}, \bibinfo {author} {\bibfnamefont {K.~G.}\ \bibnamefont {Leach}},\ and\ \bibinfo {author} {\bibfnamefont {S.}~\bibnamefont {Friedrich}},\ }\bibfield  {title} {\bibinfo {title} {{Monte Carlo Simulations of Superconducting Tunnel Junction Quantum Sensors for the BeEST Experiment}},\ }\href {https://doi.org/10.1007/s10909-022-02770-4} {\bibfield  {journal} {\bibinfo  {journal} {J. Low Temp. Phys.}\ }\textbf {\bibinfo {volume} {209}},\ \bibinfo {pages} {857} (\bibinfo {year} {2022})}\BibitemShut {NoStop}%
\bibitem [{\citenamefont {Levinger}(1953)}]{levinger1953electrons}%
  \BibitemOpen
  \bibfield  {author} {\bibinfo {author} {\bibfnamefont {J.~S.}\ \bibnamefont {Levinger}},\ }\bibfield  {title} {\bibinfo {title} {Effects of radioactive disintegrations on inner electrons of the atom},\ }\href {https://doi.org/10.1103/PhysRev.90.11} {\bibfield  {journal} {\bibinfo  {journal} {Phys. Rev.}\ }\textbf {\bibinfo {volume} {90}},\ \bibinfo {pages} {11} (\bibinfo {year} {1953})}\BibitemShut {NoStop}%
\bibitem [{\citenamefont {Robertson}\ and\ \citenamefont {Venkatapathy}(2020)}]{robertson2020shake}%
  \BibitemOpen
  \bibfield  {author} {\bibinfo {author} {\bibfnamefont {R.~G.~H.}\ \bibnamefont {Robertson}}\ and\ \bibinfo {author} {\bibfnamefont {V.}~\bibnamefont {Venkatapathy}},\ }\bibfield  {title} {\bibinfo {title} {Shakeup and shakeoff satellite structure in the electron spectrum of $^{83}\mathrm{Kr}^{m}$},\ }\href {https://doi.org/10.1103/PhysRevC.102.035502} {\bibfield  {journal} {\bibinfo  {journal} {Phys. Rev. C}\ }\textbf {\bibinfo {volume} {102}},\ \bibinfo {pages} {035502} (\bibinfo {year} {2020})}\BibitemShut {NoStop}%
\bibitem [{\citenamefont {Guo}\ and\ \citenamefont {Robertson}(2014)}]{guo2014defect}%
  \BibitemOpen
  \bibfield  {author} {\bibinfo {author} {\bibfnamefont {Y.}~\bibnamefont {Guo}}\ and\ \bibinfo {author} {\bibfnamefont {J.}~\bibnamefont {Robertson}},\ }\bibfield  {title} {\bibinfo {title} {{Oxygen vacancy defects in Ta$_2$O$_5$ showing long-range atomic re-arrangements}},\ }\href {https://doi.org/10.1063/1.4869553} {\bibfield  {journal} {\bibinfo  {journal} {Appl. Phys. Lett.}\ }\textbf {\bibinfo {volume} {104}},\ \bibinfo {pages} {112906} (\bibinfo {year} {2014})}\BibitemShut {NoStop}%
\bibitem [{\citenamefont {Dembinski}\ \emph {et~al.}(2020)\citenamefont {Dembinski}, \citenamefont {Ongmongkolkul}, \citenamefont {Deli}, \citenamefont {Schreiner}, \citenamefont {Feickert}, \citenamefont {Burr}, \citenamefont {Watson}, \citenamefont {Rost}, \citenamefont {Pearce}, \citenamefont {Geiger} \emph {et~al.}}]{iminuit}%
  \BibitemOpen
  \bibfield  {author} {\bibinfo {author} {\bibfnamefont {H.}~\bibnamefont {Dembinski}}, \bibinfo {author} {\bibfnamefont {P.}~\bibnamefont {Ongmongkolkul}}, \bibinfo {author} {\bibfnamefont {C.}~\bibnamefont {Deli}}, \bibinfo {author} {\bibfnamefont {H.}~\bibnamefont {Schreiner}}, \bibinfo {author} {\bibfnamefont {M.}~\bibnamefont {Feickert}}, \bibinfo {author} {\bibfnamefont {C.}~\bibnamefont {Burr}}, \bibinfo {author} {\bibfnamefont {J.}~\bibnamefont {Watson}}, \bibinfo {author} {\bibfnamefont {F.}~\bibnamefont {Rost}}, \bibinfo {author} {\bibfnamefont {A.}~\bibnamefont {Pearce}}, \bibinfo {author} {\bibfnamefont {L.}~\bibnamefont {Geiger}}, \emph {et~al.},\ }\bibfield  {title} {\bibinfo {title} {scikit-hep/iminuit}\ }\href {https://doi.org/10.5281/zenodo.3949207} {10.5281/zenodo.3949207} (\bibinfo {year} {2020})\BibitemShut {NoStop}%
\bibitem [{\citenamefont {{Borax high-performance computing cluster}}()}]{borax}%
  \BibitemOpen
  \bibfield  {author} {\bibinfo {author} {\bibnamefont {{Borax high-performance computing cluster}}},\ }\href@noop {} {}\bibinfo {note} {\url{https://hpc.llnl.gov/hardware/compute-platforms/borax}}\BibitemShut {NoStop}%
\bibitem [{\citenamefont {Voytas}\ \emph {et~al.}(2001)\citenamefont {Voytas}, \citenamefont {Ternovan}, \citenamefont {Galeazzi}, \citenamefont {McCammon}, \citenamefont {Kolata}, \citenamefont {Santi}, \citenamefont {Peterson}, \citenamefont {Guimar\~aes}, \citenamefont {Becchetti}, \citenamefont {Lee} \emph {et~al.}}]{voytas2001lkratio}%
  \BibitemOpen
  \bibfield  {author} {\bibinfo {author} {\bibfnamefont {P.~A.}\ \bibnamefont {Voytas}}, \bibinfo {author} {\bibfnamefont {C.}~\bibnamefont {Ternovan}}, \bibinfo {author} {\bibfnamefont {M.}~\bibnamefont {Galeazzi}}, \bibinfo {author} {\bibfnamefont {D.}~\bibnamefont {McCammon}}, \bibinfo {author} {\bibfnamefont {J.~J.}\ \bibnamefont {Kolata}}, \bibinfo {author} {\bibfnamefont {P.}~\bibnamefont {Santi}}, \bibinfo {author} {\bibfnamefont {D.}~\bibnamefont {Peterson}}, \bibinfo {author} {\bibfnamefont {V.}~\bibnamefont {Guimar\~aes}}, \bibinfo {author} {\bibfnamefont {F.~D.}\ \bibnamefont {Becchetti}}, \bibinfo {author} {\bibfnamefont {M.~Y.}\ \bibnamefont {Lee}}, \emph {et~al.},\ }\bibfield  {title} {\bibinfo {title} {{Direct Measurement of the $\mathit{L}/\mathit{K}$ Ratio in $^{7}$Be Electron Capture}},\ }\href {https://doi.org/10.1103/PhysRevLett.88.012501} {\bibfield  {journal} {\bibinfo  {journal} {Phys. Rev. Lett.}\ }\textbf {\bibinfo {volume} {88}},\ \bibinfo {pages} {012501} (\bibinfo {year}
  {2001})}\BibitemShut {NoStop}%
\bibitem [{\citenamefont {Bra\ss{}}\ \emph {et~al.}(2018)\citenamefont {Bra\ss{}}, \citenamefont {Enss}, \citenamefont {Gastaldo}, \citenamefont {Green},\ and\ \citenamefont {Haverkort}}]{echo2018shake}%
  \BibitemOpen
  \bibfield  {author} {\bibinfo {author} {\bibfnamefont {M.}~\bibnamefont {Bra\ss{}}}, \bibinfo {author} {\bibfnamefont {C.}~\bibnamefont {Enss}}, \bibinfo {author} {\bibfnamefont {L.}~\bibnamefont {Gastaldo}}, \bibinfo {author} {\bibfnamefont {R.~J.}\ \bibnamefont {Green}},\ and\ \bibinfo {author} {\bibfnamefont {M.~W.}\ \bibnamefont {Haverkort}},\ }\bibfield  {title} {\bibinfo {title} {\textit{Ab initio} calculation of the calorimetric electron-capture spectrum of $^{163}\mathrm{Ho}$: Intra-atomic decay into bound states},\ }\href {https://doi.org/10.1103/PhysRevC.97.054620} {\bibfield  {journal} {\bibinfo  {journal} {Phys. Rev. C}\ }\textbf {\bibinfo {volume} {97}},\ \bibinfo {pages} {054620} (\bibinfo {year} {2018})}\BibitemShut {NoStop}%
\bibitem [{\citenamefont {Holzschuh}\ \emph {et~al.}(1999)\citenamefont {Holzschuh}, \citenamefont {Kündig}, \citenamefont {Palermo}, \citenamefont {Stüssi},\ and\ \citenamefont {Wenk}}]{Ni1999limit}%
  \BibitemOpen
  \bibfield  {author} {\bibinfo {author} {\bibfnamefont {E.}~\bibnamefont {Holzschuh}}, \bibinfo {author} {\bibfnamefont {W.}~\bibnamefont {Kündig}}, \bibinfo {author} {\bibfnamefont {L.}~\bibnamefont {Palermo}}, \bibinfo {author} {\bibfnamefont {H.}~\bibnamefont {Stüssi}},\ and\ \bibinfo {author} {\bibfnamefont {P.}~\bibnamefont {Wenk}},\ }\bibfield  {title} {\bibinfo {title} {{Search for heavy neutrinos in the $\beta$-spectrum of $^{63}$Ni}},\ }\href {https://doi.org/https://doi.org/10.1016/S0370-2693(99)00200-2} {\bibfield  {journal} {\bibinfo  {journal} {Physics Letters B}\ }\textbf {\bibinfo {volume} {451}},\ \bibinfo {pages} {247} (\bibinfo {year} {1999})}\BibitemShut {NoStop}%
\bibitem [{\citenamefont {Holzschuh}\ \emph {et~al.}(2000)\citenamefont {Holzschuh}, \citenamefont {Palermo}, \citenamefont {Stüssi},\ and\ \citenamefont {Wenk}}]{S2000limit}%
  \BibitemOpen
  \bibfield  {author} {\bibinfo {author} {\bibfnamefont {E.}~\bibnamefont {Holzschuh}}, \bibinfo {author} {\bibfnamefont {L.}~\bibnamefont {Palermo}}, \bibinfo {author} {\bibfnamefont {H.}~\bibnamefont {Stüssi}},\ and\ \bibinfo {author} {\bibfnamefont {P.}~\bibnamefont {Wenk}},\ }\bibfield  {title} {\bibinfo {title} {{The $\beta$-spectrum of $^{35}$S and search for the admixture of heavy neutrinos}},\ }\href {https://doi.org/https://doi.org/10.1016/S0370-2693(00)00476-7} {\bibfield  {journal} {\bibinfo  {journal} {Physics Letters B}\ }\textbf {\bibinfo {volume} {482}},\ \bibinfo {pages} {1} (\bibinfo {year} {2000})}\BibitemShut {NoStop}%
\bibitem [{\citenamefont {Schreckenbach}\ \emph {et~al.}(1983)\citenamefont {Schreckenbach}, \citenamefont {Colvin},\ and\ \citenamefont {{von Feilitzsch}}}]{Cu1983limit}%
  \BibitemOpen
  \bibfield  {author} {\bibinfo {author} {\bibfnamefont {K.}~\bibnamefont {Schreckenbach}}, \bibinfo {author} {\bibfnamefont {G.}~\bibnamefont {Colvin}},\ and\ \bibinfo {author} {\bibfnamefont {F.}~\bibnamefont {{von Feilitzsch}}},\ }\bibfield  {title} {\bibinfo {title} {{Search for mixing of heavy neutrinos in the $\beta$+ and $\beta$- spectra of the $^{64}$Cu Decay}},\ }\href {https://doi.org/https://doi.org/10.1016/0370-2693(83)90858-4} {\bibfield  {journal} {\bibinfo  {journal} {Physics Letters B}\ }\textbf {\bibinfo {volume} {129}},\ \bibinfo {pages} {265} (\bibinfo {year} {1983})}\BibitemShut {NoStop}%
\bibitem [{\citenamefont {Deutsch}\ \emph {et~al.}(1990)\citenamefont {Deutsch}, \citenamefont {Lebrun},\ and\ \citenamefont {Prieels}}]{F1990limit}%
  \BibitemOpen
  \bibfield  {author} {\bibinfo {author} {\bibfnamefont {J.}~\bibnamefont {Deutsch}}, \bibinfo {author} {\bibfnamefont {M.}~\bibnamefont {Lebrun}},\ and\ \bibinfo {author} {\bibfnamefont {R.}~\bibnamefont {Prieels}},\ }\bibfield  {title} {\bibinfo {title} {Searches for admixture of massive neutrinos into the electron flavour},\ }\href {https://doi.org/https://doi.org/10.1016/0375-9474(90)90541-S} {\bibfield  {journal} {\bibinfo  {journal} {Nuclear Physics A}\ }\textbf {\bibinfo {volume} {518}},\ \bibinfo {pages} {149} (\bibinfo {year} {1990})}\BibitemShut {NoStop}%
\bibitem [{\citenamefont {Bryman}\ and\ \citenamefont {Shrock}(2019)}]{meson2019limit}%
  \BibitemOpen
  \bibfield  {author} {\bibinfo {author} {\bibfnamefont {D.~A.}\ \bibnamefont {Bryman}}\ and\ \bibinfo {author} {\bibfnamefont {R.}~\bibnamefont {Shrock}},\ }\bibfield  {title} {\bibinfo {title} {{Constraints on sterile neutrinos in the MeV to GeV mass range}},\ }\href {https://doi.org/10.1103/PhysRevD.100.073011} {\bibfield  {journal} {\bibinfo  {journal} {Phys. Rev. D}\ }\textbf {\bibinfo {volume} {100}},\ \bibinfo {pages} {073011} (\bibinfo {year} {2019})}\BibitemShut {NoStop}%
\end{thebibliography}%

\end{document}